\documentclass[UTF8,a4paper]{article}
\usepackage{amsmath}
\usepackage{graphicx}
\usepackage{subfigure}
\usepackage{epstopdf}
\usepackage{geometry}
\usepackage{indentfirst}
\usepackage{amsfonts,amssymb}
\usepackage{setspace}
\usepackage{threeparttable}
\usepackage{amsmath,mathtools,amsthm}
\usepackage{extarrows}
\usepackage{upgreek}

\makeatletter
\renewcommand*\env@matrix[1][\arraystretch]{%
  \edef\arraystretch{#1}%
  \hskip -\arraycolsep
  \let\@ifnextchar\new@ifnextchar
  \array{*\c@MaxMatrixCols c}}
\makeatother
\begin{document}


\title{Tight-binding model and $ab\ initio$ calculation of silicene with strong spin-orbit coupling in low-energy limit}
\author{Chen-Huan Wu
\thanks{chenhuanwu1@gmail.com}
\\Key Laboratory of Atomic $\&$ Molecular Physics and Functional Materials of Gansu Province,
\\College of Physics and Electronic Engineering, Northwest Normal University, Lanzhou 730070, China}

\maketitle
\vspace{-30pt}
\begin{abstract}
\begin{large}
We carry out both the tight-binding model and the $ab\ initio$ to study the layered silicene,
the spin, valley, sublattice degrees of freedom are taken into consider 
and the effects of electric field, magnetic field, and even the light in finite frequency together with its interesting optical propertice,
which are all closely related to the spin-orbit coupling and Rashba coupling
and lead to the tunable phase transitions
(between the nontrivial topological phase which has nonzero Chern number or nonzero spin Chern number and the trivial phase).
An exotic quantum anomalous Hall insulator phase are found which has nonzero spin Chern number and nonzero valley
Chern number and as a giant-application-potential spintronic and valleytronics 
in the two-ternimate devices based on the monolayer silicene for the information transmission.
In fact, the gap-out action can be understanded by analyse the Dirac mass as well as the Zeeman splitting or the external-field-induced symmetry-broken,
and the changes of gap has a general nonmonotone-variation characteristic under both the effects of electron filed-induced 
Rashba-coupling $R_{2}(E_{\perp})$ and the electron field-induced band gap,
and the band inversion related to the spin-orbit coupling which absorbs both the spin and orbital angular of momentum may happen
during this process.
The quantized Hall/longitudinal conductivity together with the optical conductivity are also explored,
we see that even in the quantum spin Hall phase without any magnetic field,
the two-terminate conductivity can be reduced to the value $e^{2}/h$ by controlling the helical edge model,
and it can be further reduced to $e^{2}/2h$ by appling the magnetic field which similar to the graphene.

\end {large}
\end{abstract}
\begin{large}

\section{Introduction}

The topological insulator (TI) silicene is investigated in this paper,
which has many attractive properties in the spin and electron transport in the general layer form
or in the nanotube or nanoribbon form which is easy to achieved due to its highly structure anisotropy,
and it has a much large spin-orbit-coupling-induced gap than the graphenes' 
thus generates the robust edge state which is helical in quantum spin Hall phase or chiral in massless case (with gapless Dirac cone),
and it also has a much large bulk gap (about 0.8 eV) in the quantum spin Hall phase as a TI which is much larger than the monolayer Bi$_{4}$Br$_{4}$ and tin film.

We firstly suppose the Wannier orbitals are unorthogonalizable so that the electrons on lattice satisfy the connectivity condition\cite{Misumi T}.
We utilize the different strength of hopping in different directions of the lattice which cause that the wide bands (with obvious dispersion) 
cross with the narrow ones.
It has been experimentally confirmed that such a crossover of wide and narrow bands will leads to the high $T_{c}$ superconductivity\cite{Kuroki K}
due to the resulting discontinuous Fermi surface which is easy to realized by, e.g., the electron doping,
especially for the layered lattice structures materials,
like the tetragonal one (LaFeAsO\cite{Kuroki K2},La$_{2-x}$Ba$_{x}$CuO$_{4}$\cite{Jorgensen J D,Moodenbaugh A R}, LaOBiS$_{2}$\cite{Usui H}) or the honeycomb one
(monolayer silicene\cite{Chen L},undoped bilayer silicene (BLS)\cite{Liu F}).
Through above discuss, we can know that the amplitude of the pair scattering from the wide band to the narrow one is suppressed by the global coherence
when it's enlarged by the strong repulsive interaction (this is also the result of unorthogonal Wannier orbitals).
In fact, once the Hubbard interaction (repulsive here) strength is enhanced to the critical value $U_{c}$,
the increased spin susceptibility will leads to the instability of long-range spin density wave (SDW)\cite{Liu F}.
While for the charge density wave (CDW) with the operator of density of number $n_{A}(n_{B})$,
the order parameter is $\mathcal{G}_{{\rm CDW}}=(1/2R)\sum_{l}^{2R}|(-1)^{l}(n_{Al}+n_{Bl})|$ for the quasi-1D atoms chain in the leg-direction with the number of
rungs $R$ is the number of rungs,
and this CDW also becomes instable by the fluctuation of susceptibility.
Note that the scattering process here doesn't need to obey the conserved laws due to the nonintegrability\cite{Wu C H}.
The superconducting pairs are mediated by the spin-fluctuation with antiferromagnetic (AFM) collinear spin density wave order and the phonon model 
(e.g., the AFM Haldane spin ladder in silicene which play a important role in the SC 
as discussed below), and it's even related to the nontrivial topological phases like the unconventional vortex phase.

In this paper, we investigate the nearest-, second nearest-, and third nearest-neighbor hopping in our honeycomb lattice tight-binding model-silicene,
which we begin with a geometrical study on the planar honeycomb lattice model.
We take the silicene as a explicit example and explore the relations between the properties of high $T_{c}$ superconductivity with these hopping
with different strength in the different bands,
as well as the related topological properties. 
But before that, it's important to clarify the essence of the existence of the narrow band (``flat band")
which has been found in the superfluid or topological insulator (TI) phase and even the vortex-fluid one\cite{Greschner S}.
A mathematical reason is given in the Ref.\cite{Misumi T} which give an explanation from the density matrix\cite{Wu C H}:
if the density matrix of
a bipartite system $V$ is a $m\times n$ matrix with $n$ orthonormal columns and $m<n$, 
where $V$ acts periodic with period of square of number of column $n$,
then the rank of $V$ is smaller than the rank of $m\times m$ one, i.e., the density matrix is reduced. 

\subsection{susceptibility in silicene lattice}
In our local model, since the local correlation can be obtained from the single-site model with the self-consistency condition
which is achieved by integrating out all the other lattice sites,
it's convenient for us to focus only on a single-particle spectrum or the dispersion and therefore the band filling 
(i.e., the number of electrons per unit cell or the spin projection of the single-particle level
which can be well described by the Landau levels in the QHE phase) described by the filling factor $v_{f}$ is also important.
We focus on the fully filling case (two electrons per site for single-band model and
note that the $n$-band model here stands the models which have $n$ maximally localized Wannier orbitals\cite{Souza I} per site) and the 
half-filling case (one electron per site for single-band model).
By using the Euclidean momentum $p$, the transverse susceptibility and longitudinal susceptibility compare to the direction of 
spontaneous magnetization,
have endow the spin-wave model the osillation and renormalized spin-wave velocity (see Appendix.A)\cite{Sachdev S,Chubukov A V}
\begin{equation} 
\begin{aligned}
\chi_{\bot}(k,i\omega)=&\frac{N_{0}^{2}}{\rho_{s}(0)p^{2}},\\
\chi_{\parallel}(k,i\omega)=&\frac{N_{0}^{2}}{\rho_{s}(0)p(p+\frac{16\rho_{s}(0)}{cN})},
\end{aligned}
\end{equation}
where $\rho_{s}(0)$ is the spin stiffness at zero temperature.
$N_{0}$ is the ground state spontaneous magnetization which being widely found in the system with antiferromagnetic singlet pairs (i.e., 
the mixed chirality 
pairs which give rise the topological chiral superconductor (it's weak BCS pairing here and breaks the U(1) conservation) that mediated by strong spin fluctuation 
of AFM SDW order in the undoped bilayer silicene\cite{Liu F,Black-Schaffer A M})
and it has $N_{0}\propto S$ where $S$ is the renormalized spin which consider the order-parameter fluctuations,
e.g., the exponential decay of the order-parameter in the integrability-broken system due to the perturbations (see, e.g., Ref.\cite{Wu C H}).
From these expressions we can see that the amplitude is mainly given by the longitudinal susceptibility $\chi_{\parallel}$,
and the strong damping appear when the term $\frac{16\rho_{s}(0)}{cN}$ is close to zero.
The multi-component field $\phi_{\alpha}$ whose fluctuation can be regulated by a local
order phase $\varphi$, i.e., $\phi_{\alpha}\sim e^{i\varphi}$, and hence contain the imaginary part.
The spin-wave fluctuation polarization
 term for $d=2$ at zero temperature with the cutoff of the momentum and frequency is 
$\Pi(k,\omega)=1/(8p)$\cite{Polkovnikov A,Chubukov A V},
which has a nonzero imaginary part when damping is possible.
In fact, except the spontaneous magnetization, the magnetic phase with such antiferromagnetic chiral $d_{x^{2}-y^{2}}\pm id_{xy}$ 
(or the $d_{xy}\pm id_{x^{2}-y^{2}}$)which is singlet pairing will
also lead to the edge current with the chiral edge states 
\cite{Greschner S,Sauls J A,Laughlin R B} and correlate with the surface scattering of the Cooper pairs\cite{Sauls J A},
spin or thermal Hall effect\cite{Read N}, and even the Majorana fermions in the bulk-edge-coupling type materials\cite{Wu C H,Diehl S,Alicea J}
and can be found in the zero model edge state in the vortex core of topological SC with nonzero Chern number which act as non-Abelian anyons,
such vortex pattern and domain structures also has been experimentally found in the $\sqrt{13}\times\sqrt{13}$ reconstructed silicene
(see, e.g., Ref.\cite{Li W} and the references therein),
while the superconductor or superfluid have a full pairing gap within the bulk\cite{Chung S B}.
The related pairing symmetrys in the lattice space are shown in the Fig.4(c).
Such $d$-wave order own the anisotropic energy gap in the FS except the Dirac-point in the annular FS or the nodal lines if exist.
And the quasiparticles which induced in the gapless state provide the Hall conductance according to the $E_{p}$\cite{Laughlin R B} and with the 
broken time reversal symmetry.
In fact, the unbound scattering which the particles are free-motion and get apart with each other in the non-interaction limit as discuss below,
would exhibit no chirality and dramatically suppress the edge current.
In this case, the perfect normal retroreflection of scattering will form a flat band with the zero model\cite{Sauls J A} 
which also leads to the
equal weight of bands and thus shows no chirality and the edge current.
While in othercase with retroreflected trajectories,
the Dirac equation is satisfied by the Fermi scalar field just like the Jackiw-Rebbi zero mode
at the domain wall with chirality.
For the AFM Cooper pair which has equal but oppsite-diraction spins when undoped,
the pairing strength is associate with the $k$-dependent Fermi velocity in the FS sheetw.
For silicene, this Fermi velocity $v_{F}=|q|(E_{F}\pm \varrho)/\hbar $\cite{Cahangirov S},
where $\varrho$ is a small amount due to the ambipolar of ,
and it's very large that reaches $5\times 10^{5}\sim 1\times 10^{6}m/s$ by the method of LDA and PBE\cite{Drummond N D,Cahangirov S}.
Such a large Fermi velocity is also favorable to make the high speed spintronic devices.
We use $v_{F}=5.5\times 10^{5}$ in this paper which is the same as the HgTe/CdTe quantum well.

Since in $k$-space, the exchange effect of spin fluctuation between the such antiferromagnetic cooper pairs which with pairing energy 
$E_{p}=k_{B}T_{c}$\cite{Sauls J A}
where $T_{c}$ is the temperature of the phase transition,
will give rise to the unconventional superconductivity,
$T_{c}=1.13\hbar w_{D}e^{-1/\lambda}$\cite{Sato M,Ynduráin F} in the weak coupling limit where $\lambda$ is the eigenvalue of the gap function (Eq.(17))
which describe the electron-phonon and electronelectron interactions and $w_{D}$ is the cutoff of the interaction.
$\hbar w_{D}$ is the Debye energy (also the low energy bandwidth of the two-band model\cite{Liu F})
this $T_{c}$ has been found that related to the isotopic mass by the Debye temperature $\theta_{D}$\cite{Pei Q X}.
Such spin fluctuation also give rise to the unconventional $s$-wave pairing with a sign reversal of the order parameters (like gap) 
between the different Fermi surface sheets,
(or the $d$-wave one with repulsive interactions which is also mediated by the spin fluctuations)
that not only affects the stability of the pair scattering but also lead to a nodeless superconducting gap in the Fermi surface\cite{Kuroki K,Mou D}.
This amplitude model is closely relate with the spin-wave model as well as the phonon model with the order parameter consist of the multi-component quantum field 
(see Eq.(2)), e.g., the spin fluctuation near the superfluid-insulation transition with critical repulsive interaction and critical hopping amplitude 
between neighbor sites.

Consider the two-dimension wave vector and Matsubara frequency $\omega_{M}=(2n+1)\pi/\beta$ into the renormalized Green's function 
\begin{equation} 
\begin{aligned}
G(\tau,\tau')=&\frac{1}{\beta}\sum_{i\omega_{M}}e^{i\omega_{M}(\tau'-\tau)}G(i\omega_{M}),\\
G(i\omega_{M})=&\int^{\beta}_{0} d\tau e^{i\omega_{M}\tau}G(\tau,0),
\end{aligned}
\end{equation}

The local self-energy can be obtained by the lattice Dyson equation 
$G_{k}(i\partial_{t}+\mu-\epsilon_{k}-\Sigma)=1$,
where $\mu$ is the chemical potential and $\epsilon_{k}$ is the $k$-dependent band energy (mainly the kinetic energy) 
which has a important effect on the continuity of Fermi surface.
We note that the Fermi chemical potential $\mu_{F}$ is setted as zero in this paper,
thus it's particle-hole symmetry in the Fermi level.
This lattice Dyson equation resulting in a momentum- and frequency-dependent Green's function $G(k)\equiv G(k,i\omega_{M})
=-\langle \mathcal{T}_{\tau}c_{\sigma}(k,\tau)c_{\sigma}(-k,\tau')\rangle$, where $\mathcal{T}_{\tau}$ is the imagnary-time ordering operator
and $\langle\cdot\cdot\cdot\rangle$ denotes the thermodynamical average.
the diagonalized spin irreducible susceptibility matrix is\cite{Graser S,Kuroki K} 
\begin{equation} 
\begin{aligned}
\chi(p)=\frac{1}{N\beta}\sum_{k}G(k+p)G(k)
\end{aligned}
\end{equation}
base on the standard random-phase-approximation (RPA)
approach,
where $N$ is the number of lattice sites (and also the $K$-point meshes in Brillouin zone) and $\beta=1/k_{B}T$.
This spin susceptibility also describes the spin fluctuation as well as the spin polarizarion of the charge carriers,
and it's important in the fluctuation exchange approximation\cite{Takimoto T} and the exact diagonalization for the impurity or cluster model
which with the boundary vertex function.

\section{Geometrical studies of the isotropic honeycomb lattice with particle-hole symmetry}

We consider only the nearest($t$)- and second nearest($t'$)-neighbor hopping in this subsection and ignore the diagonal one $t''$ which is qualitatively unimportant
in discussion of vortex and the chiral current.
Here we suppose the isotropic honeycomb case which the $t$ and $t'$ are both invariant for three directions on one site on the lattice-plane
(e.g., $t_{AB}=t_{AA'}=t_{A'B'}$ in Fig.1).
Then the transfer Hamiltonian $H_{t}$ can be written as (note that here we made the simplified: $k_{x}=k_{x}a,\ k_{y}=k_{y}a$)
\begin{equation}
\begin{aligned}
H_{t}=&\begin{pmatrix} H_{AB'}&H_{AA'}\\ H_{A'A}& H_{A'B}\end{pmatrix},\\
H_{AB'}=&\varepsilon_{A}+t'_{AB'}(4{\rm cos}\frac{\sqrt{3}k_{x}}{2}{\rm cos}\frac{k_{y}}{2}+2{\rm cos}k_{y}),\\
H_{AA'}=H^{*}_{A'A}=&t_{AA'}[{\rm exp}(i\frac{\sqrt{3}k_{x}}{3})+{\rm exp}(i(-\frac{\sqrt{3}}{6}k_{x}+\frac{k_{y}}{2}))
+{\rm exp}(i(-\frac{\sqrt{3}k_{x}}{6}-\frac{k_{y}}{2}))],\\
H_{A'B}=&\varepsilon_{A'}+t'_{A'B}(4{\rm cos}\frac{\sqrt{3}k_{x}}{2}{\rm cos}\frac{k_{y}}{2}+2{\rm cos}k_{y}),
\end{aligned}
\end{equation}
where $\epsilon_{A}$ is the on-site energy.
The upper and lower bands energies of this model are
\begin{equation} 
\begin{aligned}
E_{+}(k_{x},k_{y})=&\varepsilon+t{\rm cos}\frac{\sqrt{3}k_{x}}{3}
+2t{\rm cos}\frac{\sqrt{3}k_{x}}{6}{\rm cos}\frac{k_{y}}{2}+4t'{\rm cos}\frac{\sqrt{3}k_{x}}{2}{\rm cos}\frac{k_{y}}{2}+2t'{\rm cos}k_{y},\\
E_{-}(k_{x},k_{y})=&\varepsilon+t{\rm cos}\frac{\sqrt{3}k_{x}}{3}-t{\rm cos}\frac{\sqrt{3}k_{x}}{6}{\rm cos}\frac{k_{y}}{2},
\end{aligned}
\end{equation}
where $t'$ is the second nearest-neighbor hopping for one site towards the six possible directions,
while $t$ is the nearest-neighbor hopping for one site towards the three possible directions.
It's remarkable that the lower and upper bands energies here is full-interaction-dependent.
The $\epsilon$ here is the on-site energy for different sites.
Thus a exactly flat band is obtained only when $t'=0$, which is differ from the requirement of the ladder lattice model to yields a flat band which is 
$t'=\pm t_{l}$ where $t_{l}$ is the nearest-neighbor hopping in leg direction\cite{Kuroki K} but the same to the tetragonal one\cite{Misumi T},
and we can modulate the energy of the flat band as well as the gap between the flat band and the dispersion band by adjusting $t$ and $t'$
in this system.
It's also clearly that the lower and upper bands are split by an energy which related to the on-site energy difference, $t$ and $t'$,
and indeed this split is origin from the hybridization of the eigenstates 
with plane-wave states which populating along the two leg-directions ($l_{1}$ and $l_{2}$).
Theoretically, we can also obtain that the bands width are related to the $t$ and $t'$.
Then the energy of such flat-band configuration within a single-particle picture is obtained as
\begin{equation} 
\begin{aligned}
\varepsilon^{2}_{k}=&3t^{2}+2t^{2}[{\rm cos}(-\frac{\sqrt{3}}{2}k_{x}+\frac{1}{2}k_{y})+{\rm cos}(-\frac{\sqrt{3}}{2}k_{x}-\frac{1}{2}k_{y})+{\rm cos}(\frac{\sqrt{3}k_{y}}{3})]\\
&+6t'^{2}+2t'^{2}[{\rm cos}(k_{y})+{\rm cos}(-k_{y})+
{\rm cos}(\frac{\sqrt{3}}{2}k_{x}-\frac{k_{y}}{2})+{\rm cos}(\frac{\sqrt{3}}{2}k_{x}+\frac{k_{y}}{2})\\
&+{\rm cos}(-\frac{\sqrt{3}}{2}k_{x}-\frac{k_{y}}{2})+{\rm cos}(-\frac{\sqrt{3}}{2}k_{x}+\frac{k_{y}}{2})]\\
&+2t'^{2}[{\rm cos}(\sqrt{3}k_{x})+{\rm cos}(-\sqrt{3}k_{x})+
{\rm cos}(-\frac{\sqrt{3}}{2}k_{x}+\frac{3k_{y}}{2})+{\rm cos}(-\frac{\sqrt{3}}{2}k_{x}+\frac{3k_{y}}{2})\\
&+{\rm cos}(\frac{\sqrt{3}}{2}k_{x}-\frac{3k_{y}}{2})+{\rm cos}(\frac{\sqrt{3}}{2}k_{x}-\frac{3k_{y}}{2})],
\end{aligned}
\end{equation}
where the $\varepsilon_{k}$ also describe the dispersion of such hopping configuration which contain nine different hopping directions.
The charts of $E_{+}$ and $\varepsilon^{2}_{k}$ with different $t'$ and $E_{-}$ are shown in the top and bottom panel of Fig.2, respectively.
We can see that the fluctuation of both the upper bands energies and the single-particle spectrum in the momentum space
is increase with the enhancement of $t'$.
That means the system is become less stable.
Note that base on the process which taking account the above band energies $E_{\pm}$,
we can obtain the two-band model with the two energy bands which are intersect the FS.
Note that here we ignore the bulking distance for simplify the calculation,
the tight-binding results considering the bulking distance are presented in the Appendix.B.
While for the pairing scattering process which mentioned above, the induced new band energies $E_{\pm}$ will have more complicate form,
but the splitting interval and widths are still related to the $t$ and $t'$.
In this case the effective interaction has the similar form\cite{Graser S,Liu F}
\begin{equation} 
\begin{aligned}
U_{{\rm eff}}=\frac{1}{N}\sum_{ab,kk'}\Gamma'^{l_{1}l_{2}}(k,k',\omega)c^{\dag}_{l_{1}}(k)c^{\dag}_{l_{1}}(-k)c_{l_{2}}(-k')c_{l_{2}}(k'),
\end{aligned}
\end{equation}
with the effective interaction vertex $\Gamma'^{ab}$ between two Cooper pairs near the FS
\begin{equation} 
\begin{aligned}
\Gamma'^{ab}(k,k',\omega)={\rm Re}\sum_{ab,kk'}\Gamma^{a_{1}a_{2}}_{a_{3}a_{4}}(k,k',0)\Lambda^{l_{1}*}_{a_{1}}(k)\Lambda^{l_{1}*}_{a_{2}}(-k)\Lambda^{l_{2}}_{b_{1}}(-k')\Lambda^{l_{2}}_{b_{2}}(k').
\end{aligned}
\end{equation}

Since we suppose the isotropic honeycomb model here, it's gapless in the $K$-point 
unless we do the electron doping or the hole doping which produce a band gap between the orbital $a_{1}$ and the orbital $a_{2}$ in the $K$-point which
also the Dirac-point for silicene
(see below which we take the silicene as a explicit example).
When $\epsilon_{A}$ and $\epsilon_{A'}$ equal zero, then the two flat bands formed above have zero energy too.
Through the electron-doping, as we discuss below, the special Dirac-point 2D Dirac semimetal may be dip below the Fermi level and 
lead to a metallic band,
and there may exist a charge-tansfer insulating band at the same time in hard-core-boson system.
That also give rise to the SC.


\section{silicene}
The silicene which is a $3p$-orbital-based materials with the noncoplanar low-buckled (with a buckle about $0.51$ \AA\ due to the hybridization between the 
$sp^{2}$-binding 
and the $sp^{3}$-binding (which the bond angle is $109.47^\text{o}$) and 
that can be verified by thr Raman spectrum which shown in the Fig.9(f) with the intense peak at 578 cm$^{-1}$ larger than the planar one and the $sp^{3}$-binding one
\cite{Tao L},
and thus approximately forms two surface-effect like the thin ferromagnet matter) lattice structure.
The bulked structure not only breaks the lattice inversion symmetry,
but also induce a exchange splitting between the upper atoms plane and the lower atom plane and thus forms a emission geometry which allows the
optical interband transitions, which for the graphene can happen only upon a FM substrate\cite{Rader O}.
The FM or AFM order can be formed in monolayer silicene by the magnetic proximity effect
that applying both the perpendicular electric field and in-plane FM or AFM field.
Silicene has a small intrinsic band gap 
and the energy difference $\Delta=\epsilon_{A}-\epsilon_{B}=\epsilon_{A'}-\epsilon_{B'}$ is proportional to
the bulking distance $\overline{\Delta}$ which reduced the crystal symmetry from the point symmetry group $C_{6v}$ (like graphene) to the 
wave-vector symmetry $C_{3v}$
and while for the projected surface irreducible BZ ($\overline{\Gamma}-\overline{M}-\overline{K}$),
the crystal symmetry is $C_{2v}$ the same as the graphene.
The Dirac-point $K$($K'$) which is also the charge neutrality point where has minimum conductance
has zero electronic density of states (DOS) when it's undoped,
and obey the relativistic Dirac equation with maximum Fermi velocity around these points under low energy.
The buckled structure of silicene also give rise to the out-of-plane acoustic phonon scattering as well as the electron-phonon interaction,
and thus lower the charge mobility.
Both the graphene and silicene contains the degrees of freedom of spin, sublattice, valley in the low-energy band strucure (flavor or species degrees of freedom),
which is distincted from the weyl semimetal with the Landau bands.
Such gap in Dirac-point will becomes obvious when apply a perpendicular electric-field (normal to the silicene layer) which won't shift the FS
and this low-buckled structure also split the nodal lines along $K-K'$ which also require the Kramers degenerate in momentum space with time-reversal-invariant (TRI)
and thus the nodal lines or Weyl points are nonexistent in the present model.
The sizable band gap of silicene opened under the electric reveal enormous value of silicene in the application of field effect transitor and the nanoelectronics.
This is because that the Dirac-point is sensitive to the perturbation which is proportional to the $\sigma_{z}$.
Recently, it's also been found that the 
fully hydrogenated graphene (graphane) can open a band-gap of 5.4 eV\cite{Cudazzo P}.

We also know that this kind of two-dimensional materials have the inspired modifiable electronic structure,
like the characteristics of 2D Dirac semimetal
and the ambipolar of excess electrons which is due to the electron-hole symmetry,
and the modifiable electron-phonon interactions by hole- or electron-doping which will shift the FS.
The silicene is similar to the graphene which with the hybridized $\pi$ and $\pi^{*}$ bands ($\sigma$ and $\sigma^{*}$ bands) cross (near) the FS 
at $M(\Gamma)$-point (see the band structure),
even when the intralayer-symmetry is broken by the external field, and they are contributed by the $3s$ and $3p$ orbitals.
The direct hybridization between the $\pi$ bands and the $\sigma$ bands is allowed by the low-buckled
The electron-phonon interaction is enhanced by the hole- or electron-doping into the $sp^{3}$-bonding bands
near the Fermi surface and thus give rise the another possible Brillouin zone which is diamond-shape.
This kinds of hexagonal lattice like the silicene, germanene, stanene, SnSi, etc. have weak $\pi$-bonding due to the crinkled structure,
but for silicene or germanene, they obtain the stability from the $sp^{3}$ hybridization (with the overlap between the $\sigma$-band
and $\pi$-band) 
and the low bulked structure
which is induced by the $sp^{2}$ dehybridization.
For electron-doping, the Dirac-point of both the silicene and graphane may be moved under the FS and leads to the metal. 
In Ref.\cite{Chen L}, the Fermi level was found to be lifted above the Dirac point as 0.5 eV due to the charge transfer from Ag(111) to silicene
with a superconducting gap of 35 meV,
while in Ref.\cite{Vogt P}, the Dirac point which is 0.3 eV below the Fermi level was found with a gap about 0.6 eV due to the interaction with the Ag(111) substrate.
Such charge transfer (provided by the conducting substrate especially the noble metal
and thus brings the large screening effect) also exacerbated the asymmetry of silicene, and the topological SC gap (whose pairing symmetries break the TRI
and breaks the U(1) charge conservation by the SC proximity)
within the SC Hamiltonian expression with a AFM Cooper pair 
\begin{equation} 
\begin{aligned}
H_{SC}=\sum_{i,\mu_{ij}=A,B}[\Delta_{SC}c^{\dag}_{i\uparrow}(k)c^{\dag}_{i\downarrow}(-k)+
\Delta^{*}_{SC}c_{i\uparrow}(k)c_{i\downarrow}(-k)]_{\mu_{ij}}+(k\rightarrow k')
\end{aligned}
\end{equation}
also leads to the absence
of the Dirac fermion and the 2D electron characteristics.
If we apply a electron field, 
the sublattices symmetry may be broken and then the two on-site energy $\epsilon$ within the above two band energy equations (Eq.) become unequal,
and the sublattice with lower energy becomes more closer to the Fermi surface and therefore play the main role in the low-energy model.
Further, the low energy bands is flattened near the Fermi surface with their increased values of the dispersionand and the
spin correlations within the lower energy sublattice which with a ordered ferromagnetic-like phase favor the triplet f-wave pairing\cite{Zhang L D}.
In the strong coupling regime, the particle scattering is becomes maximal in the saddle points of the stable Fermi surface while
it's much weaker in the diagonals of Brillouin zone.
In the case of the $t'$ is not too small, there is always a relatively strong coupling regime near the saddle points
which are close to the FS\cite{Honerkamp C}.
Althought the saddle-point is not well obsrved in the silicene,
the strong interaction regime which with the remarkable AF spin-fluctuation is been found.

\section{Tight-binding model}
We begin with the single-particle tight-binding model 
\begin{equation} 
\begin{aligned}
H=&H_{0}+V
=&\sum_{\mbox{\tiny$\begin{array}{c}
ij\\
\sigma=\uparrow,\downarrow\end{array}$}}t_{ij}c^{\dag}_{i\sigma}c_{j\sigma}+U\sum_{i}n_{i\uparrow}n_{i\downarrow},
\end{aligned}
\end{equation}
where $t_{ij}$ is a
hopping matrix, $U$ is the on-site pairwise interaction (intraorbital Coulomb), and note that we consider the intraorbital interaction only here 
and ignore the interorbital interaction since it's very small in our discussing system.
Similar treating has also applied in the iron-base high $T_{c}$ superconductor ${\rm LaFeAsO}_{1-x}{\rm F}_{x}$\cite{Kuroki K2}
which take account the five Fe $d$-bands (not including the highly hybridized one $d_{xz-yz}$)
has a more stronger spin fluctuation than the orbital one and thus the spin susceptibility is also more effective (due to the Fermi level crossing) 
than the orbital one.
Thus the Hund's rule coupling (including the intraorbital pairwise exchange $J$ and the pair hopping $t_{p}$ within one cell) is considerable (although not shown in the Eq.(7)).
For the spin-singlet state and spin-triplet state,
the on-site interaction pair of the Hamiltonian are
\begin{equation} 
\begin{aligned}
H^{(s)}_{{\rm int}}=&\sum_{\mbox{\tiny$\begin{array}{c}
k\\
\sigma=\uparrow,\downarrow\end{array}$}}
\frac{U}{4}[c^{\dag}_{\sigma}(k+k')c_{\sigma}(k)]^{2},\\
H^{(t)}_{{\rm int}}=&-\sum_{\mbox{\tiny$\begin{array}{c}
k\\
\sigma,\sigma' =\uparrow,\downarrow\end{array}$}}
\frac{U}{12}[c^{\dag}_{\sigma}(k+k'){\pmb \sigma}_{\sigma\sigma' } c_{\sigma' }(k)]^{2},\\
\end{aligned}
\end{equation}
where ${\pmb \sigma}$ is the Pauli operator.
The Hubbard repulsion here (like the electron-electron interaction) would lead to the renormalization of tight-binding parameter,
and the pairing repulsion also give rise to the $d$-wave superconducting gap $\Delta_{d}\sim {\rm cos}k_{x}-{\rm cos}k_{y}$
in the quasiparticle spectra as shown in the Ref.\cite{Yang X}.
There are four sites (sublattices) $A,B,A',B'$ with antiferromagnetic SDW order in unit cell and form a four-band tight-binding model (see Fig.1),
and we define nearest-, second nearest-, and third nearest-neighbor hopping as $t,t',t''$, respectively.
The Berry phase of the neighbor Dirac-point is $\pi$ and $-\pi$, respectively, thus it's totally zero for one unit cell,
which is consistent with the case of 2D electron gas and equivalent to that of the bilayer graphene of $2\pi$\cite{Park C H2}.
To obtain the single-particle spectrum (and dispersion), we first mapping the Hamiltonian (fermion system) to the momentum space through the transformation:
$c_{r}(k)=\frac{1}{\sqrt{N}}\sum_{k}c_{k}e^{ikr}$ ($N$ is the sample size)
for even parity which correspond to antiperiodic boundary conditions $\psi(r+N)=-\psi(r)$ with essential vectors $k=\pi (2n-1)/N$, 
and 
$c_{r}(k')=\frac{1}{\sqrt{N}}\sum_{k'}c_{k'}e^{ik'r}$ 
for odd parity which correspond to periodic boundary conditions $\psi(r+N)=\psi(r)$ with essential vectors $k'=2\pi n/N$\cite{Wu C H}.
Here $N$ is the system size, and left and right current are correspond to the antiperiodic boundary conditions and the periodic boundary conditions, respectively.
In unit cell, the vector $c_{r}(k)=(c_{A},c_{B},c_{A'},c_{B'})^{T}$,
the Hamiltonian in momentum space is $H_{0}=\sum_{k>0,\sigma}c^{\dag}_{k\sigma}{\bf H}_{k\sigma}c_{k\sigma}$ in Fourier representation,
which is even-parity-type and can be diagonalized into (through the Bogoliubov rotation $\bf R$) 
\begin{equation} 
\begin{aligned}
H_{d}=\sum_{\mbox{\tiny$\begin{array}{c}
ij\\
\sigma=\uparrow,\downarrow\end{array}$}}\epsilon_{k}c^{\dag}_{k\sigma}c_{k\sigma},
\end{aligned}
\end{equation}
where $\epsilon_{k}$ is the band energy which equal to the on-site energy here and it related to the hopping amplitude and the single-particle dispersion.
Then we obtain the free fermions operator (Bogoliubov fermions) $A_{k\sigma}={\bf R}c_{k\sigma}$ with zero interaction (U=0).
Note that the Bogoliubov rotation here has ${\bf R}\sigma_{z}{\bf R}^{\dag}=\sigma_{z}$.
In this zero-$U$ case (or the case with strong Coulomb coupling\cite{Khveshchenko D V}), the spin susceptibility and charge susceptibility are equivalent and the $4\times 4$ free susceptibility (Eq.(6)) can be rewritten as
\cite{Liu F,Zhang L D}
\begin{equation} 
\begin{aligned}
\chi_{a_{3}a_{4}}^{(0)a_{1}a_{2}}(k,i\omega_{M})\equiv \frac{1}{N}\sum_{k_{1}k_{2}}\langle T_{\tau}c^{\dag}_{a_{1}}(k_{1},\tau)c_{a_{2}}(k_{1}+k,\tau)
c^{\dag}_{a_{3}}(k_{2},0)c_{a_{4}}(k_{2}+k,0)\rangle,
\end{aligned}
\end{equation}
where $a_{1},a_{2},a_{3},a_{4}=1,2,3,4$ are the indices of the unit cells with four orbits (bands or sublattices).
The Hermitian static homogeneous susceptibility (noninteracting susceptibility, i.e., the susceptibility with zero Dirac mass as discussed below) matrix is
$\chi_{a_{1}a_{2}}^{(0)}(k)=\chi_{bb}^{(0)aa}(k,0)$ with the strong momentum-dependence due to the strong Coulomb coupling, 
and its largest eigenvalue is the static homogeneous susceptibility which will becomes more homogeneous due to the neglect of some unimportant orbitals,
and the eigenvector which corresponding to the largest eigenvalue determines the dominant spin fluctuations\cite{Zhang L D}.
At zero temperature, the Pauli susceptibility which is proportional to the total density of states (DOS) at the Fermi surface is\cite{Graser S} 
$\chi(0)=\chi_{a_{2}a_{2}}^{(0)a_{1}a_{1}}(k,0)=\sum_{ab}n_{ab}(0)$ where $n_{ab}$ is the single-spin DOS at the Fermi surface for the bands in cell $a$ and $b$.
When such Hubbard interaction $U\neq 0$,
the charge and spin renormalized
 susceptibilities which enhanced by RPA are\cite{Liu F}
\begin{equation} 
\begin{aligned}
\chi^{(s)}(k,i\omega_{M})=[I-\chi^{(0)}(k,i\omega_{M}){\bf U}^{(s)}]^{-1}\chi^{(0)}(k,i\omega_{M}),\\
\chi^{(c)}(k,i\omega_{M})=[I+\chi^{(0)}(k,i\omega_{M}){\bf U}^{(c)}]^{-1}\chi^{(0)}(k,i\omega_{M}),
\end{aligned}
\end{equation}
where $I$ is the identity matrix, ${\bf U}$ is the $16\times 16$ matrix and there are only 40 nonzero elements for the spin 
susceptibility
and 28 nonzero elements for the charge 
susceptibility:
\begin{equation} 
\begin{aligned}
{\bf U}^{(s)a_{m}a_{m}}_{a_{m}a_{m}}=&U,\ {\bf U}^{(s)a_{m}a_{m}}_{a_{n}a_{n}}=\frac{J}{2},\ {\bf U}^{(s)a_{m}a_{n}}_{a_{m}a_{n}}=\frac{J}{4},\ {\bf U}^{(s)a_{n}a_{m}}_{a_{m}a_{n}}=t_{p},\\
{\bf U}^{(c)a_{m}a_{m}}_{a_{m}a_{m}}=&U,\ {\bf U}^{(c)a_{m}a_{n}}_{a_{m}a_{n}}=\frac{3J}{4},\ {\bf U}^{(c)a_{m}a_{n}}_{a_{n}a_{m}}=t_{p}.\\
(m,n=&1,2,3,4)
\end{aligned}
\end{equation}
We can see that except the four intraband elements ${\bf U}^{(s)a_{m}a_{m}}_{a_{m}a_{m}}$ there are also some off-diagonal elements (interband) ia nonzero,
there is the result of considering the Hund's rule coupling,
and note that the charge and spin
 susceptibility matrices here are also $16\times 16$.
Through the charge/spin fluctuations (or the charge/spin susceptibility),
the pairing scattering between the cooper pairs between different cells through the spin or charge fluctuations, 
i.e., $(ka_{1},-ka_{2})\rightarrow (k'a_{3},-k'a_{4})$ which also scatter to a new FS sheet,
is govern by the interaction-Hamiltonian in the momentum-space
\begin{equation} 
\begin{aligned}
H_{{\rm int}}=\sum_{a_{1}a_{2}a_{3}a_{4},\sigma\sigma',kk'}\frac{t_{p}}{2}c^{\dag}_{a_{1}\sigma}(k)c^{\dag}_{a_{2}\sigma' }(-k)c_{a_{3}\sigma}(-k')c_{a_{4}\sigma' }(k'),
\end{aligned}
\end{equation}
where $t_{p}$ is the pair-hopping,
give rise to the effective interaction in RPA level
\begin{equation} 
\begin{aligned}
U_{{\rm eff}}=\frac{1}{N}\sum_{a_{1}a_{2}a_{3}a_{4},kk'}\Gamma^{a_{1}a_{2}}_{a_{3}a_{4}}(k,k',\omega)c^{\dag}_{a_{1}}(k)c^{\dag}_{a_{2}}(-k)c_{a_{3}}(-k')c_{a_{4}}(k'),
\end{aligned}
\end{equation}
with effective pairing interaction vertex from the generalized RPA $\Gamma^{a_{1}a_{2}}_{a_{3}a_{4}}(k,k',\omega)$ in
spin-singlet and spin-triplet representations in momentum-space are
\begin{equation} 
\begin{aligned}
\Gamma^{(s)a_{1}a_{2}}_{a_{3}a_{4}}(k,k',\omega)=&\left[\frac{3}{2}{\bf U}^{(s)}\chi^{(s)}(k-k',\omega)\chi^{(s)}-\frac{1}{2}{\bf U}^{(c)}\chi^{(c)}(k-k',\omega)\chi^{(c)}
+\frac{1}{2}{\bf U}^{(s)}+\frac{1}{2}{\bf U}^{(c)}\right]^{a_{1}a_{2}}_{a_{3}a_{4}},\\
\Gamma^{(t)a_{1}a_{2}}_{a_{3}a_{4}}(k,k',\omega)=&\left[\frac{-1}{2}{\bf U}^{(s)}\chi^{(s)}(k-k',\omega)\chi^{(s)}-\frac{1}{2}{\bf U}^{(c)}\chi^{(c)}(k-k',\omega)\chi^{(c)}
+\frac{1}{2}{\bf U}^{(s)}+\frac{1}{2}{\bf U}^{(c)}\right]^{a_{1}a_{2}}_{a_{3}a_{4}},
\end{aligned}
\end{equation}
respectively.

In 
the case of strong on-site interaction which may cause the multiple particles occupying the same site simultaneously, will leads to a pairing interaction shift\cite{Tai M E}.
The superconducting critical temperature $T_{c}$ is determined by the temperature where the largest eigenvalue $\lambda$ which mentioned above reaches unity,
and we can form the linearized gap equation (Eliashberg equation) as follow
\begin{equation} 
\begin{aligned}
\lambda \Lambda_{a_{1}a_{2}}(k)=-\frac{1}{N\beta}\sum_{a_{1}a_{2}a_{3}a_{4},kk'}U_{{\rm eff}}G_{a_{1}a_{2}}(k')G_{a_{3}a_{4}}(-k')\Lambda_{a_{3}a_{4}}(k'),
\end{aligned}
\end{equation}
where $\Lambda_{a_{1}a_{2}}$ is the corresponding eigenvector.
This gap equation cause the sign change of gap between the flat band and the dispersion band
due to the effective effect of the pair hopping (scattering), but note that there are no sign change within the bands.

In spectral representation, the irreducible susceptibility can be rewritten as\cite{Graser S}
\begin{equation} 
\begin{aligned}
\chi_{a_{3}a_{4}}^{a_{1}a_{2}}(k,i\omega_{M})= \frac{1}{N}\sum_{q,l_{1}l_{2}}
\frac{\Lambda^{a_{1}}_{l_{1}}(q)\Lambda_{l_{1}}^{a_{2}}(q)\Lambda_{l_{2}}^{a_{3}}(k')\Lambda_{l_{2}}^{a_{4}}(k')}{i\omega_{M}+\lambda_{l_{1}}(q)-\lambda_{l_{2}}(k')}
[f(\lambda_{l_{2}}(k'))-f(\lambda_{l_{1}}(q))],
\end{aligned}
\end{equation}
where $k'=k+q$ and $l_{1}$ and $l_{2}$ are the band indices and $f$ is the Fermi-Dirac distribusion function.
Here the eigenvectors $\Lambda$ connect the orbital spaces and the band spaces here.

\subsection{Effects of SOC and other quantities}
Considering the even parity for the AFM $d$-wave singlet chiral Cooper pair here,
and the $d$-wave spin pair have the same amplitude in the case of undoped and zero external electric field,
which has even-parity has the chiral pairing amplitude as 
\begin{equation} 
\begin{aligned}
\mathcal{A}_{p}=&\left\langle\frac{N}{2}-1\bigg|-\psi(r+N)\psi(r)\bigg|\frac{N}{2}\right\rangle\\
=&\mathcal{A}(r)\Delta(k)(d_{1}+ id_{2})\cdot\vec{y},
\end{aligned}
\end{equation}
where $N/2$ is the number of the Cooper pairs or the dimer bosons (or moleculars),
and the gap function $\Delta({\bf k})={\bf d}({\bf k})\cdot{\pmb \sigma}$ which is in a coordinate independent representation is a spin-dependent term, 
since 
${\bf d}({\bf k})=[t'_{SOC}{\rm sin}k_{x},t'_{SOC}{\rm sin}k_{y},M_{z}-2B(2-{\rm cos}k_{x}+{\rm cos}k_{y})]$ 
is the $k$-dependent unit vector which related to the SC order parameter
where $t'_{SOC}$ is the next-nearest hopping strength due to the intrinsic SOC.
The $\Delta({\bf k})$ here is also the ${\bf H}_{k\sigma}$ mentioned above which contains both the spin-independent hopping and the spin-dependent one.
$B$ is a Bernevig, Hughes, and Zhang (BHZ) model
-dependent parameter and $M_{z}$ is the Zeeman field which dominate the surface magnetization
but can be ignore when a strong external electric field (or external magnetic field to obtain the QHE) is applied.
The two continuously Landou-level of the two neighbor sublattices are mixed up by the intrinsic Zeeman field,
For the untiled AFM/FM helical spin edge states,
the fluctuation of CDW and the fluctuation of $z$-component SDW is conjugate with each other
and with gapless single-particle edge-excitation dispersion in the thermodynamic limit ($N\rightarrow\infty$)\cite{Zheng D}.
The gapless helical edge states will be gap out by the magnetic order which can be measured by the AFM factor,
and resulting the losing of the masslesss Dirac fermions characteristics (or the marginal Fermions\cite{Volovik G E2}).
In fact, in the case of large $U$ at half-filling (and naturly the chemical potential is large), 
the electrons is highly localized and degrees of freedom are highly suppressed except the directions of spin,
and the relatively small pairing potential order parameter $\mathcal{G}$ which is spatial homogeneous can be viewed as real,
i.e.,
$\mathcal{G}_{p}=\mathcal{G}_{p}^{0}$ but no $\mathcal{G}_{p}\approx\mathcal{G}_{p}^{0}e^{-i\phi_{k}})$, which can be implemented by choosing a gauge,
where the SC phase factor $\phi_{k}={\rm tan}^{-1}(k_{x}/k_{y})$,
and then resulting a low-energy pairings which is similar to the spinless $p+ip$ SC which with a pair of helical band in the Dirac cone.
Thus planar AFM state can be realized as $|{\bf d}\rangle=(1/\sqrt{2})(|\uparrow\downarrow\rangle+|\downarrow\uparrow\rangle)$
for the spin-polarization along the $z$-direction.
That despict a scenario that the unparallel momentums aligned along the $x$-direction
and polarized along the $\pm z$-direction (by the Zeeman field),
which also has Ising spin effect with local spin fluctuation (or interaction) (cf.Ref.\cite{Wu C H}).
But note that the ${\bf d}$ is not always along the $z$-direction,
it may also perpendicular to the $z$-direction which corresponds to the case of
the order $d_{x^{2}-y^{2}}=d_{xy}$ as discuss below.
In fact, considering the SC effect,
the antiferromagnetic SDW order may become incommensurate under a collective effect
of the Rashba-coupling and the weak Zeeman field,
and provides a possible that the ${\bf d}({\bf k})$ rotate along the lattice sites (cf. Ref\cite{Farrell A}) if the $t'_{SOC}$ is large enough,
which can be induced by the external electric field for silicene\cite{Ezawa M} just similar to the case of graphene\cite{Tse W K}.
Although the electric field induced next-nearest-neighbor Rashba-coupling breaks the above planar AFM state and the spin of the $z$-component is no more conserved 
(and thus we can't judge the amount of spin flipped and the unflipped interband scttering or the particle-hole scttering process
in a TI (note that the spin flipped transition can only occur in a TI)),
but the $z$-component spin is nearly a good quantum number since $R_{2}$ is very small.
In this case, the Chern number (or TKNN index\cite{Thouless D J}) is related to the $Z_{2}$ topological number,
which is one of the symmetry protected topological invariant similar to the topological charge
and it's ill-defined when the TRI is broken,
and the single spin Chern number is identical to the two-band Pontryagin topological index
$C_{s}=\frac{1}{4\pi}\int dk_{x}dk_{y}\hat{{\bf K}}\cdot(\partial_{k_{x}}\hat{{\bf K}}\times\partial_{k_{y}}\hat{{\bf K}})$ where $\hat{{\bf K}}$\cite{Haldane F D M2}
is a unit vector operator in the 3D space (the $z$-component can be described by the Dirac mass)
as $\hat{{\bf K}}={\bf K}/|{\bf K}|
=({\bf k}^{2}{\rm cos}^{2}\theta,{\bf k}^{2}{\rm sin}^{2}\theta,m_{D}^{2})/\sqrt{{\bf k}^{2}{\rm cos}^{2}\theta+{\bf k}^{2}{\rm sin}^{2}\theta+m_{D}^{2}}$,
which reflects the topological invariant of the ground state (or the vacuum) as a nonlocal quantum number,
other quantity like the Berry phase, topological charges\cite{Volovik G E2}, 
and the pseudospin winding number\cite{Park C H2}.
The induced SC is voluerable against the disorder unless the Rashba-coupling is large-enough,
since the Rashba-coupling is help to robust against weak electron-electron interactions (like the backscattering) and disorder.

In 3D space, the $\vec{d}$ is along the nodal direction and it's parallel to the orbital angular momentum $\hbar\hat{L}$ which is along the $z$-axis
due to the nuclear dipolar which is important 
for the gapless excitation\cite{Leggett A J},
the local spin density $\hat{I}$ for this 3D model has
\begin{equation} 
\begin{aligned}
\hat{I}_{x}=&\frac{1}{2}(\psi^{\dag}_{\uparrow}\psi_{\downarrow}-\psi^{\dag}_{\downarrow}\psi_{\uparrow}),\\
\hat{I}_{y}=&\frac{1}{2}(\psi^{\dag}_{\uparrow}\psi_{\downarrow}+\psi^{\dag}_{\downarrow}\psi_{\uparrow}),\\
\hat{I}_{z}=&-i\psi_{\uparrow}\psi_{\uparrow},
\end{aligned}
\end{equation}
thus in unepitaxial case, the nonmetallic surface state is possible when the local perturbation coup to the $\hat{I}_{z}$
(i.e., the component of local spin density which is normal to the surface [100]).
Here such perturbation here may caused by the external magnetic field or the internal spin interaction,
in fact, for the thermodynamic quantitys in our tight-binding model, including the interband interaction and the orbital or spin susceptibility, etc.,
their time evolution is associate with these perturbations which may induce the quench effect(cf.Ref.\cite{Wu C H}) as well as the band energy spectrum\cite{Piéchon F}.
While the pairing potential order parameter
$\mathcal{G}_{p}(k)=\Delta(k)\mathcal{G}_{o}$ where the orbital order parameter for the chiral pair is $\mathcal{G}_{o}(k)=\mathcal{G}_{p}(d_{1}+id_{2})\frac{l_{F}}{\hbar}$
where $l_{F}$ is the Fermi wavelength.
This can be used the investigate the fluctuation-behavior of spin in the chiral states with edge current.
In fact, the $d_{xy}$-wave may disappear if there exist the spatial inversion symmetry respect to the $\langle {\rm 100}\rangle$ axis
and the time-reversal invariant (TRI) was broken for the ,
while the reflection symmetry about the $z$-axis will also suppress the mixture of the $d_{xz}$ bands or the $d_{yz}$ bands with the $d_{xy}$ bands.
It is not doubt that this $d_{xy}$ band which is quasi-2D in the FS is more stubborn that the other two which is quasi 1D for silicene,
and the reflection symmetry was broken and the TRI keeps.
But in the case that the pair hopping between the $d_{xz}$ bands or $d_{yz}$ bands with the $d_{xy}$ band is precise,
the triplet SC also can be realized. 
Then a silicene system expression for the four-band tight-binding model is given by the Ref.\cite{Liu C C,Ezawa M2,Ezawa M4,Ezawa M3,Ezawa M,Zhang J} as
\begin{equation} 
\begin{aligned}
H=&t\sum_{\langle i,j\rangle ;\sigma}c^{\dag}_{i\sigma}c_{j\sigma}
+i\frac{\lambda_{{\rm SOC}}}{3\sqrt{3}}\sum_{\langle\langle i,j\rangle\rangle ;\sigma\sigma'}\upsilon_{ij}c^{\dag}_{i\sigma}\sigma^{z}_{\sigma\sigma'}c_{j\sigma'}
- i\frac{2R}{3}\sum_{\langle\langle i,j\rangle\rangle ;\sigma\sigma'}c^{\dag}_{i\sigma}(\mu_{ij}\Delta({\bf k}_{ij})\times {\bf e}_{z})_{\sigma\sigma'}c_{i\sigma'}\\
&+iR_{2}(E_{\perp})\sum_{\langle i,j\rangle;\sigma\sigma'}c^{\dag}_{i\sigma}(\Delta({\bf k}_{ij})\times {\bf e}_{z})_{\sigma\sigma'}c_{i\sigma'}
-\frac{\overline{\Delta}}{2}\sum_{i\sigma} c^{\dag}_{i\sigma}\mu_{ij} E_{\perp}c_{i\sigma}\\
&+M_{s}\sum_{i\sigma}c^{\dag}_{i\sigma}\sigma_{z}c_{i\sigma}
+M_{c}\sum_{i\sigma}c^{\dag}_{i\sigma}c_{i\sigma}
+U\sum_{i\sigma}\mu_{ij}n_{i\sigma},
\end{aligned}
\end{equation}
where $t$ is the nearest-neoghbor hopping, which been measured as $1.12$ eV for the $\pi$ bands\cite{Liu C C} and $1.6$ eV\cite{Ezawa M3} for both the $\pi$ band 
and $\sigma$ band.
$\langle i,j\rangle$ and $\langle\langle i,j\rangle\rangle $ are the nearest-neighbor pairs and the next-nearest-neighbor pairs, respectively. 
$\mu_{ij}\pm 1$ denote the $A$ ($B$) sublattices. 
${\bf k}_{ij}=\frac{{\bf d}_{ij}}{|{\bf d}{ij}|}$ is the next-nearest-neighbor vector.
$R$ is the small instrinct Rashba-coupling due to the low-buckled structure, which is relate to the helical bands
(helical edge states) and the SDW in silicene, and it's disappear in the Dirac-point ($k_{x}=k_{y}=0$) due to the special geometry of Dirac-point.
The existence of $R$ breaks the U(1) spin conservation (thus the $s^{z}$ is no more conserved) and the mirror symmetry of silicene lattice.
$M=M_{s}+M_{c}$ is the exchange field which breaks the spatial-inverse-symmetry and the $M_{s}$ is related to the out-of-plane FM exchange field
with parallel alignment of exchange magnetization
and $M_{c}$ is related to the CDW,
which endows sublattice pseudospin the $z$-component\cite{Ezawa M6}.
While for the out-of-plane AFM exchange field $M_{s}^{AFM}$ which is not contained here with antiparallel alignment of exchange magnetization.
Here the $M$ is applied perpendicular to the silicene, and it can be rised by proximity coupling to the ferromagnet\cite{Ezawa M}.
Thus the induced exchange magnetization along the $z$-axis between two sublattices-plane is
related to the SOC, Rashba-coupling, and even the Zeeman-field since it will affects the magnetic-order in $z$-direction.
In fact, if without the exchange field and only exist the SOC, the spin-up and spin-down states won't be degenerates but will mixed around the crossing
points between the lowest conduction band and the highest valence band just like the spin-valley-polarized semimetal (SVPSM).
$\upsilon_{ij}=({\bf d}_{i}\times{\bf d}_{j})/|{\bf d}_{i}\times{\bf d}_{j}|=1(-1)$ 
when the next-nearest-neighboring hopping of electron is toward left (right),
with ${\bf d}_{i}\times{\bf d}_{j}=\sqrt{3}/2(-\sqrt{3}/2)$.
The term contains the exchange field $M$ is the staggered potential term induced by the buckled structure
which breaks the particle-hole symmetry.
Here the coordinate-independent representation of the Rashba-coupling terms is due to the broken of inversion symmetry as well as the mirror symmetry,
and the last term is the Hubbard term with on-site interaction $U$ which doesn't affects the bulk gap here but affects the edge gap.
Thus the $U$ is setted as zero within the bulk but nonzero in the edge, 
which is also consist with the STM-result of silicene that the edge states have 
hihger electron-density than the bulk.

the Hamiltonian of exchange field can be divided into the spin one and charge one,
\begin{equation} 
\begin{aligned}
H_{M}=&H_{M_{s}}+H_{M_{c}}\\
=&J_{s}\left[\sum_{i}(c^{\dag}_{i}\sigma_{z}c_{i})_{\eta=A}-(A\rightarrow B)\right]+
J_{c}\left[\sum_{i}(c^{\dag}_{i}c_{i})_{\eta=A}-(A\rightarrow B)\right],
\end{aligned}
\end{equation}
where $J_{s}$ is the coupling strength and the $J_{s}$ here breaks the TRI while $J_{c}$ not,
thus when $J_{s}=0$ the TRI keeps and the states with up-spin and down-spin are degenerated but with opposite Chern numbers if the $J_{c}\neq 0$,
when $J_{c}=0$ the magnetizations have $M_{A}+M_{B}=1$ for Hubbard model
and the ferromagnetism is increase with the value of $\sum_{i}|M_{Ai}+M_{Bi}|$.
The term of $H_{M_{s}}$ split the spin-degenerate states and lifts the anisotropic chiral edge model.
The above Hamiltonian can be diagonized into 
\begin{equation} 
\begin{aligned}
H_{M}(k)=&\sum_{k,\sigma}c^{\dag}_{k\sigma} H_{M\sigma}c_{k\sigma},\\
H_{M}=&{\rm diag}[J_{s}+J_{c},-J_{s}-J_{c},-J_{s}-J_{c},J_{s}-J_{c},-J_{s}+J_{c},-J_{s}+J_{c}].
\end{aligned}
\end{equation}
We found that the term $H_{M_{c}}$ doesn't breaks the gapless edge state which protected by the TRI,
but breaks the spatial inverse symmetry,
and it can't be able to split the spin-degenerate state unlike the $H_{M_{s}}$ and $E_{\perp}$.
In the absence of the perpendicular electric field and CDW-order-like exchange magnetization,
the quantum spin Hall (QSH) phase emerges when $M_{s}<3.23$ meV which satisfies the $\sqrt{3}\lambda_{SOC}-M_{s}>\sqrt{M_{s}^{2}+4R^{2}}$
and also well agrees with our computing results.
And it has perfect four-fold degenerate (spin and valley) in the zero-field limit.
In fig.3, we plot the band gap evolution in the valley K under the effect of $M_{s}$ (group 1) and $M_{c}$ (group 2) and both of them (group 3),
which are have great agreement with the results discussed above. Note that for non-spin-degenerate states in group 1 and 3,
the spin-dependent-exchange field $M_{s}$ lift the up-spin band upward and lower the down-spin band downward, respectively, while
the charge-dependent-exchange field $M_{c}$ doesn't open up the gap between the TR pairs,
i.e., the spin degenerate is not broken.
The quantum anomalous Hall (QAH) phase is center in the region of $M_{s}=7.8$ meV in the group 1,
while the group 3 in $M_{s}=7.8$ meV has QAH phase with opened band gap.
We can also see that the $M_{c}$ has a unconspicuous effect on the bulk-gap-opening,
but it shifts the valley upward which may cause the valley-polarization and 
it reaches the trivial band insulator when $M_{c}=7.8$ meV as shown in the plot
(while for the out-of-plane AFM exchange field $M_{s}^{AFM}$, the trivial band insulator will emerges in the same intensity).
In the following, except with special instructions, we treat $M$ as only the spin-dependent part $M_{s}$.
In fig.3, we set the on-site interaction U as zero, i.e., the Hubbard repulsive is zero, thus the electron
excitation-energy of edge states is maximum.

Similar to the interlayer displacement field which acts on the bilayer or multilayer structures,
the external potential between the two kinds of sublattices is $eE_{\perp}\overline{\Delta}$.
This potential difference also provides the prerequirement for the excitons-the bound states of a electron-hole pair-
just by the gated silicene and don't need to construct the bilayer form like graphene\cite{Park C H,Ju L}.
Since the bulked structure provides valley degree of freedom,
it's possible to obtain the different masses in the neighbor Dirac-point ($K$ and $K'$) by the external field, laser beam,
or some other electromagnetic radiations\cite{Ezawa M3} and 
the synthesization with the metallic substrate\cite{Liu H,Zhang X L} which may dramatically affect the electronic band stucture and disturb the band topology
of silicene 
and intensify the symmetry-broken and the valley-polarization 
due to its sensitive upper and bottom surface and the active 3$p_{z}$-orbit,
and also can be implemented by imposing the tensile strain\cite{Wan W} like the graphene\cite{Zhang D B}.
The different masses also results in the different low-energy dynamics in the neighbor Dirac-point 
and different velocities of spin polarizations $\sigma$ and $\sigma'$ which are
carriered by the edge states along $k_{x}$ direction that distincted by the periodic boundary phase $\phi$ (modulo 2$\pi$) 
(see the Schematic diagram in Fig.4(e)).
$\lambda_{SOC}=3.9$ meV and $R_{2}(E_{\perp})=\frac{eE_{\perp}\overline{\Delta}}{3V^{(1)}_{sp\sigma}}\xi$\cite{Min H} 
is the extrinsic Rashba-coupling induced by the electric field.
The stength of the SOC is $\xi=0.775$ meV for the buckled silicene whose 
Hamiltonian can be expressed as $H_{{\rm SOC}}=\frac{\hbar}{4m^{2}c^{2}}{\pmb \sigma}\cdot(\nabla_{{\bf r}}U\times {\bf p})
=\lambda_{{\rm SOC}}\hbar^{2}\hat{L}\hat{S}$
\cite{Kane C L,Liu C C,Liu C X,Min H} with $m$ the free electron mass
where $\hat{S}$ is the spin angular momentum and
with the effective force (nearest-neighbor or next-nearest-neighbor) ${\bf F}\equiv\nabla_{{\bf r}}U=\partial_{t}{\bf k}_{c}$
acting on the center of mass,
where ${\bf k}_{c}$ is the momentum of the center of mass,
and it also has ${\bf F}=\int \dot{{\bf A}}\cdot d{\bf r}$ where ${\bf A}$ is the gauge vector potential, according to the Faraday's law.
The contributions from both the $\hat{L}$ and $\hat{S}$ are guaranteed by the broken of the inversion symmetry in silicene due to the next-nearest-neighbor hopping
which can be can be presented by the basis of the all three Pauli matrices\cite{Liu C X} (thus it's robust against the perturbation)
\begin{equation} 
\begin{aligned}
H_{{\rm SOC}}=\frac{\lambda_{{\rm SOC}}\hbar^{2}}{2}
\begin{pmatrix}[1.5] 0&i\sigma_{z}&-i\sigma_{y}\\
-i\sigma_{z}&0&i\sigma_{x}\\
i\sigma_{y}&-i\sigma_{x}&0
\end{pmatrix}.
\end{aligned}
\end{equation}
The last term of above equation decides the sign (the direction) of the half-instrinct-SOC by the components of $p$-orbit.
The zero $M$ corresponds to the zero Chern number $\mathcal{C}$ and zero Hall conductance $\sigma_{xy}$,
and it
leads to the normal band insulator which with two spin-degenerate Dirac-cone in the quantum spin Hall state.
And the $M$ may arised due to the interaction with the ferromagnetic commensurate substrate.
The SOC and the variational band structures lead to the anisotropic exchange coupling in silicene,
thus the exchange magnetization is sublattice-dependent (or orbital-dependent),
but here we takes only the effective two bands model and hence ignore the 
exchange anisotropy together with the magnetic anisotropic term.
Distinct phases are achieveble under the effects of both the $M$ and $E_{\perp}$(cf. Refs.\cite{Ezawa M,Ezawa M3,Ezawa M4})
in the sample of a strip of silicene (nanoribbon),
which 
including four kinds of phase that own a gapless nontrivial (with nonzero Chern number) topological flat band 
(which formed by the degenerated edge states in opposite directions within the bulk gap and with weak band dispersion or even dispersionless)
edge model: spin-polarized semimetal (SPSM) phase, spin-valley-polarized semimetal (SVPSM) phase,
quantum anomalous Hall (QAH) phase, and the non-polarized semimetal phase (NSM).
The helical currents are into the bulk in these phases,
and the topological flat band has the energy twice as the pure spin current
in helical edge model.
In SPSM phase, the topological flat band connects the two Dirac-point within the spin texture 
which are not in the same plane (but staggered by a distance as $\overline{\Delta}$),
and with a length as $k_{F}{\rm cos}\theta$
where $k_{F}=E_{F}/(\hbar v_{F})$ is the Fermi momentum and $\theta$ is the angle of the phase of spin texture and has the relation
$H(M)=\Theta H(M)\Theta^{-1}$ (see below) in $\theta=0,\pi$.
Even in the strong electron-interaction ot hard-core bosons system with filling factor's\cite{Watanabe H,Lian B} constrainsand the resulting fractional QHE,
the topological flat band is still possible than formed by the highly degenerate Landau level\cite{Yang S,Wang Y F}.
And these four phases with topological flat band have a common characteristic:
they all exsit in the case which at least has one of the $M$ and $E_{\perp}$ equals zero,
e.g., the QAH phase with the nonzero $U$, $M$, SOC , $R$ and nearly zero $E_{\perp}$ in the doped\cite{Zhang J} and undoped\cite{Ezawa M} silicene,
and the gapless nearly flat edge states in QAH splitted by the exchange field (or the magnetic field in the $z$-direction\cite{Wang F}) 
is different from the double degeneracy completely flat bands
which are pertected by the TRI;
while in the other case,
the particle-hole symmetry protected zero mode cannot exist due to the perturbations,
and results the phases
like the spin quantum anomalous Hall insulator (SQAH) and the spin-polarized quantum anomalous Hall insulator (SPQAH).
In fact, the topological phase transitions are dominated by the
spin-valley-dependent Dirac mass $m_{D}$ which is half of the band gap.
Furthermore,
for the increasing CDW and AFM SDW order, which with conjugate fluctuation in the helical edge states, 
the trivial band insulators also emerge in the region of zero $M$
\cite{Ezawa M4} and the AFM order take the region of $U/t>4.3$\cite{Meng Z Y} for the Hubbard model on honeycomb lattice
at half-filling (chemical potential $\mu=0$ and with the particle-hole symmetry (at the $M$-point where wih the TRI $\Theta=i\sigma_{y}\mathcal{K}$))
with only the nearest-neighbor-hopping, i.e.,$t'=0$.
And there is a spin liquid in $t'=0$ region between the semimetal phase (also the $d$-wave SC region;
and with a nernst region which as a transition zone to the superconductor) and the AFM Mott insulator (i.e., in $4.3>U/t>3.7$;
(cf. the phase diagram in Fig.5 or the Refs.\cite{Wu C,Zheng D,Meng Z Y,Honerkamp C,Laubach M}).
The strong intrinsic SOC makes silicene be a good material to implement the Kane-Mele-Hubbard (KMH) model.
As shown in the Fig.5 where the QMC sign problem is vanishing,
the metallic phase emerges in the critical ratio $|R|/|\lambda_{SOC}|=1$.
And according to the result of Ref.\cite{Ezawa M}, the metallic phase emerges in the point $M=\lambda_{SOC}$ with zero external field.
The 2D TI phase of silicene has significant difference
with the ordinary insulators by the gapless helical edge states which with odd number of channels (the charge or spin channel;
and thus have the odd number of TR pairs),
and this gapless helical edge states is gaped out by the magnetic order (magnetic impurity scattering or concentration\cite{Bernevig B A2})
and resulting a lattice mismatch and the broken of TRI.
The 2D TI will be unstable in the AFM SDW phase since AFM SDW phase favors the large $U\ (\le {\rm half}\ {\rm filling})$
and thus gives rise to the strong repulsive effect.

To increasing the (Rashba) SOC to achieve more applications on the spintronics,
a direct way is the uniaxial pressure which can changes the $U$ within the above expression of $\lambda_{{\rm SOC}}$,
or carry out the ploydirection-pressure to reduce the atom radius and then increse the strength of SOC (through the relativistic effect),
another way is through the impurity adatoms' surface deposition.
Here the nearest-neighbor SOC is vanish due to the mirror symmetry respect to the single bands,
but the perpendicular electric field allow the exist of Rashba SOC between the nearest neighbors by broken the mirror symmetry (thus the inversion symmetry is broken).
In fact, the next-nearest-neighbor hopping breaks the particle-hole symmetry (which requires the homogenate on-site potential) 
between the $K(K')$-point and $M$-point, 
but the mirror symmetry along the $\Gamma-M$ are preserved and that's why there is not exist Dirac-point in the $M$-point 
unlike the $K(K')$ one.
The preserved mirror symmetry allowed the spin dependent term only exist in the midpoint of the BZ periodic boundary (or the disorder-induced twist phase;
i.e, along the $k_{y}$ direction) with a TRI-protected (Kramers two-fold) degeneracy
by the spin degreed of freedom coupling in such point with the TRI even in the presence of Haldane model\cite{Sheng D N}.
That means that, the degeneracy in these points are robust against the TRI perturbations.
Due to the strong intrinsic SOC for silicene, the TRI operator is antiunitary: $\Theta^{2}=(i\sigma_{y}\mathcal{K})^{2}=-1$,
which has a robust confine by a two-time thermal average form (the Keldysh Green's function)
$G_{ij}(t,t')=-i\langle \mathcal{T}\psi_{i\uparrow}(t)\psi^{\dag}_{j\downarrow}(t')\rangle=\delta(t-t')/4\pi$
which been found that decay as $1/\Delta t^{K+\frac{1}{K}}$ in the edge excitation\cite{Zheng D} 
where $K=\sqrt{\frac{1}{1+\frac{2J_{z}a}{\pi\hbar v_{F}}(1-{\rm cos}(2k_{F}a))}}$ is the Luttinger parameter\cite{Lancaster J},
where $k_{F}=v_{f}\pi/a$ is the Fermi momentum,
$J_{z}$ is the exchange interaction with the forwaid scattering in the XXZ-type coupling
and it vanishes for the gapless XX model with $K=1$.
While for the case that without SOC, $\Theta^{2}=1$ and it suggest a conventional insulator and don't has TR polarization\cite{Fu L2} in the edge state spectra.
There are four Kramers degeneracy points in the Helical edge model of the 2D TI
that both have $H(M)=\Theta H(M)\Theta^{-1}$ or $|u_{n{\bf k}}\rangle=\Theta|u_{n{\bf k}}\rangle$
(corresponds the four points $(0,0)$, $(0,\pi)$, $(\pi,0)$, $(\pi,\pi)$ in the projected square BZ,
while at the points $(0,\pi/2)$, $(\pi/2,0)$, $(\pi/2,\pi)$, $(\pi,\pi/2)$
have obviously different results as discuss in the following) with $\Theta^{2}=1$ (while $\Theta^{2}=0$ correponds the broken TRI),
and the elastic backscattering is forbidden in these points.
Each Kramers degeneracy pair carrie the totally zero spin (two orthogonal spins) and opposite momentums
while the other places in edge states are robust against elastic backscattering and with the conserved spin,
e.g., the helicity of edge states for QSHI phase that the electrons carrier the same spin can only move in one direction,
and form the anisotropic helicals.
At the TRI particle-hole symmetric points, the nonchiral umklapp backscattering term (not contains the forward scattering exchange $J_{z}$) which is\cite{Wu C2}
\begin{equation} 
\begin{aligned}
J_{z}a\int dx e^{-i\phi}\psi^{\dag}_{L\uparrow}(x)\psi^{\dag}_{L\uparrow}(x+a)\times
e^{-i\phi}\psi^{\dag}_{R\downarrow}(x)\psi^{\dag}_{R\downarrow}(x+a)+h.c.
\end{aligned}
\end{equation}
in Fermion language
is allowed, where $\phi=\varphi x$ with the left move and right move N\'{e}el order $\varphi=\pi$ 
(half of the phase of Wigner-Seitz unit cells) and with the scaling dimension 
just be one Luttinger parameter $K$\cite{Zheng D} at commensurate filling\cite{Wu C2}
where the phase transition to a insulator with gap happen. $g$ is the scattering strength factor. 
And here the Fermi velocity has $v_{F}=1+\frac{J_{z}a}{\pi\hbar v_{F}}(1-{\rm cos}(2k_{F}a))$.
While the chiral term renormalize the Fermi velocity in the homogenerate system without the domain wall,
the inhomogenerate case will be discuss below.
For the random AFM Kondo singlet with $K\ll 1$ disorder after the quenching, the renormalization group flux in the edge states
is randomly shifted, and the random AFM fixed points are occupied by the localized spinless fermions (noninteraction) 
which rise the insulating phase.
Except that, in $K\ll 1$ region, the band gap power law damping with the increase of $K$.
The umklapp backscattering term conserve the momentum
(which is important for the electrons or photons scattering and the redistribution of edge state and even the resistance) 
and change the U(1) spin/pseudospin rotation symmetry to the $Z_{2}$ invariant
but makes the helical edge state become unstable.
The umklapp backscattering may rised from the anisotropic spin or electron interaction
with the broken U(1) rotation symmetry and quasimomentum conservation.
These places which obey the TRI follows the relation $H({\bf k})=\Theta^{-1} H(-{\bf k})\Theta$\cite{Fu L} or $|u_{n-{\bf k}}\rangle=\Theta|u_{n{\bf k}}\rangle$,
and the former can also be represented by $H({\bf k})=U H^{*}(-{\bf k})U^{\dag}$ where $U$ is the unitary matrix which with $UU^{*}=-1$,
or be represented by the S-unitary matrix\cite{Kane C L}: $\Psi_{{\rm out}}=S\Psi_{{\rm in}}$, $\Psi^{*}_{{\rm in}}=S\Psi^{*}_{{\rm out}}$, 
$S^{\dag}S=I$, $S^{*}S=I$, $S=\sigma_{y}S^{T}\sigma_{y}=S^{T}$, where $I$ is the identity matrix.
The cell periodic Bloch eigenstate is $|u_{n{\bf k}}\rangle=e^{-i{\bf k}\cdot{\bf r}}|\psi_{n{\bf k}}\rangle$ 
where $\psi_{n{\bf k}}$ is the $n$th band Bloch wave function.
The periodic TRI diagonal Bloch Hamiltonian which communicate with $\Theta$ is $H_{d}=e^{i{\bf k}\cdot{\bf r}}He^{-i{\bf k}\cdot{\bf r}}$.
For ${\bf k}\rightarrow{\bf k}+{\bf G}$,
the Bloch wave function which with integer filling band structure has 
\begin{equation} 
\begin{aligned}
|\psi_{n{\bf k}}\rangle=e^{i{\bf k}\cdot{\bf r}}|u_{n{\bf k}}\rangle
=e^{i{\bf k}\cdot{\bf r}}\cdot e^{i{\bf G}\cdot{\bf r}}|u_{n{\bf k}+{\bf G}}\rangle=-e^{i{\bf k}\cdot{\bf r}}|u_{n{\bf k}+{\bf G}}\rangle,
\end{aligned}
\end{equation}
and the Bloch eigenstates in each Kramers degeneracy pair will adiabatically switch place between the two Kramers TRI momentums as shown in the Fig.4(e).
For different bands $n$ and $n$', it has 
\begin{equation} 
\begin{aligned}
\langle u_{n'{\bf k}'}|e^{-i({\bf k}'-{\bf k})\cdot r}iv|u_{n{\bf k}}\rangle
=\delta({\bf k}'-{\bf k})\langle u_{n{\bf k}}|iv|u_{n{\bf k}}\rangle,
\end{aligned}
\end{equation}
where $v=\frac{\partial}{\hbar \partial {\bf k}}$ is the velocity operator. 
This is also reminiscent to the geometry of cylindric which threaded by a flux (edge states
along the zigzag-direction) $\phi=\frac{nhc}{2e}\cite{Byers N},\ n\in N$ which act as the circumference
and localized by one lattice constant in the $k_{x}$ direction with open boundary condition
which support a integrable model in the homogenate and conserved case.
Except that, the zigzag edges also found that have higher thermal conductance and stronger spin-polarized current in the quantum Hall insulating phase
than the armchair one.
Their comparation is shown in the Fig.6, which the width (atom number) is setted as 32.
And the zigzag panel is in the trivial phase without the edge states and bandgap and 
the bands cross at $\pi$.
From Fig.6, we can see the edge gap is about 0.16 eV in the presence of $\lambda_{SOC}$, and
it's well know that the zero-energy-model can't emerge in the spectrum of the finite length armchair silicene along the periodic $k_{y}$
because its ground state energy is negative and the edge gap exponentially approaches to the zero\cite{Ezawa M7}
similar to that of the hexagonal nanodisk od the graphene\cite{Ezawa M8}.
In the third panel of Fig.6, we shows the amplitude of wave function of function from the zero-energy-state to the negative-energy-state,
where the minimum period is setted as $N=40$, $m_{D}=0.2t=0.32$ eV, in this case we can obtain a much longer penetration length of armchair silicene than the 
zigzag one, which is $l_{{\rm arm}}=$3.272 \AA\ for the armchair which is very close to the lattice constant and $l_{{\rm zig}}=$0.656 \AA\ for the zigzag,
and the ratio of $l_{{\rm arm}}/l_{{\rm zig}}$ increase with the decrease of $m_{D}$.
It's found that the wave function is square integrable for the bonding state of armchair.

The width of one lattice constant is since the kinetic energy is typically provided by the strong intrinsic SOC in silicene.
It can also be imaginated as a generalized cylinder spanned by the vectros $k_{x}$ and $k_{y}$
(or the reciprocal lattice vectors)
and form a torus to see the $Z_{2}$ invariant in the 
$n$-field configration of silicene\cite{Liu C C2,Huang Z Q}.
Net magnetic flux together with the twist angle rise the ground state spin stiffness $\rho_{s}^{0}=N\frac{\partial^{2}E_{0}}{\partial \phi^{2}}$
for the homogenate case.
Also, the approximated-Chern number (lattice-spin Chern number\cite{Kalesaki E} which is still a integer and mod 2)
can be obtained by calculating the $n$-field,
and it becomes spin Chern number in the thermodynamic limit, i.e.,
$\mathcal{C}_{s}=\lim_{N=\infty}\frac{1}{4\pi i}\sum_{{\bf k}}F({\bf k})$\cite{Fukui T,Kalesaki E},
where
\begin{equation} 
\begin{aligned}
F({\bf k})={\rm ln}[U_{x}({\bf k})U_{y}({\bf k}+\hat{{\bf x}})U^{-1}_{x}({\bf k}+\hat{{\bf y}})U^{-1}_{y}({\bf k})],
\end{aligned}
\end{equation}
with the gauge potential $A^{\mu}={\rm ln}U_{\mu}
={\rm ln}\frac{\langle u_{n{\bf k}}|u_{n{\bf k}+\hat{{\bf x}}}\rangle}{|\langle u_{n{\bf k}}|u_{n{\bf k}+\hat{{\bf y}}}\rangle|}
={\rm Tr}\langle u_{n{\bf k}}|v|u_{n{\bf k}+\mu}\rangle$
where $\mu=\hat{{\bf x}},\hat{{\bf y}}=k_{x}/N,k_{y}/N$ in a $N\times N$ grid BZ
and the gauge potential is invariant under the period of $N$.
While in more generate case with the Coulomb potential $V({\bf r})$ and SU(2) symmetry with the
$\lambda_{SOC}=0$ (since the SOC will modifies the spin rotation and the charge carriers
by generating AFM anisotropic exchange, but it's still possible for the SU(2) to exist with nonzero SOC when that strength of $R$
equals the Dresselhaus coupling\cite{Li Z},
or when the external polarized-light is aboves the critical frequency (see below)
and thus the contributions from interband scattering to the longitudinal conductivity becomes zero),
the spin stiffness is given as\cite{Moon K}
\begin{equation} 
\begin{aligned}
\rho_{s}^{0}({\bf r})=\frac{1}{16\pi^{3}}\int dkd^{3}r\ V({\bf r}){\bf K}\cdot(1-g({\bf r}))e^{-2i{\bf k}\cdot{\bf r}},
\end{aligned}
\end{equation}
with the universal function $g(r)\sim 1/e^{-\sqrt{r}}$ and Coulomb potential $V_{c}({\bf r})=\frac{e^{2}}{4\pi\epsilon|{\bf r}|}$.
The final resulting band gap can be obtained by diagonalizing the above tight-binding Hamiltonian as
\begin{equation} 
\begin{aligned}
\Delta_{\pm}(E_{\perp},\overline{\Delta})=|2\eta s_{z}\lambda_{{\rm SOC}}-eE_{\perp}\overline{\Delta}+2Ms_{z}|,\ M=\frac{1}{2}|M_{A}-M_{B}|,
\end{aligned}
\end{equation}
where $e$ is the elementary charge and we omit it in the following.
$\eta=\pm 1$ denotes the K (K') valleys.
This final band gap contains both the effective first- and second-order SOC, 
which is $\lambda_{{\rm SOC}}\approx \frac{(\Delta_{\rm SOC}^{0})^{2}|\epsilon_{3s}-\epsilon_{3p}|}{18(V^{(1)}_{sp\sigma})^{2}}$\cite{Liu C C2,Yao Y}
where $\epsilon_{3s}$ is the energy of 3$s$-orbit and $\epsilon_{3p}$ is the energy of 3$p$-orbit (mainly the $3p_{z}$),
$\Delta_{\rm SOC}^{0}$ is the intrinsic SOC which is $34$ meV\cite{Liu C C2},
$V^{(1)}_{sp\sigma}=2.54$\cite{Liu C C} is a parameter about the $\sigma$-bond formed by $s$ and $p$ orbits.
In this case, the quantized chiral charges in the massless Dirac cones which are exist in pairs can't be exist due to the SOC-gap-induced mass term,
but the spin eigenstates are remain chiral due to the Rashba coupling.

The nonzero next-nearest-neighbor hopping $t'$ would shift the Dirac cones and breaks the particle-hole symmetry 
even in the case of zero-chemical-potential since the SU(2) symmetry is broken, (similar to the $d/l>0$ case of the quantum bilayer Hall system\cite{Moon K})
except for the case that $t'$ is purely imaginary for the case which with a 1D single-particle chiral edge model
and thus keeps the particle-hole symmetry and with zero charge- and spin-edge current 
(which are odd under the particle-hole transition nomatter in $k_{x}$-direction or $k_{y}$-
direction)\cite{Zheng D},
like the next-nearest-neighbor hopping $i\lambda_{{\rm SOC}}$.
As a example, the perfect flat band and the particle-hole symmetry and the TRI coexist in the QAH state of Kane-Mele-Hubbard (KMH) model, 
which contains only the purely imaginary next-nearest-neighbor hopping.
Furthermore, the notorious sign problem of the Quantum Monte Carlo (QMC) simulation in our tight-binding model $H'$ 
can be avoided (Ref.\cite{Zheng D} or see Appendix.C) due to the exact particle-hole symmetry
just like the case of spin-1/2 Hubbard model in a half-filling or a biparticle lattice,
and the helical edge states is unstable under the strong Andreev-scattering and therefore the Tomonaga-Luttinger liquid can't emerge.
There are two conduction bands and two valence bands both have the spin chirality symmetry by the chiral spin angular momentum in QAH phase.
With the broken SU(2) symmetry, we assume the XXZ-type exchange interaction $J_{z}=J_{\|}=J$ in the renormalization group (RG),
i.e., without the Kondo impurities, then the RG equaltion is $dJ/d({\rm ln}\Delta)=-\rho_{F}J^{2}$
with equivalent forward scattering and backward scattering with a finite Hubbard interaction, 
where $\Delta$ is the band gap
and $\rho_{F}$ is the DOS at Fermi level.
Then the nearest-neighbor exchange is isotropic as $J\sum_{\langle i,j\rangle}(s^{x}_{i}s^{x}_{j}+s^{y}_{i}s^{y}_{j}+s^{z}_{i}s^{z}_{j})$
while the next-nearest-neighbor exchange is anisotropic as 
$J'\sum_{\langle\langle i,j\rangle\rangle}(s^{x}_{i}s^{x}_{j}+s^{y}_{i}s^{y}_{j}-s^{z}_{i}s^{z}_{j})$.

The Zeeman field which is against to the Rashba-coupling is closely related to the exchange magenetization $M$ and it may tilted a AFM order as 
$|\theta;\nearrow\swarrow\rangle=e^{i\theta\sum \sigma_{y}}|\uparrow\downarrow\rangle$
especially for the isotropic Hubbard model, and even turn it into the FM one if the Zeeman field
strong enough.
For the helical gapless edge modes (i.e., the zero model with the zero-energy states through the unitary transformation
$\psi_{\eta}(k)={\rm exp}(-i\phi_{k}\sigma_{z}/2)\frac{\psi_{\uparrow}(k)+\eta\psi_{\downarrow}(k)}{\sqrt{2}}$) in the QSH insulator phase,
or in a variational wavw function form
$|\psi_{\eta}(k)\rangle=\prod_{k}{\rm exp}(-i\phi_{k}\sigma_{z}/2)\frac{\psi_{\uparrow}(k)+\eta\psi_{\downarrow}(k)}{\sqrt{2}}|0\rangle$.
The strong Zeeman field may tilted the helical bands toward the $z$-direction with a angle $\theta={\rm tan}^{-1}\frac{M_{z}}{\sqrt{M_{z}^{2}+R^{2}{\bf k}^{2}}}$
which control the phase and leads to
$\psi_{\pm}(k)={\rm exp}(-i\phi_{k}\sigma_{z}/2-i\theta\sigma_{y}/2)\frac{\psi_{\uparrow}(k)\pm\psi_{\downarrow}(k)}{\sqrt{2}}$.
Further, when both the Rashba-coupling and the Zeeman field take effect in the system,
a Majorana bound state would appear at the edge\cite{Black-Schaffer A M} and thus support the Majorana fermion.
We commend that the divided bands by the Zeeman splitting has a totally zero Chern number
like the $\pm 1$ Chern number in the Haldane model
and shows the integer quantum Hall effect (QHE).

\subsection {The rivalry effect between the electric field and SOC}

The electric-field induced phase transition has been reported in Ref.\cite{Zhang L D} 
which turn doped silicene from $d_{1}+id_{2}$-pairing-wave superconducting phase
to the triplet $f$-wave superconducting phase, and the
pairing symmetries is change from he nearest-neighbor pairing into the next-nearest-neighbor one.
The low-buckled silicene has more remarkable QSHE than the planar one (see the band gaps as shown 
in the band structure of silicene), which is consistent with the result that the intrinsic SOC of graphene is found that to be
too weak to realize the QSHE unless under the extremely low temperature\cite{Yao Y}.
For the pure band-gap insulating phase induced by the electric field (without the SOC-effect like the $t-J$ model\cite{Farrell A}
and was applied vertically to avoids the relaxation of the electron spin in the TI surface state\cite{Chung S B}), 
the two-band model which mentiond above has the (here we ignore the gap opened by the crystal field and the effect of Rashba coupling)
following dispersion
\begin{equation} 
\begin{aligned}
\varepsilon_{\pm}=\pm\sqrt{\frac{\Delta_{\perp}}{2}^{2}+v_{F}^{2}k_{\eta}^{2}},
\end{aligned}
\end{equation}
where $k_{\eta}(\eta=\pm {\rm for\ the\ two\ valleys})$ is the valley momentum which is $k_{\eta}=\{k_{+},k_{-}\}$ corresponds to $K$- and $K'$-point respectively,
and $v_{F}=\frac{\sqrt{3}}{2}at$ is the Fermi velocity of the charge carriers which is consisted with the free massless fermions (mainly the $\pi$-electron here)
and the Dirac electrons in the Dirac-cone are in a large number due to the strong SOC in silicene (about 3.9 meV which is much larger than the 0.07 meV of graphene).
The mass of Dirac electrons provided by the SOC breaks the chiral symmetry and 
leads to the single-Dirac-cone state (SDC) which is allowed for the chiral fermion in the 2D lattice\cite{Ezawa M3,Wu C4}
with the special spin helical states as 
$\psi_{\pm}(k)=\frac{\psi_{\uparrow}(k)\pm e^{-i\phi_{k}}\psi_{\downarrow}(k)}{\sqrt{2}}$
and it's impossible for the fermion doubling to emerge
in the Dirac-cone\cite{Ezawa M} and thus the BEC regime which with the tightly bound dimers and the bose liquids can't be found. 
The SDC state can be achieved by applying both the perpendicular circular polarized light and electric field\cite{Ezawa M3}
or by just applying the out-of-plane AFM exchange field in the critical intensity which is about 3.9 meV,
or by turning the $R_{2}(E_{\perp})$ (see below).
Naturely, such a SDC state will exhibits the half-integer Hall conductivity.
The edge models within the bulk gap in SDC state is fully polarized,
i.e., has only two edge within the bulk gap which carry the same spin-direction.
For SDC state under both the effect of out-of-plane AFM exchange field and the CDW-like charge-dependent exchange field,
the bulk gap can be obtained as $|M_{c}+M^{AFM}_{s}|$.

The
$\Delta_{\perp}$ is proportional to difference of onsite-energy between the sublattice A(A') and B(B') in a unit cell,
but also is the electric field indeuced-gap between these two dispersions here. 
These two dispersions are symmetry with each other respect to the transverse middle line (i.e., the Fermi level for undoped case),
and the $k_{+}=((\frac{2\sqrt{3}}{3}k_{x})^{2},0)$, $k_{-}=((-\frac{2\sqrt{3}}{3}k_{x})^{2},0)$.
Thus this dispersion for silicene can also be represented as
\begin{equation} 
\begin{aligned}
\varepsilon_{\pm}=\pm\sqrt{(\frac{\Delta}{2})^{2}+t^{2}[{\rm exp}(i\frac{\sqrt{3}k_{x}}{3})+{\rm exp}(i(-\frac{\sqrt{3}}{6}k_{x}+\frac{k_{y}}{2}))
+{\rm exp}(i(-\frac{\sqrt{3}k_{x}}{6}-\frac{k_{y}}{2}))]^{2}},
\end{aligned}
\end{equation}
which we consider only the nearest-neighbor hopping here.
Nextly we consider the SOC-induced gap in the case which the gapless edge state (currents) is protected by the nontrivial topology.
Then the dispersions is similar to the form of other semimetallic energy band structure,
\begin{equation} 
\begin{aligned}
\varepsilon_{A,\pm}=\pm\sqrt{(\frac{\Delta_{{\rm SOC}}}{2}+\frac{\eta\Delta_{\perp}}{2})^{2}+v_{F}^{2}k_{\eta}^{2}},\\
\varepsilon_{B,\pm}=\pm\sqrt{(\frac{\Delta_{{\rm SOC}}}{2}-\frac{\eta\Delta_{\perp}}{2})^{2}+v_{F}^{2}k_{\eta}^{2}}.
\end{aligned}
\end{equation}
A topological phase transition, which accompanied by sign-changing of the mass term,
between the topological insulating ($\Delta_{{\rm SOC}}>\Delta_{\perp}$ where the QSHE emerge) and the band-gap insulating 
($\Delta_{{\rm SOC}}<\Delta_{\perp}$ where the trivial band insulator emerge)
is possible by turning the external electric field 
(the critical external electric field $E_{c}$ which has $\Delta_{{\rm SOC}}=\Delta_{\perp}$ is in a range of 17 meV/ \AA$\sim$
19 meV/ \AA\ experimentally
which is close to the theoretical result as 16 mev/\AA\ from
$\frac{\Delta}{2}E_{c}=\lambda_{SOC}$ and it's independent of the Rashba-coupling. The semimetallic state (SVPSM) emerge in this critcal area).

We show the energy of unit cell and the band gap at different lattice constant which may affected by the pressure of the external fields,
in Fig.7.
The effect of perpendicular electric field is shown in the two inset where the SOC are considered and not considered, respectively.
The decrease of the value of the critical electric field due to $R_{2}(E_{\perp})$ can be ignored since it's too small\cite{Ezawa M5}.
And the evolustion of spin Chern number in the absence of exchange field and Rashba coupling is
$\mathcal{C}_{s}=1/2\sum_{\eta}(\mathcal{C}^{\eta}_{\uparrow}-\mathcal{C}^{\eta}_{\downarrow})$ 
is from 1 (QSHE) to 0 (SVPSM) to 0 (band insulator).
For dc-transport, we don't consider the thermal broadenand and the scattering factor,
the spin Hall conductivity and the valley Hall conductivity can be written in the following general form
$\sigma_{xy}^{s}=\sum_{\eta}\sigma_{xy}^{\eta\uparrow}-\sigma_{xy}^{\eta \downarrow}$,
$\sigma_{xy}^{v}=\sum_{s_{z}}\sigma_{xy}^{\eta=1,s_{z}}-\sigma_{xy}^{\eta=-1,s_{z}}$,
while the charge Hall conductivity is always $\sigma_{xy}^{c}=\sum_{\eta=\pm 1}\sigma_{xy}^{\eta\uparrow}+\sigma_{xy}^{\eta \downarrow}=0$
both for the dc and ac conductivity\cite{Vargiamidis V}.
We show in Fig.8 (top panel) the spin Hall conductivity and the valley Hall conductivity
together with the spin Chern number and the valley Chern number in zero temperature and for the case that Fermi level ($\omega=0$) 
is lies within the band gap,
i.e., the $\varepsilon_{1}>0$ above the Fermi level and $\varepsilon-\hbar\omega$ below it and it's split beside the $\varepsilon=0$
level if under a magnetic field.
As shown in the figure, the spin Chern number is change with $\Delta_{SOC}/\Delta_{\perp}$,
while the valley Chern number is always zero due to the zero exchange field presented here.
The spin and valley Hall conductivity for the Fermi level within the conduction band in shown in the lower panel of Fig.8.

Due to the Bloch states in the particle-hole (Nambu) space,
the Bloch wave function with two distinguish sublattice states $A=|\uparrow\rangle$, $B=|\downarrow\rangle$
which with the AFM spin order,
can be represented as
\begin{equation} 
\begin{aligned}
\psi_{\Uptheta}(r-r_{\Uptheta})=\frac{1}{\sqrt{N}}\sum_{r_{\Uptheta}}e^{ikr_{\Uptheta}}\phi_{p_{z}}(r-r_{\Uptheta}),\ \Uptheta=A,B
\end{aligned}
\end{equation}
where $N$ is the number of cells and
$\phi_{p_{z}}$ is the wave function of the normalized $p_{z}$-orbitals which is along the direction of order parameter
and this Bloch wave function linearly combines the wave functions of two spin states for two distinguish sublattice states A and B.
And it has the following relations due to the particle-hole symmetry
\begin{equation} 
\begin{aligned}
\psi^{A}_{\eta}(-r)=\eta\psi^{B}_{\eta}(r),\ \psi^{A}_{\eta}(r)=\eta\psi^{B}_{-\eta}(r)\\
\psi^{B}_{\eta}(-r)=\eta\psi^{A}_{\eta}(r),\ \psi^{B}_{\eta}(r)=\eta\psi^{A}_{-\eta}(r),
\end{aligned}
\end{equation}
which is in a similar form with the Eq.(39).
From this Bloch wave function, we can also obtain the order paramater of the of the tight-binding two-band model,
which is $\Psi(r)=\sqrt{M}\sum_{r_{\Uptheta}}\mathcal{A}(r_{\Uptheta})\psi_{\Uptheta}(r-r_{\Uptheta})$.
For this $3p_{z}$-orbit wave function $\phi_{p_{z}}$ of the silicon atoms,
the effective mass can be obtained as (for the conduction band near the Dirac-point in the path of $K-\Gamma$($H-A$),
which has smaller mass that the ones in other paths) (use the unit of $\hbar=a=1$)
\begin{equation} 
\begin{aligned}
\frac{1}{m^{*}}=\frac{\partial^{2}\epsilon_{p_{z}}}{\hbar^{2}\partial k^{2}_{x}},
\end{aligned}
\end{equation}
where the energy of $3p_{z}$-orbit $\epsilon_{p_{z}}$
\begin{equation} 
\begin{aligned}
\epsilon_{p_{z}}=\frac{t}{eE_{\perp}\overline{\Delta}}(
{\rm cos}\frac{\sqrt{3}k_{x}}{3}
+{\rm cos}(\frac{\sqrt{3}k_{x}}{6}+\frac{k_{y}}{2})+{\rm cos}(\frac{\sqrt{3}k_{x}}{6}-\frac{k_{y}}{2})),
\end{aligned}
\end{equation} 
thus the effective mass is 
\begin{equation} 
\begin{aligned}
\frac{1}{m^{*}}=\frac{\partial^{2}\epsilon_{p_{z}}}{\partial k^{2}_{x}}=-\frac{t}{eE_{\perp}\overline{\Delta}}(\frac{1}{3}{\rm cos}(\frac{\sqrt{3}k_{x}}{3})+
\frac{1}{12}{\rm cos}(\frac{\sqrt{3}k_{x}}{6}+\frac{k_{y}}{2})+
\frac{1}{12}{\rm cos}(\frac{\sqrt{3}k_{x}}{6}-\frac{k_{y}}{2}))
=\frac{5t}{12eE_{\perp}\overline{\Delta}}.
\end{aligned}
\end{equation}
It's obvious that the effective mass of silicene is much smaller than that of the silicon, which is range from 0.26 $m_{0}$ (for conduction band) to 1.08 
$m_{0}$ (for DOS)\cite{van Vliet C M},
due to its very small band gap when without SOC. 
We can also know that the effective mass is the same for both the $k_{x}$ and $k_{y}$ directions since the lattice constants have $a=b$.
And the effective kineic energy is $E_{{\rm kin}}^{\rm eff}=\hbar^{2}k^{2}/(2m^{*})$,
which arrives the minimum in the Dirac-point due to the reduction of $v_{F}$.

The gap function $\Lambda$ (the eigenvector) has
\begin{equation} 
\begin{aligned}
\Lambda(k_{+},k_{-})=\Lambda((k_{A},k_{B}),(k_{B},k_{A}))
\end{aligned}
\end{equation}
Then for s-wave state, it has $\Lambda(k_{A},k_{B})=\Lambda(k_{B},-k_{A})=\Lambda(-k_{B},k_{A})=\Lambda(k_{B},k_{A})$;
for $d_{x^{2}-y^{2}}$-wave state, it has $\Lambda(k_{A},k_{B})=-\Lambda(k_{B},-k_{A})=-\Lambda(-k_{B},k_{A})=-\Lambda(k_{B},k_{A})$;
for $d_{xy}$-wave state, it has $\Lambda(k_{A},k_{B})=-\Lambda(k_{B},-k_{A})=-\Lambda(-k_{B},k_{A})=\Lambda(k_{B},k_{A})$.
The proximity effect which to an ordinary SC (like the ordinary $s$-wave SC) won't affect these rules.
It's obviously that the $k_{+}=k_{-}$ for s-wave case, and the $k_{+}=-k_{-}$ for $d_{x^{2}-y^{2}}$ or $d_{xy}$-wave case.
Particularly, when the z-direction is along the [001]
(i.e., the case we discuss above) and parallel to the surface $\{100\}$,
the $d_{x^{2}-y^{2}}$-wave is equivalent to $d_{xy}$-wave since $k_{x}^{2}-k_{y}^{2}=\mp 2k_{x}k_{y}$ now\cite{Hu C R}.
And the surface state is $H_{sur}=\hbar v_{F}{\pmb \sigma}\cdot({\bf d}\times\hat{{\bf K}})$ according to the discussion in Sect.2.
Typically, for the Andreev reflection, 
a off-diagonal pairing potential (also the eightvector of Eq.(19)) may change an electron which with momentum $k$ into the hole which with $-k$
(with the Bogoliubov spinors in the Nambu space),
which can be reflected by the Bogoliubov equation\cite{Sauls J A}($\alpha,\beta=l_{1},l_{2}$) in the inhomogeneous case
\begin{equation} 
\begin{aligned}
(-\frac{\hbar^{2}\nabla^{2}}{2m^{*}}-\mu)_{\sigma}u_{\alpha}(r)+\Delta_{\alpha\beta}(r,k)v_{\beta}(r)=\varepsilon u_{\alpha}(r),\\
-(-\frac{\hbar^{2}\nabla^{2}}{2m^{*}}-\mu)_{\sigma'}v_{\alpha}(r)+\Delta^{\dag}_{\alpha\beta}(r,k)u_{\beta}(r)=\varepsilon v_{\alpha}(r),
\end{aligned}
\end{equation}
where $u_{\alpha}(r)$ and $v_{\alpha}(r)$ are the particle and hole operator, respectively, and 
with opposite spin direction.
They satisfy $v_{\alpha}(r)=\Theta u_{\alpha}(r)$.
$\Delta_{\alpha\beta}(r,k)$ is the pairing potential which breaks the TRI in a single valley and thus lifts the SC pairing from the different valley,
and it's related to the gap function by $\sigma_{y}\Delta_{\alpha\beta}(r,k)=\Delta(k)$.
$\mu$ is the chemical potential.
$m^{*}$ is the effective mass.
$\varepsilon$ is the energy and it is related to the quasiparticle spectrum which can be obtained from the Bogoliubov-de Gennes (BdG) Hamiltonian
\begin{equation} 
\begin{aligned}
H_{{\rm BdG}}=\begin{pmatrix}\Delta(E_{k})&0&-\frac{\Delta l_{F}}{\hbar}k_{-}&\frac{\Delta l_{F}}{\hbar}k_{z}\\
0&\Delta(E_{k})&\frac{\Delta l_{F}}{\hbar}k_{z}&\frac{\Delta l_{F}}{\hbar}k_{+}\\
-\frac{\Delta l_{F}}{\hbar}k_{+}&\frac{\Delta l_{F}}{\hbar}k_{z}&-\Delta(E_{k})&0\\
\frac{\Delta l_{F}}{\hbar}k_{z}&\frac{\Delta l_{F}}{\hbar}k_{-}&0&-\Delta(E_{k})\\
\end{pmatrix},
\end{aligned}
\end{equation}
where $\Delta(E_{k})=E_{{\rm kin}}(k)-E_{F}$ with $E_{{\rm kin}}(k)=\hbar^{2}k^{2}/(2m)$ is the kineic energy of the free Fermions (or the charge carriers)
and $E_{F}$ is the free particle Fermi energy $E_{F}=\hbar^{2}k_{F}^{2}/(2m)$ 
or by using the Nambu matrix ${\bf \tau}$ in 3D particle-hole space
\begin{equation} 
\begin{aligned}
\mathcal{H}={\bf \tau}\cdot \hat{{\bf K}}=\Delta(E_{k})\tau_{3}+\tau_{1}\Delta_{\alpha}k+\tau_{2}\Delta_{\beta}k,
\end{aligned}
\end{equation}
the gap functions $\Delta_{\alpha}=d_{x^{2}-y^{2}}+s_{1}$, $\Delta_{\beta}=d_{xy}+s_{2}$
which consist the chiral pair which contains both the $id$- and $is$-component by $\Delta_{\beta}=i\Delta_{\alpha}$ for the nodeless $d$-wave SC.
A small amount of joining of the $s$-state in the above two expression can be seem in the band structure of silicene that the $\pi$- and $\pi^{*}$-band
combined with the $s$-band.
And the basis of Bogoliubov Hamiltonian (also the essential Nambu vector) is
${\bf c}(k)=((c_{\uparrow}(k),c_{\downarrow}(k)),(c^{\dag}_{\uparrow}(k),c^{\dag}_{\downarrow}(k)))^{T}$.
Note that the particle and hole excitation here is hard to distinguish when the spin fluctuation within a singlet pair is equal.
Eqs.(42,43) not only shows the scenario that the unparallel momentums aligns along the $x$-direction
and polarized along the $\pm y$-direction as mentioned above,
which for the quasiparticles with opposite spin direction in each pairs in the surface state have the energy dispersions as 
$\pm(k_{x}\mathcal{G}_{o}/k_{F})$,
but also shows that the coupling between the spin and orbital degreed of freedom within the $x-y$ plane.
Note that here the orbital order parameter $\mathcal{G}_{o}$
consider only the gapless part (gapless SC) due to the surface state.
While for the case that without the constrain of TRI,
the gapless only emerge on the points which have $\Delta_{\alpha}=0,\Delta_{\beta}=0$ and $E_{k}=E_{F}$.
Due to the invariance under the simultaneously rotatation of the spin and orbital around $z$-direction,
which is also the result of the vanished interorbital Coulomb interactions,
the coupling between the spin and orbital degreed of freedom is always maintain a stable strength.
In the BCS limit of the fermions, the $k_{F}=(3\pi^{2}n)^{1/3}$ for the dilute electron gas\cite{Adhikari S K}(2D or 3D)(or the Bogoliubov quasiparticle), 
and the density $n$ will exhibit exceptionally spatially uniform
and becomes zero in the boundary, the spatially uniform orbital angular momentum also leads to a density of the no center-of-mass superfluid as 
$\rho=\nabla\times(\frac{1}{4}n\hbar\hat{L})$ at $T=0$\cite{Sauls J A}
in which case the system is dominated by the ground state topology with a topological integer rather that the pairing symmetry.

\subsection{Mean-field approximation of 2D silicene in optical field and the possible chiral currents in the 2D silicene with semiclassical dynamic}
We focus on the geometry property of the 2D silicene under a Landau gauge vector potential in optical field (optical lattice) in this section.
The recoid energy as $E_{R}=h^{2}/2m\lambda^{2}$ with $\lambda$ the wavelength of the laser,
and the effective force acting on the center of mass is ${\bf F}=E_{R}/\lambda$.
The interwell barrier energy $sE_{R}$ can be controlled by regulating the laser frequency\cite{Cataliotti F S}.
Under the mean-field approximation, the bound state exists only for the potential larger that one $E_{R}$,
i.e., the frequency of laser need to $\gtrsim 300$ THz.
The laser-induced ($<1000$ THz, i.e., in the resonant region) microscopic Haldane model in $v_{f}=1$ (with the particle-hole symmetry and spin polarization):
\begin{equation} 
\begin{aligned}
H=t\sum_{\langle i,j\rangle\sigma}(c^{\dag}_{i\sigma}c_{j\sigma}+h.c.)+t'\sum_{\langle\langle i,j\rangle\rangle\sigma}(e^{i\phi_{ij}}c^{\dag}_{i\sigma}c_{j\sigma}+h.c.)
+U\sum_{i}n_{i\sigma}n_{i\sigma'},
\end{aligned}
\end{equation}
which with Wigner-Seitz unit cells on the honeycombs with sixfold rotation symmetry and zero net flux\cite{Haldane F D M} and the
phase factor $\phi_{ij}=2\pi\frac{\Upphi}{\Upphi_{0}}$ where $\Upphi$ is the total fluxes through the unit cell and $\Upphi_{0}=h/2e\approx 2.07\times 10^{-15}$ Wb
is the basic fluc quantum,
and they are related to the vector potential by $\phi_{ij}=\sum_{ij}(\frac{1}{\Upphi_{0}}\int^{r_{j}}_{r_{i}}{\bf A}({\bf r})d{\bf r})$.
The ferromagnetism occur when the $U>\Delta$\cite{Fukuyama H}.
And in the nonlinear tight-binding MF-approximation is (here we assume t'=0.1t)
\begin{equation} 
\begin{aligned}
H_{{\rm MF}}=\sum_{\langle i,j\rangle\sigma}(z^{\dag}_{i\sigma}(t)z_{j\sigma}(t)+h.c.)+0.1\sum_{\langle\langle i,j\rangle\rangle\sigma}(e^{i\phi_{ij}}z^{\dag}_{i\sigma}(t)z_{j\sigma}(t)+h.c.)
+\frac{U}{t}\sum_{i\sigma}|z_{i}(t)|^{4},
\end{aligned}
\end{equation}
where $z_{i\sigma}(t)$ is the discrete amplitude under MF-approximation: 
$\sum_{i}|z_{i\sigma}(t)|^{2}=1$ which has $\dot{z_{t}}=\partial H_{{\rm MF}}/\partial(i\hbar z^{*}_{i\sigma}(t))=\varphi_{i}e^{i\phi_{i}}$, i.e.,
the $i\hbar z^{*}_{i\sigma}(t)$ and $z_{i}(t)$ are a pair of the canonical conjugate variables.
And there exist the follwing relation\cite{Trombettoni A,Smerzi A2,Xue J K}:
\begin{equation} 
\begin{aligned}
\frac{U}{t}=\frac{2\pi \hbar^{2}a_{s}N}{tm}\int d{\bf r}\varphi^{4}_{i}
\end{aligned}
\end{equation}
where $\varphi_{i}$ is the condensed Wannier wave function of the lowest band 
which describe the continue amplitude with the normalization $\int d{\bf r}\varphi_{i}^{2}=1$ and $a_{s}\approx 5.5$ nm is the scattering length of the $s$-wave.
The flux around the cell yields a nontrivial phase which analogous to the Aharonov-Bohm phase and the gauge-invariant Berry phase which is mentioned above.
To deeply analyses the dynamics, we consider the 2D Gaussian profile wave packet as 
\begin{equation} 
\begin{aligned}
\psi_{G}=\psi_{Gr}\psi_{Gr'}=\sqrt{\frac{1}{\pi\omega_{r}\omega_{r'}}}{\rm exp}[-\frac{(r-\xi_{r})^{2}}{2\omega_{r}^{2}}-\frac{(r'-\xi_{r'})^{2}}{2\omega_{r'}^{2}}
+ik_{r}(r-\xi_{r})+ik_{r'}(r'-\xi_{r'})],
\end{aligned}
\end{equation}
where $\xi$ is the centers of the wave packet (also the center-of-mass coordinates)
and here we denote $r$ the position on the nearest-hopping directions and $r'$ the position on the next-nearest-hopping directions,
and the $\omega_{r}$ and $\omega_{r'}$ are the corresponding Gaussian wave packets width.
And the norm of Gaussian profile satisfy $\int |\psi_{Gr}|^{2}dr=\int |\psi_{Gr'}|^{2}dr'=1$.
According to the properties of Haldane model, we know that the motions in $r$-direction keeps the TRI and the momentum of the center of wave packet 
$k_{cr}$ is periodic.
And the equation of motion in this direction is\cite{Trombettoni A,Smerzi A}
\begin{equation} 
\begin{aligned}
\dot{k_{cr}}=&-\frac{\partial}{\partial \xi_{r}}\frac{1}{\pi\omega_{r}}\int^{\infty}_{-\infty}d{\bf r} 
\frac{\rho_{r}}{2t} e^{-2\frac{(r-\xi_{r})^{2}}{2\omega_{r}^{2}}},\\
\rho_{r}=&\int d{\bf r}[\frac{\hbar^{2}}{2m}(\partial^{2}\varphi_{i}(r)+V\varphi_{i}^{2})],\\
V=&\frac{\hbar}{2im}\frac{\varphi^{*}\partial\varphi-\varphi\partial\varphi^{*}}{\varphi^{*}\varphi},
\end{aligned}
\end{equation}
And the group velocity in this direction is 
$v_{g}=\partial H_{{\rm MF}}/(\hbar\partial k_{cr})=2{\rm sin}k_{cr}e^{-\frac{1}{4\omega_{r}^{2}}-\frac{\phi_{ij}^{2}\omega_{r}^{2}}{4}}/\hbar$
and with the effective mass of the charge carriers $1/m^{*}=\partial^{2} H_{{\rm MF}}/(\hbar^{2}\partial k_{cr}^{2})
=2{\rm cos}k_{cr}e^{-\frac{1}{4\omega_{r}^{2}}-\frac{\phi_{ij}^{2}\omega_{r}^{2}}{4}}/\hbar^{2}$.

The momentum $k_{cr}$ and center of wave packet $\xi$ which as the dynamical variables\cite{Trombettoni A} have the follow relations\cite{Zhang A X}
\begin{equation} 
\begin{aligned}
\frac{dk_{cr}}{d\xi_{r'}}=&-\phi_{ij},\\
\frac{dk_{cr}}{d\xi_{r}}=&\frac{-\phi_{ij}\ {\rm sin}(k_{cr'}+\phi_{ij}\xi_{r})
{\rm exp}(\frac{1}{4\omega_{r}^{2}}-\frac{1}{4\omega_{r'}^{2}}-\frac{\phi_{ij}^{2}\omega_{r}^{2}}{4})}{{\rm sin}(k_{cr})},\\
\frac{dk_{cr'}}{d\xi_{r'}}=&\frac{dk_{cr'}}{d\xi_{r}}=0,
\end{aligned}
\end{equation}
where variational parameters $k_{c}$ and $\xi$ and satisfy the Euler-Lagrangian function
$\frac{\partial \mathcal{L}}{\partial q_{i}}-\frac{d}{dt}\frac{\partial \mathcal{L}}{\partial \dot{q_{i}}}=0,\ q_{i}=k_{c},\xi$\cite{Trombettoni A},
with the Lagrangian 
$\mathcal{L}=\sum_{ij}\frac{1}{2}(i\dot{z_{i}}(t)z_{i}^{*}(t)-iz_{i}(t)\dot{z}_{i}^{*}(t))-H_{{\rm MF}}$,
Thus the $r$-component momentum is change with time unlike the $r'$-component one due to the special gradient of artificial magnetic field,
and in fact, such a field coupling the $r$-component and the $r'$-component by $\varphi$ and suppress the diffusion of the wave packet.

\subsection{Hopping chiral current}
The magnetic field together with the $t$ and $t'$ hopping give rise the hopping current in silicene rather than the diverging current.
In one hexagon lattice of silicene, the next-nearest-neighbor hoppings (in $r'$-direction) can form
two closed triangle vortices with opposite directions, thus the formed non-trivial net phase (flux)
is constrained by $\phi_{{\rm net}}=2\pi(\frac{\Upphi}{\Upphi_{0}}-n)$ and the $n$ is the occupancy number of the vortices in one honeycomb regiem.
When the net phase is zero (zero flux), the dynamics along the $r$-direction vanish, i.e., there are not interaction between particles in $r$-direction
but may have strong hopping.
Such situation may happen in the case that $t\gg t'$, and the current configuration in this case is certainly chiral.
As shown in the Fig.9(a)，we isolate a row of the honeycomb-lattice-ladder,
although the currents in $r$-direction (along the legs direction $l_{1}$ and $l_{2}$ in Fig.9) produce the delocalization and against the chiral symmetry, 
the currents in the $r'$-direction which is contribute to the localization within one honeycomb is still asymmetry with each other
with respect the center rung and thus exhibits the Meissner phase.
Note that the pairwise on-site interaction $U$ along leg-direction
is important for the chiral characteristic since these interactions will increase the currents in leg-direction and
break the current-symmetry between the two neighbor honeycombs with respect the center rung.
If without these interactions, 
the system won't shows the chiral characteristic even if it's under a gauge field.
There's because the bands weight for the 
The special chiral (Meissner-like) phase of honeycomb requires the two on-site interactions on each lungs small enough 
that make sure the currents in these lung-directions nearly vanish (see Fig.9(a)). 
The estimator of the chiral currents (especially for the thermodynamic limit) is
\begin{equation} 
\begin{aligned}
j_{c}=\frac{it}{4R}\sum_{r}[(c^{\dag}_{r',r+2}c_{r',r})-(c^{\dag}_{r'+1,r+2}c_{r'+1,r})].
\end{aligned}
\end{equation}
For the current-conserved case like for the ground-state current-configration,
this chiral currents also can be represented as $\frac{\partial E_{0}}{\partial\varphi}$.
In fact, this expression (Eq.) taking account for the net dynamics for the upper and lower
legs (in directions $l_{1}$ and $l_{2}$, see Fig.9) which depend on the gauge,
A recent experiment\cite{Tai M E} reveals that such net dynamics can be realized by the Raman beams in opital lattice with a artificial magnetic field
and mapped to the Fourier space,
and related to the artificial flux as $({\bf k}_{1}-{\bf k}_{2})\cdot{\bf e}_{y}/|{\bf k_{{\rm 2D}}}|=\Phi/\pi$
with the 2D lattice $k_{{\rm 2D}}=\pi/a$ where $a$ is the crystal constant. 

The obvious chiral (Meissner-like) phase shown in the Fig.9(a) is relys on the strong pairwise on-site interaction $U$,
especially for these on-site interaction which populate in the two endpoints of lungs,
like the insulating spin liquid for ground state in the hard-core boson limit.
The different directions of currents for upper legs and lower legs is also depend on strong on-site interaction and the energy-difference between
$A(A')$ and $B(B')$ atoms, and therefore it's also related to the band energies of upper one and the lower one,
but such energy difference only make sure the incompatibility of direction for upper and lower leg currents
and can not decide the currents flow rightward or leftward,
which is indeed decided by the $\Phi$ and $t'$. 
While in the vortex phase which with much smaller $U$ don't has such pattern (the currents which through the lungs may flows leftward or rightward 
depending on the intensity distribution of $U$ along the two leg-direction $l_{1}$ and $l_{2}$).
When $t$ is close to the $t'$, the small $U$ in two leg-directions may small and give rise to the vortex-phase (Fig.9(b)),
and in fact, for all the case of $U/t<\infty$,
there will be a Mott-insulator-superfluid transition\cite{Piraud M,Wu C H}.

We define two quantum states $|\alpha\rangle$ and $|\beta\rangle$ according to flow directions of the upper legs and the lower legs in chiral pattern, 
and have $\sigma^{z}|\alpha\rangle=|\alpha\rangle$, $-\sigma^{z}|\beta\rangle=|\beta\rangle$.
Then for an isolated lung ($t=t'=0$ except the hopping within the lung), the superpositions can be presented as $\frac{1}{\sqrt{2}}(|\alpha\rangle+|\beta\rangle)$,
where $|\alpha\rangle=\frac{1}{\sqrt{2}}(c^{\dag}_{u}+c^{\dag}_{l})|{\rm vac}\rangle$ and 
$|\beta\rangle=\frac{1}{\sqrt{2}}(c^{\dag}_{u}-c^{\dag}_{l})|{\rm vac}\rangle$ with the creation operator for the upper site $c^{\dag}_{u}$ and the lower one $c^{\dag}_{l}$,
the kinetic Hamiltonian in momentum space through the $H_{t}$ which is defined in above can be written as 
\begin{equation} 
\begin{aligned}
H=\sum_{k\sigma}(u^{\dag}_{k\sigma},l^{\dag}_{k\sigma})H_{t}\begin{pmatrix}u_{k\sigma}\\l_{k\sigma}\end{pmatrix},
\end{aligned}
\end{equation}
where $u^{\dag}_{k}$ and $l^{\dag}_{k}$ are the creation operators for the upper leg and the lower leg respectively.
Since the rungs in our model are full-filled, i.e., has two atoms per rung,
the quantum state of a rung in the unbounding-limit\cite{Tai M E} can be view as the superposition of the two half-filled band (single-particle)
states $\alpha|{\rm vac}\rangle$ and $\beta|{\rm vac}\rangle$ as $c^{\dag}_{u}c^{\dag}_{l}|{\rm vac}\rangle=\frac{1}{2}(\alpha^{2}-\beta^{2})|{\rm vac}\rangle$.
The above procedure is enlightening to dealing with the superconductor of superfluid.
The geometry Hamiltonian $H_{t}$ can be replaced by the above Bogoliubov-de Gennes Hamiltonian to deal with the SC problem,
which has $\Xi H_{{\rm BdG}}\Xi^{-1}=-H_{{\rm BdG}}=\tau_{y}\sigma_{y} H^{*}_{{\rm BdG}}\sigma_{y}\tau_{y}$ 
under the particle-hole symmetry where $\Xi=\sigma_{y}\tau_{y}\mathcal{K}$ is the particle-hole operator.

Since the Dirac-point are protected by the symmetry but vulnerable for the spin-orbit coupling (SOC) which will open a gap in the Dirac-point
and leading to the phase of topological insulators.
The Bloch states of each band which can be choosen as their eighstates in the momentum space under the present gauge field
which produce two patterns of chiral edge current and
which lead to a shift in the momentum space, can be expressed by the Hamiltonian\cite{Tai M E}
\begin{equation} 
\begin{aligned}
H=E_{+}(k_{x}+\frac{\Phi}{2},k_{y}-\frac{\Phi}{2})+E_{-}(k_{x}+\frac{\Phi}{2},k_{y}-\frac{\Phi}{2})\sigma_{z}+H_{t}\sigma_{x},
\end{aligned}
\end{equation}
here we consider the spin degrees-of-freedom and the $H_{t}\sigma_{x}$ describe the coupling between upper legs and lower legs. Due to the particle-hole symmetry,
it satisfy the anti-commute relation with the mass term $\{H-E_{+}(k_{x}+\frac{\Phi}{2},k_{y}-\frac{\Phi}{2}),i\sigma_{y}\}=0$.
and with a dispersion under the effect of gauge field as
\begin{equation} 
\begin{aligned}
\epsilon_{\pm}=E_{+}(k_{x}+\frac{\Phi}{2},k_{y}-\frac{\Phi}{2})+\sqrt{E_{-}^{2}(k_{x}+\frac{\Phi}{2},k_{y}-\frac{\Phi}{2})+H_{t}^{2}},
\end{aligned}
\end{equation}
and in fact, the local magnetization of per hexagonal zone has $M=\frac{hc}{\tau e}\frac{\phi}{2\pi}$\cite{Wu C}
where $1/\tau$ is the filling fraction in nesting fermi surface.
For silicene
in the invariant space with the spatial inversion which protect the Dirac-point by the feature of inversion symmetry here and exchange of the state
$|\alpha\rangle$ and $|\beta\rangle$ 
simultaneously with the particle-hole symmetry in our half-filling model,
which can be express as 
\begin{equation} 
\begin{aligned}
H_{+}(-r)|\alpha\rangle=H_{+}(r)|\beta\rangle,\ H_{-}(-r)|\alpha\rangle=-H_{-}(-r)|\beta\rangle,\\
H_{+}(-r)|\beta\rangle=H_{+}(r)|\alpha\rangle,\ H_{-}(-r)|\beta\rangle=-H_{-}(r)|\alpha\rangle,
\end{aligned}
\end{equation}
where $r$ stands for the position in the momentum space.
That's also consistent with the conclusion that the Dirac-point can't exist for a single symmetry protection case in 2D model\cite{Young S M}
and this particle-hole symmetry will also broken due to the rise of Landau-level mixing.

\section{Dirac mass in the silicene with spin and orbit fluctuation}

We carry out the first-principle (FP) density functional theory (DFT) calculations 
using the QUANTUM ESPRESSO package\cite{Paolo} with the local-density approximation (LDA)
the plane wave energy cutoff is setted as 250 eV in our calculation,
and the Hellmann-Feynman force on the atoms was relaxed to below 0.01 eV/\AA\ .
For silicene which is one kind of the heavy boson superconductors with the crystal constant $a=b=3.86$ \AA and $c=15.31$ \AA
which are close to the result of Ref.\cite{Pflugradt P},
the band structure with the pairing symmetry open the door to modulating the SC,
which is essentially because the different hoppings in $r$-direction and $r'$-direction, e.g., from the $d_{1}+id_{2}$ chiral symmetry SC to the $f$-wave symmetry SC.
As the interaction $U$ increase with the spin fluctuation under low-doping which keep the electron and hole pockets comparable\cite{Qi X L},
the susceptibility will diverging from the peak $\Gamma$-point which is related to the 
antiferromagnetic SDW instability due to the RPA, and get close to the FS which may be slightly modified by the electron-doping.
The FS nesting, which accompanied with magnetic interaction, 
also give rise to the repulsive Cooper-pair interaction within the band dispersion of $K-K'$ (see Fig.4(b))
and gives a gap with sign reversing between the undoped (non-interaction) FS and the electron-doped FS, which is related to the $t$.

If don't apply a external electric field, the pairing configuration is mostly dominated by the $d_{1}+id_{2}$-wave pairing in low-doping 
since the FS is not shifted too much by the electron-doping and therefore the connection lines between $\Gamma$-point and the center $M$-points
don't cross the FS.
But a electric field may cause the ferromagnetic-like intraorbital spin interaction and the density of states are also enhanced remarkably.
The intraorbital Coulomb interaction will enhanced by the further electron-doping and thus lead to the large spin or charge (orbital) fluctuation, and also cause the
instability of the long-range SDW and the pairing-state (including the singlet and triplet pairing).
For the phase transition of singlet-pairing state to triplet-pairing state which mentioned above, 
when we implete a large doping (e.g., $>0.2$),
the static susceptibility obviously spread from the center ($\Gamma$-point) of the hexagonal first Brillouin zone of silicene\cite{Zhang L D} 
to the $M$-point (midpoints) and becomes close to the FS as shown in the Fig.4(b), 
and thus the long-range SDW order becomes instability\cite{Zhang L D}.
The rise of spin-triplet SC here is due to the increase of the anisotropy AFM fluctuation
which is isotropy when for undoped case.
Except that, when $U>U_{c}$, the proportion of the spin-fluctuation-mediated superconducting-induce-pairing is decrease
while the phonons- and charge-fluctuation-mediated one is increase, and that also leads to the weakening of $d_{1}+id_{2}$-pairing wave superconducting 
and makes the $f$-wave superconducting becomes a good candidate.
The domination of $d$-wave SC anisotropic gap which is due to the repulsive (with sign reversing)
pairing interaction in low doping case is because the diagonal nodes do not intersect the FS.


The comparable electron and hole pockets (no extraneous) with low-doping will leading to the two classes of FS and the pairing symmetries
with the shifted FS topology by the excess electrons. 
We show the ``correct zone" of silicene within the two-orbital model in Fig.4(b)
which we can clearly see that each FS sheets is separated into four sheets by the boundary or the diagonal of the correct zone,
the electron pockets (FS sheet) which are marked in red are generally in a larger energy when the Dirac-point emerge compare to the
hole pockets (FS sheet) which are marked in blue because the absence of the Dirac-point in the $M$-point
which is due to the gapping in these places by the distortion.
But the energies of the neighbor $\Gamma$-point ($\Gamma$ and $\Gamma'$) are surely equal due to the mirror structure relate the $M$-point.

For the 2D honeycomb lattice, considering the full-interaction Hamiltonian with the sublattice degrees-of-freedom $\tau$ (pseudospin; which also induce the mass term) 
and the influence of SOC
\begin{equation} 
\begin{aligned}
H=&\tau_{x}(t{\rm cos}\frac{\sqrt{3}k_{x}}{3}+2t{\rm cos}\frac{\sqrt{3}k_{x}}{6}{\rm cos}\frac{k_{y}}{2})+
4t'{\rm cos}\frac{\sqrt{3}k_{x}}{2}{\rm cos}\frac{k_{y}}{2}+2t'{\rm cos}k_{y}\\
&+t'_{{\rm SOC}}\tau_{z}\sigma_{z}\\
&+t'_{R}\tau_{z}(4\sigma_{y}{\rm sin}\frac{\sqrt{3}k_{x}}{2}-(2\sigma_{x}{\rm sin}k_{y}+4\sigma_{x}{\rm sin}\frac{k_{y}}{2})),
\end{aligned}
\end{equation}
where next-nearest-neighbor $t'_{{\rm SOC}}$ is proportional to energy difference between the $3s-$ and $3p$-band
and defined as $t'_{{\rm SOC}}=\frac{\lambda_{{\rm SOC}}}{3\sqrt{3}}$ while $t'_{R}=\frac{2R}{3}$,
the $t'_{{\rm SOC}}$ breaks the degeneracy on the boundary of BZ except the $K,K',M$ points, and thus also breaks the particle-hole symmetry.
Specially, when the we ignore the fourth term since it's quite small, the particle-hole symmetry can be obtained as the anti-commute relation 
$\{H',\Xi\}=0$ similar to Eq.(32), where $H'=H-(4t'{\rm cos}\frac{\sqrt{3}k_{x}}{2}{\rm cos}\frac{k_{y}}{2}+2t'{\rm cos}k_{y})$
and the particle-hole operator $\Xi=\sigma_{y}\tau_{y}\mathcal{K}$ satisfy $\Xi^{\dag}H'\Xi=H'$,
while the spin-rotational symmetry is broken by the SOC,
and $\mathcal{K}$ is the complex conjugation which with 
$\mathcal{K}H'(\sigma_{y})\mathcal{K}=H'^{*}=H'(-\sigma_{y})$.

The external field together with the staggered potential as well a sthe Haldane term would also breaks the particle-hole symmetry 
(by the quantum Hall effect which is due to the Landau-level quantization mixing).
And the TRI here in the case of the broken inversion symmetry (due to the exist of Rashba-coupling) reads
$[H',i\sigma_{y}\mathcal{K}]=0$ near the $M$-point,
and $[H',i\tau_{y}\sigma_{y}\mathcal{K}]=0$ near the $K$-point.
And the chiral symmetry operator by combining the particle-hole symmetry and TRI is $i\tau_{y}$ near $M$-point and $i$ near $K$-point.
Because of the distortion along diagonal direction and hence breaks the $C_{2}$ symmetry but preserve the mirror-line as we mention above,
the Dirac-point at $M$-point vanishes but the other two Dirac-point at the adjacent sites is equivalent.
And this distortion is proportional to the term $\tau_{x}({\rm sin}\frac{\sqrt{3}k_{x}}{3}+2{\rm sin}\frac{\sqrt{3}k_{x}}{6}{\rm cos}\frac{k_{y}}{2})$.
Comparing these two commutation relations, we can obtain that the vanish of $\tau_{y}$ for the $M$-point shifts the sublattice pseudospin symmetry
as well as the particle-hole symmetry.
A new form which satisfy the particle-hole symmetry can be written as\cite{Ezawa M4,Ezawa M3}
\begin{equation} 
\begin{aligned}
\mathcal{H}=\eta\hbar v_{F}(\tau_{x}k_{x}+\tau_{y}k_{y})+\eta\lambda_{{\rm SOC}}\tau_{z}\sigma_{z}+aR\eta\tau_{z}(k_{y}\sigma_{x}-k_{x}\sigma_{y})
-\frac{\Delta}{2}E_{\perp}\tau_{z}+\frac{R_{2}(E_{\perp})}{2}(\eta\sigma_{y}\tau_{x}-\sigma_{x}\tau_{y}).
\end{aligned}
\end{equation}
The Pauli matrix here ${\pmb \sigma}=\{\sigma_{x},\sigma_{y},\sigma_{z}\}$
are act on the electron-amplitude (not the pairing amplitude pairing amplitude) of the orbitals of the consisdered bands.
This SOC also give rise to the coupling of the $\pi$-band and $\sigma$-band.
Indeed, the above equation contains two kinds of hopping, the first one $t$ is spin-independent and the spin is conserved in this case,
the second one $t'_{{\rm SOC}}$ is spin-dependent and the up-spin and down-spin are variant with time and accompanied by a spin flip,
which is necessary for the QSHE and note that for silicene, the QSHE is more robust than the Rashba-coupling.
We can also see that the second Rashba coupling $R_{2}(E_{\perp})$ is orbital (sublattice)-dependent while the first one $R$ is orbital-independent.
The in-plane spin texture ($\sigma_{x}$ and $\sigma_{y}$) can be obtained bt the
spin expectation values as\cite{Li Z} $s_{x\pm}=\pm\frac{\hbar}{2}\frac{\mathcal{R}}{\sqrt{\mathcal{R}^{2}+\mathcal{R}_{2}^{2}}}$,
$s_{y\pm}=\pm\frac{\hbar}{2}\frac{-\mathcal{R}_{2}}{\sqrt{\mathcal{R}^{2}+\mathcal{R}_{2}^{2}}}$,
with $\mathcal{R}=(a\eta\tau_{z}R k_{y}-\frac{R_{2}(E_{\perp})}{2}\tau_{y})$,
$\mathcal{R}_{2}=(-a\eta\tau_{z}Rk_{x}+\frac{R_{2}(E_{\perp})}{2}\tau_{x})$.
While the out-of-plane spin texture is related to the Dirac mass and the Zeeman splitting.
The group velocity through the above Dirac Hamiltonian as
$v_{gx}=\frac{\partial \mathcal{H}}{\hbar\partial k_{x}}=v_{F}\eta\tau_{x}-\frac{1}{\hbar}a\eta\tau_{z}R\sigma_{y}$,
$v_{gy}=\frac{\partial \mathcal{H}}{\hbar\partial k_{y}}=v_{F}\tau_{y}+\frac{1}{\hbar}a\eta\tau_{z}R\sigma_{x}$.

Thus in the presence of both the $E_{\perp}$ and the first-order and second-order Rashba-coupling,
the system can be described by
$H=\Psi^{\dag}H^{\pm}_{{\rm eff}}\Psi/2$,
the BCS-like effective Hamiltonian of the 
neighbor valleys by the low-energy Dirac theory in the basis of $\{\tau\otimes\sigma\}$ which reflected in the two-component 
spinor-valued field operators as 
$\Psi=[(\psi_{\uparrow}^{A},\psi_{\downarrow}^{A},\psi_{\uparrow}^{B},\psi_{\downarrow}^{B}),((\psi_{\uparrow}^{A\dag},\psi_{\downarrow}^{A\dag},\psi_{\uparrow}^{B\dag},\psi_{\downarrow}^{B\dag}))]^{T}$, are
\begin{equation} 
\begin{aligned}
H^{+}_{{\rm eff}}=&
\begin{pmatrix} \mathcal{H}({\bf k},\sigma_{z})&\Delta({\bf k},\sigma_{y})\\\Delta^{\dag}({\bf k},\sigma_{y})&\mathcal{H}({\bf k},-\sigma_{z}) \end{pmatrix},\\
\mathcal{H}({\bf k},\sigma_{z})=&\lambda_{{\rm SOC}}\sigma_{z}\tau_{z}+ a R(k_{y}\sigma_{x}-k_{x}\sigma_{y}\tau_{z})
+M\tau_{z}\sigma_{z}-\frac{\overline{\Delta}}{2}E_{\perp}\tau_{z}+\frac{R_{2}(E_{\perp})}{2}(\sigma_{y}\tau_{x}-\sigma_{x}\tau_{y}),\\
\Delta({\bf k},\sigma_{y})=&\begin{pmatrix} i\Delta_{A}&0\\0&i\Delta_{B}\end{pmatrix},\\
\Delta_{A}=&k_{y}\sigma_{y}-ik_{x}\sigma_{x},\\
\Delta_{B}=&-k_{y}\sigma_{y}-ik_{x}\sigma_{x},
\end{aligned}
\end{equation}
and 
\begin{equation} 
\begin{aligned}
H^{-}_{{\rm eff}}=&
\begin{pmatrix} \mathcal{H}({\bf k},-\sigma_{z})&-\Delta({\bf k},-\sigma_{y})\\-\Delta^{\dag}({\bf k},-\sigma_{y})&\mathcal{H}({\bf k},\sigma_{z}) \end{pmatrix},
\end{aligned}
\end{equation}
$\Delta_{A}$ and $\Delta_{B}$ are the pairing gaps of two sublattices.
In the case for valley-polarized metal phase (i.e., the SDC state) which is achieveble under the effect of both the 
vertical electric field\cite{Ezawa M} or magnetic field\cite{Xiao D} and the exchange magnetization
especially under the such a strong SOC which will further intensifys the particle-hole asymmetry between the two valleys.
with the broken sublattice-pseudospin symmetry but remain the chiral symmetry between two valleys,
and the valley-hybridization-term $\Delta({\bf k},\sigma_{y})$ in $H^{+}_{{\rm eff}}$ can be replaced by
\begin{equation} 
\begin{aligned}
\mathcal{V}({\bf k})=&\begin{pmatrix} \sqrt{\mathcal{V}_{1}}&-\sqrt{\mathcal{V}_{2}}(k_{x}+ik_{y})\\-\sqrt{\mathcal{V}_{2}}(k_{x}-ik_{y}) &\sqrt{\mathcal{V}_{1}} \end{pmatrix},\\
\end{aligned}
\end{equation}
which is proportional to the exchange effect between two sublattices (or the potential differentce
between two sublattices) with $\mathcal{V}_{1}$ the hybridization gap which is also proportional to the potential differentce 
and $\mathcal{V}_{2}$ the parabolic band dispersion which with a opened gap.
When we ignore the small Rashba coupling,
the Chern number which summation over the Berry curvature in momentum space of 
all the occupied bands below the gap and with both the spin and valley degrees of freedom is
\cite{Thouless D J}
\begin{equation} 
\begin{aligned}
\mathcal{C}=&\sum_{s_{z},\eta=\pm 1}\mathcal{C}_{s_{z}}^{\eta}=\frac{1}{2\pi}\sum_{{\bf \Omega}}\int_{BZ}dk_{x}dk_{y}{\bf \Omega}(\hat{{\bf K}})=M/|M|,\\
{\bf \Omega}(\hat{{\bf K}})=&\nabla_{\hat{{\bf r}}}\times{\bf A}(\hat{{\bf K}})=\hat{{\bf K}}\cdot(\partial_{k_{x}}\hat{{\bf K}}\times\partial_{k_{y}}\hat{{\bf K}}),\\
{\bf A}(\hat{{\bf K}})=&i\langle u_{n{\bf k}}|\partial_{{\bf k}}|u_{n{\bf k}}\rangle,\\
\hat{{\bf K}}=&{\bf K}/|{\bf K}|,
\end{aligned}
\end{equation}
where ${\bf \Omega}(\hat{{\bf K}})$ is the momentum-pseudospin 
space Berry curvature in units of $e^{2}/h$ which is in a similar distribution with the orbital magnetic moment
which couples to the magnetic field in $z$-direction\cite{Xiao D} as
\begin{equation} 
\begin{aligned}
m({\bf k})=\frac{3ea^{2}m_{D}t^{2}}{2\hbar c(4m_{D}+3(k_{x}^{2}+k_{y}^{2})a^{2}t^{2})}=\frac{e}{\hbar}(\varepsilon_{\eta}({\bf K})-\mu){\bf \Omega}(\hat{{\bf K}}),
\end{aligned}
\end{equation}
and it's shown in the Fig.10.
${\bf A}(\hat{{\bf K}})$ is a gauge-dependent Berry connection\cite{Fukui T,Price H M}
which display as a diagonal matrix element of the velocity operator in the above equation,
and $u_{n{\bf k}}$ is the $n$th band Bloch state.
And the Hall conductivity is $\sigma_{xy}=\frac{e^{2}}{h}\mathcal{C}$.
The above Berry curvature is applicable for both the antisymmetry (trivial band insulator) and symmetry (TI) case with the spin and sublattice (pseudospin) degrees of freedom.
When $|V|<|M|$ where $V$ is the potential difference which is $\propto \sqrt{\mathcal{V}_{1}}$,
the $\mathcal{C}$ and the Hall conductance is nonzero and it's in a QAH phase
which has that $|\mathcal{C}|$ equals to the number of chiral edge state in he tight-binding model (i.e., two chiral edge states for the QAH)
and the TRI is broken in this case.
While for $|V|>|M|$, 
it becomes a quantum valley Hall insulator which with zero chern number and Hall conductance, and keeps the TRI.
As mentioned above, the exchange magnetization is relys on the spin rotation which is by the Rashba-coupling 
because the Berry curvature is nonzero only at the places where $s_{z}$ change sign.
That's why the Berry curvature-dependent orbital susceptibility vanishes under the inversion symmetry.
This spin rotation also generating a Skyrmion spin texture
with inversed spin unlikes the hedgehog-type one.
As a example, for the QAH state in a system which without inversion symmetry, 
the spin rotate across the anticrossing point\cite{Ezawa M} by the nonlinear Rashba-coupling 
and induce the AFM (N\'{e}el-type) Skyrmion spin texture, 
which is one kind of topological defect like the vortices and vortex lines,
 near the Dirac-point through the Dzyaloshinskii-Moriya interaction
and favors a magnetic structure with chiral symmetry even without applying a magnetic field.
Similar AFM Skyrmion spin texture (meron) also has been found in the graphene\cite{Tse W K}, 
or the 3D TI which with a FM domain wall\cite{Wakatsuki R}.
For QAH state,
the topologically nontrivial Chern number of a pair of the valley 
becomes $\mathcal{C}=2$ for the top valence band near the $K$ and $K'$ Dirac-point with a reopended nontrivial gaps by the anticrossing phenomenon
and a symmetry Berry curvature distribution (with same sign in the $K$ and $K'$-point and the berry curvature 
${\bf \Omega}(\hat{{\bf K}})$ becomes a even function under the TR in the momentum space) due to its
special spin configuration although the 2D inversion symmetry was broken by the Rashba-coupling.
A band inversion which connects the two inverted spatial symmetries and exchange both the orbits and corresponding parities with the change of the 
Chern number from 2 to -2 can be realized by turning the electric-field-dependent nearest-neighbor Rashba coupling,
and since the intrinsic next-nearest-neighbor Rashba coupling $R$ is 0.7 meV,
and the critical ratio between $R_{2}(E_{\perp})$ and $R$ is about 0.294 and 0.143 for the valley K and K'\cite{Zhang X L}, respectively,
the critical $R_{2}(E_{\perp})$ is 0.2058 meV and 0.1 meV for K and K', respectively, which requires the external electric field about 50 V/ 300 nm
\cite{Liu C C,Kane C L}
and 25 V/300 nm, respectively.
The critical region is dominated by a valley-polarized semimetal phase
which with same Berry curvature with the SVPSM phase, 
with antisymmetry peaks between K and K' 
thus results in the totally zero Chern number ($\mathcal{C}_{K}=1$, $\mathcal{C}_{K'}=-1$) (thus becomes a trivial band insulator),
and it's achieveble by just tuning the external electric field to the critical value which has been identified above.
The band inversion happen by the exchange-site of the two counter particles through the spatial inversion,
and the corresponding orbits forms the bonding and antibonding states with negative and positive energy for the case of symmetry and antisymmetry
band rotation, respectively.

A strong exchange field and the $R_{2}(E_{\perp})$ is easy to obtained by the adsorption of transition metals, like the Nb-silicene, Ru-silicene\cite{Zhang J}
or V-silicene\cite{Zhang X L}.
It has been proved\cite{Ezawa M} that the QAHE state will replaced by the trivial band insulator by applying the electric field which
can give rise the long-range CDW with the Haldane-Hubbard potential term.
Note that the discussion in above doesn't consider the intervalley scattering induced by the short-range disorder\cite{Pan H}
due to thier spatially separation.
Next we plot three configrations of the valley-polarization in the QAHE as shown in the Fig.11(a)-(c).
The valley-difference is estimated by the valley Chern number as $\mathcal{C}_{v}=|\mathcal{C}_{K}-\mathcal{C}_{K'}|$.
In Fig.11, the Chern-number of each valley $\mathcal{C}_{K}$ and $\mathcal{C}_{K'}$ are labeled,
so the following Chern-number (total) in the valley-polarization case may appear: 
$\mathcal{C}=-3,-1,0$, with the corresponding valley-Chern-number $\mathcal{C}_{v}=1,3,2$ respectively.
The Chern number $\mathcal{C}_{K'}=-2$ only appear when the gap is reopen after the conduction band and the valence band are anticrossing (band invert).
The insect of Fig.11(b) reveals the linear-dependence of $R_{2}(E_{\perp})$ on the electron field,
and we obtain the approximate result as $R_{2}(E_{\perp})=1.12\times E_{\perp}$.
Base on this, we then plot the evolution of the band gap in valley K under the effect of electric field and effctive SOC
without and with electric field-induced Rashba coupling $R_{2}(E_{\perp})$ in the third group of Fig.12.
We see that the critical electric field where the gap close is lower down by the effect of $R_{2}(E_{\perp})$,
whoch is also agree with the previou discussion.
For the non-spin-degenerate case in the first and second groups of Fig.12,
the divided branches of energy bands which close to the E = 0 level is corresponds to the
spin-up component, while the others corresponds to the spin-down one,
and opposite for the valley K'.
while it's not for the third group where $M\neq 0$,
and we can see the obvious vlley-polarized features in this group.
We also find a anticrossing of the bands in the third group when $E_{\perp}=0.00918$ eV
which suggests the existence of the QAH phase with valley-polarization.

Except that,
we also found a new type of QAH phase which with the characteristic of single-valley-spin-degeneracy,
i.e., has one valley in the QSH phase pattern.
This single-valley-spin-degeneracy QAH phase, which we denoted as QAH$^{*}$ in the following,
it's fully polarized within the bulk gap\cite{Wang J} similar to the SDC state as we mentioned above,
but it remains the valley helical state in the boundaries.
In such phase, the Chern numbers of one of the valley is $\mathcal{C}_{K}=1$ while the other is zero,
thus it has the valley Chern number $\mathcal{C}_{v}=1$ and the spin Chern number $\mathcal{C}_{s}$ is 1/2.
Such pattern of single-valley-spin-degeneracy can be reached theoretically by applying both the out-of-plane AFM exchange field and the perpendicular electric field
with almost identical strength (to generate a AFM domain)
or by mixing the out-of-plane AFM exchange field into the QSH phase in the nanoribbon or nanotube of silicene,
which that has been proved that the helical edge state is persist between the QSH part and the others (trivial part)\cite{Rachel S,Xu Y},
and here we comment that 
the AFM Zeeman term is small compared to this out-of-plane AFM exchange magnetization term which is essentially a spin exchange term $M_{s}$.
We comment that if simply apply the out-of-plane AFM exchange field ($M_{s}^{AFM}$) in a strength larger than nearly 7.8 meV and 
has neither the QSH phase nor the perpendicular electric field,
the silicene becomes a trivial band insulator similar to the CDW one.
We plot the Berry curvature of the QAH phase, QAH$^{*}$ phase, 
and the trivial band insulator dominated by the $M_{c}$ and $M_{s}^{AFM}$, respectively, in Fig.13.

Specially, the spin-valley Chern number of such phase is 1 unlike the others whose the spin-valley Chern number is zero,
thus implies that it's a spin-valley TI\cite{Wang J}
which with broken U(1), TRI, and spatial inversion symmetry,
and it has giant possible application potentials in the spintronics and valleytronics.
For the first groups of Fig.12,
since $R=0$,
the applied perticular electric field cannot gap out the edge states (not shown),
unless apply a in-plane one,
But for the second group and third group with $R\neq 0$,
it's possible to gap out the gapless edgeb states by only applying the out-of-plane field.
Using the topologically protected interface (generated by the perpendicular electric field) 
of such a spin-valley TI in the single-valley-spin-degeneracy QAH phase (QAH$^{*}$) in the three-terminal model
(see, e.g., Ref.\cite{Wang J}),
the devices of monolayer silicene based on a SC electrode (for the nearly 1 transmission coefficients),
with two strong perpendicular electric field which with large electrostatic potential (can be generated by the STM probe) applied on the two leads.
Then by increasing the chemical potential, the exact Andreev retroreflection can be implemented
to induce the zero-edge-model with nonchirality and broken TRI,
the zero-energy valley helical edge appear,
and since the existence of the anisotropic chiral edge in QAH phase,
the junction structure may give rises the single chiral edge model that only the electrons with spin-up (or spin-down) can be transported
and thus lifts the half-integrer Hall conductivity.
For the QAH$^{*}$ phase, the band inversion may leads to the Chern number -1 in the critical point (middle point)
which is nomore a trivial phase like the QAH, and the Chern number is still -1 when the inversion finish.

\subsection{$ab\ initio$ calculation}
In Fig.14, we plot the band struture of silicene contributed by both the $\pi$-band and $\sigma$-band
in the projected surface BZ (the irreducible one).
In a large energy range, the energy band dispersion is nomore linear but becomes parabolic,
and the $1/m^{*}$ is independent of the electron wavevector in this case.
But We focus on the low-energy region.
In Fig.15, we show the band structure of silicene in the 3D hexagon first BZ 
with LDA (a), the SOC (b) and the LDA+U (c) and we set $U=1.48$ eV and magnetic moment $\mu=1.7\mu_{B}$
here where the Bohr magneton $\mu_{B}=e\hbar/(2mc)=5.78\times 10^{-2}$ meV/T.
The partial density of states (PDOS) of low-buckled silicene is shown in the Fig.15(b).
For silicene, through the DOS diagram, we can find that the two bands which are most close to the Fermi level are mainly contributed by the $3p$-orbtial
(the contrucutions have $3p_{x}+3p_{y}<3p_{z}$),
while some high-energy part may be mainly contributed by the $3s$-orbtial,
so we can also consider these two important orbitals only ($3p_{x}+3p_{y}$ and $3p_{z}$).
Then this two-band model has smaller band-filling compare to the four-band model which has the half-filling band structure 
and with the electron-hole symmetry.
The approximation of this two-orbital model also lead to the electron pockets (FS) move along the connection line of $\Gamma-\Gamma'$
like the case discussed in Ref.\cite{Graser S}.
But the 3s-orbit is dominate below the -5 eV due to its large electronegativity.

Note that results for the generalized gradient approximation (GGA) using the exchange-correlation functional of
Perdew, Burke and Ernzerhof (PBE) function\cite{Perdew J P} base on a plane wave basis set are differ from the present results less than 1\%,
so we no show here.
Above projects for the plnar-version and high-buckled version of silicene is also presented.
We obtain different effect of SOC and we found that the SOC-splitting is most obvious for the high-buckled silicene as shown in the Fig.15(i).
It's obviously that the $\pi$ and $\pi^{*}$ bands intersect in the FS in $K(H)$-point and form a Dirac-point with a band gap 0.000545 meV 
though the existing buckled structure
and they also support the two-orbital model
by the corresponding bonding and antibonding energy which split the other two bands away from the Fermi level
(the two bands that are relatively flat and touch with the $\pi$ and $\pi^{*}$ bands near the $\Gamma$-point),
while the $M(L)$-point doesn't have a Dirac-point and replaced by a gap as 1.533 eV.
That's due to the distorting in the direction of $\Gamma-\Gamma'$ of Fig.4(b)\cite{Young S M} which is also the mirror line of the correct zone.
Except that, the nonzero $t'$ breaks the particle-hole symmetry of $H$ and results in the unequal energies of $K(K')$ and $M$ Dirac-point,
if it's undistorted.
The part of $H-K$ or $M-L$ is due to our cubit BZ struture and we found that curves within these two path shows little difference for low energy part
while larger for the hight energy part of the conduction bands.
The SOC is take into consider in the Fig.15(c) and with a band gap as 1.5 meV which is consist with the valley spliting in silicon quantum well
and in good agree with the result of Ref.\cite{Ezawa M,Liu C C2,Drummond N D} and a more large band gap as 
7.9 meV is obtained in the tight-binding model\cite{Liu C C}.
That's due to the neglect of the screening effect which is increse with the electric field in the tight-binding model.
Thus the buckled structure together with the SOC dosen't obviously open a gap in the Dirac-point
which is a feature of the 2D Dirac semimetals,   
and the $\pi$ and $\pi^{*}$ bands in the band structures are mainly contributed by the $3p_{z}$ orbital and with the contribution relation $3p_{z}>3p_{x}+3p_{y}$
and are exhibit a linear features near the cross points (Drac-point; i.e., dependent linearly on the momentum) 
which is agree with the results of, e.g., Refs.\cite{Young S M,Son D T},
and it's been found that is associated with the multi-dimension representation of space group\cite{Kogan E}.
From the plot of band struture, we can see that the $\sigma$-valence-band along the $\Gamma-M$ and $\Gamma-K$ have less symmetices 
compared to that of graphenes'\cite{Kogan E2},
which also reflects the lower crystal symmetry.

For a comparision, we also form the planar silicene which with zero buckling distance and the bond length is 2.229 \AA\ Si-Si and bond angles
 is 119.97 $^\text{o}$.
Its band structures are shown in the Fig.15(e-g) with a band gap of 0.000272 meV,
which is more narrow that the low-buckled one.
The gap opened by the effective SOC for the planar silicene is much smaller than the low-buckled one which is only 0.142 meV.
That also consistent with the result that the the effective SOC at the Dirac points open a much large gap in the silicene or germanium and even the stanene
than the graphene due to the stronger atomic intrinsic SOC in the silicene, etc. 
Further, we also perform the impurity atoms doping which may give rise the band gap and transite to the band insulator and enhance the 
Rashba-coupling\cite{Zhang J,Tse W K}.
As shown in the Fig.16, we perform the 50$\%$ and 25$\%$ H atome doping and 
the 50$\%$ sulphur atome doping leads a 0.036 eV band gap.
The H doping reduced the gap in $\Gamma$-point and lift that in the $K$-point,
which is similar ti the result with a nonmetallic substrate\cite{Huang Z Q}
and it's known that the H may passivate the hybridization between the silicene layer (or multi-layer) with the metallic substrate,
thus it's possible to recover the inversion symmetry by the H-dopping.

For the distorting case which loss the protection from the inversion and nonsymmorphic symmetries (like the zigzag nanoribbon of graphene), 
the nontrivial TI phase or trivial insulating phase are possible when the center atom of the diamond zone is displaced from the original location as  
show in the Fig.4(d). 
The former is corresponds to the $Z_{2}=1$ for the $Z_{2}$ topological order which with the time reversal invariant
in BZ, and the latter one is corresponds to the $Z_{2}=0$.
Here this zero untopological $Z_{2}$ is also observable for the vertical electric field larger than the critical value which corresponds to a 
gapless intermediate state and results in a trivial band insulator for 
$\Delta_{\perp}>\Delta_{{\rm SOC}}$.
This phase transition is indeed due to the lowering of the structure symmetry by the perturbation vector
\begin{equation} 
\begin{aligned}
\delta H({\bf k})=\tau[r_{1}{\rm sin}(\frac{k_{x}}{2}+\frac{k_{y}}{2}),r_{2}{\rm sin}(\frac{k_{x}}{2}-\frac{k_{y}}{2}),2r_{3}({\rm cos}(\frac{k_{x}}{2}+\frac{k_{y}}{2})
+{\rm cos}(\frac{k_{x}}{2}-\frac{k_{y}}{2})-2)].
\end{aligned}
\end{equation}
Such change of the $Z_{2}$ topological number would also happen in the BHZ model due to the sign changing of the band mass 
and accompanied by a change of the Hubbard-U,
which is the result of the phase transition of a conventional topological insulating to a QSH one (by increasing the layer thickness)
That also provides a possible for transfer silicene to a 2D insulating substrate.
For a manifestation, we form distorted structures of silicene 
with different band lengths and the corresponding induced band gaps as presented in Table A.
As we show in the Table.A (we show the case of distortion-indeced trivial band insulator only,
and it's obviously that it has a large band gap (more larger than the TI one),
that's also why the band insulator has stronger stability against the interaction-perturbation than the TI).
Such displacement also affect the magnetic moment that the atom which close to the hexagonal center has larger magnetic moment 
and the one which far away from the center has lower magnetic moment due to the electron transition\cite{Cahangirov S}.
One another proof for the nontrivial topological in silicene is the spin (anomalous) Hall effect with SOC
and provide the Hall conductivity $\sigma_{xy}=e^{2}/(hn)$ where $n$ is the number of bands splited from the Landau level in one cell\cite{Thouless D J}.

The phonon spectrum of the low-buckled and planar silicene are shown in the (a) and (b) of Fig.17, respectively,
by the method of the energetics of finite displacements along the mode eigenvectors.
For the low-buckled one, we see that the three optical branches and the three acoustical branches are separated,
and are real at all momenta which corresponds to a stable structure ,while for the planar one, the
optical branches and acoustical branches are cross and the bottom dispersion has imaginary frequency which 
means that the planar structrue has less stability that the low-buckled one.
We also found that, for the low-buckled silicene, the upper and lower acoustical branches are linear near the $\Gamma(A)$-point,
while the middle one exhibit a two-order dispersion near the $\Gamma(A)$-point.
The linear behavior of this phonon dispersion together with the $E_{\perp}$-dependent and gapis indeed due to the nature of the tight-binding model.
Since the silicene is more close to the $sp^{3}$-bonding that the pristine graphene, it has larger electron-phonon coupling 
which may lead to the nonlinear of the Schr\"{o}dinger equation of the lattice system
and therefore enhance the phonon-mediated interactions.
Note that for both the singlet or the triplet SC (e.g., after the phase-transition caused by a sublattice-symmetry-breaking), 
there is a strongest eletron-phonon coupling (peak) $\mathcal{M}$ is position of the $k(k_{x}=0,k_{y}=0)$ especially for the
electron-doped case, but it may be broadened due to the anharmonic scattering,
and this eletron-phonon coupling strength has been confirmed that depends mainly on the phonon frequencies rather than the electronic properties\cite{McMillan W L}. 
Since for the nesting in two-band model,
there will be more electron states at the FS connected by a wave vector $k$ and thus increase the AFM fluctuation
due to both the interband interaction and the eletron-phonon interaction. 
The electron-phonon interaction here for the two-band model can be represented as 
\begin{equation} 
\begin{aligned}
H_{e-ph}=\sum_{k,k';s,s'}\mathcal{M}_{l_{1},l_{2}}c_{-k,s}^{\dag}c_{k,s'}^{\dag}c_{-k',s}c_{k',s'},
\end{aligned}
\end{equation}
where $\mathcal{M}$ is\cite{McMillan W L}
\begin{equation} 
\begin{aligned}
\mathcal{M}=&2\int\frac{d\omega_{k}\alpha^{2}(\omega_{k})F(\omega_{k}))}{\omega_{k}},\\
\alpha=&\frac{1}{2}-\frac{1}{2}(\mu^{*}{\rm ln}\frac{\theta_{D}}{1.45T_{c}})^{2}\frac{1+0.62\mathcal{M}}{1+\mathcal{M}},
\end{aligned}
\end{equation}
where $F(\omega_{k})$ is the DOS of the phonon, $\alpha^{2}$ is the average electron-phonon interaction and $\mu^{*}$ is the effective Coulomb repulsion (or the 
Coulomb pseudopotential).
The phonons which dominate the low-frequency zone of the phonon spectrum makes a lager conribution to the $\mathcal{M}$ than others.
The $\theta_{D}$ here has been found as 645K for the crystalline silicene\cite{Mertig M}.
In ideal case which with the harmonic potential and without the impurity scattering (including magnetic one and non-magnetic one and the inelastic backscattering), 
the phonons undergo a ballistic transport without scattering\cite{Wang Z}, and thus delay the relaxation of phonon. 
However at the anharmonic case due to the detuning, the scattering will happen,
and the anharmonic phonon-phonon scattering also damage the edge current since the rule of specular-scattering was violated.


In short, the intuitive 2D Dirac semimetal attributes of buckled silicene with the nonsymmorphic symmetry
is contributed by the touching conduction band and the valence band with the massless charge carriers in the bottom even in the presence of the SOC
which provides mass to the electrons and breaks the chiral symmetry of Dirac-cone pairs.
Due to the distorted structure diagonal direction,
there is no nodes along the edge of BZ due to the unparallel gap functions for the two bands, i.e., $\Delta_{\alpha}\nparallel\Delta_{\beta}$, 

\section{Inhomogeneous electromagnetic wave and circular polarized light}
The inhomogeneous electromagnetic wave and circular polarized light (left or right polarization) 
can also control the topological phase of silicene in the tight-binding model with $\pi$ bands.
The spin angular momentum $\hat{S}$ can be carrired by the nonlinear circular polarization light by $\pm \hbar$ per photon 
which with the photon frequency $\omega_{l}$ (also can be viewed as a boson frequency in constrast with the $\omega$
as shown in the follwoing conductivity's expressions).
Here we take the off-resonant circular polarized laser beam as a example firstly,
which with the frequency $\omega\gg t$ and can relativistic self-focusing and thus overcome its diffraction when it's in the plasma channel,
also, thus it's off-resonant where electrons cannot directly
absorb the photons\cite{Kitagawa T},
and the electron mass can be controlled by the laser intensity by drive the electrons to quiver with a determined velocity,
which is similar to the case of vertical electric field.
Here the plasma frequency is $\omega_{p}=\frac{2eE_{F}}{\hbar}\sqrt{\frac{1}{3\pi\hbar v_{F}\epsilon_{0}\epsilon_{s}}}$,
where the static dielectric constant of silicene $\epsilon_{s}=34.33$ (set $\epsilon_{0}=1$),
which is much larger than the silicon (11.9), SiO (3.9)\cite{Mohan B} and SiO$_{2}$ (4)
and thus has a stronger screening effect due to its stronger charge polarization,
and this screening effect also suppress the opening of the band gap
and breaks the Coulomb long-range order, 
but the band gap is still changes linearly with the $E_{\perp}$ even under this screening effect.
While for the electromagnetic wave or the on-resonant circular polarized light,
the screened interaction within the excited electron-hole pairs (i.e., the excitons which also the one kind of mang-body effect) 
cohesive the singlet and triplet states by the strong attractive effect in the optically active model.
Since the excitation is protected by the suppression of the interband scattering (between the particles and holes) which is massive for the narror flat band,
and hence enhance the pairing instability of two bands, the rise of the interband scattering will 
support the freestanding silicene and
the zero chemical potential with particle-hole symmetry will leads to the excitonic instabilities,
and with the nonlinear gap function just like the undoped graphene\cite{Khveshchenko D V2}.

Here the scattering matrix can be consisted by the two pairs: transmission (including the normal scattering (specular one or the backscattering) 
and Andreev one with a s-wave superconductor) and 
reflection (including the specular scattering and Andreev one) of the electrons,
and the scatterings are odd parity for the particle-hole transformation,
e.g., $|h{\bf k}\rangle=e^{2i\phi_{k}}|e{\bf k}\rangle$,
where $|h{\bf k}\rangle$ and $|e{\bf k}\rangle$ are the electron state and hole state, respectively,
and $e^{2i\phi_{k}}$ is the pseudospin(valley)-dependent odd parity scattering factor
(which is easy to proved by carry out the particle-hole transition as $c_{i\uparrow}\rightarrow c_{i\uparrow},c_{i\downarrow}\rightarrow (-1)^{i}c_{i\downarrow}$
in a AFM ordered spin pattern. see the below text).
We can represent it in the single-terminal travelling model as
\begin{equation} 
\begin{aligned}
\begin{pmatrix}[1.5] |h{\bf k}\rangle\\
|h{\bf k}\rangle^{\dag}\end{pmatrix}
=
\begin{pmatrix}[1.5] 
0&-1\\
-1&0
\end{pmatrix}
\begin{pmatrix}[1.5] 
|e{\bf k}\rangle\\
|e{\bf k}\rangle^{\dag}
\end{pmatrix}
=
\begin{pmatrix}[1.5] 
|-e{\bf k}\rangle^{\dag}\\
|-e{\bf k}\rangle
\end{pmatrix}
\end{aligned}
\end{equation}
or for the four-terminal one,
\begin{equation} 
\begin{aligned}
\begin{pmatrix}[1.5] |h_{1}{\bf k}\rangle\\
|h_{1}{\bf k}\rangle^{\dag}\\
|h_{2}{\bf k}\rangle\\
|h_{2}{\bf k}\rangle^{\dag}\end{pmatrix}
=
\begin{pmatrix}[1.5] 
0&-1&0&0\\
-1&0&0&0\\
0&0&0&1\\
0&0&1&0
\end{pmatrix}
\begin{pmatrix}[1.5] 
|e_{1}{\bf k}\rangle\\
|e_{1}{\bf k}\rangle^{\dag}\\
|e_{2}{\bf k}\rangle\\
|e_{2}{\bf k}\rangle^{\dag}
\end{pmatrix}
=
\begin{pmatrix}[1.5] 
|-e_{1}{\bf k}\rangle^{\dag}\\
|-e_{1}{\bf k}\rangle\\
|e_{2}{\bf k}\rangle^{\dag}\\
|e_{2}{\bf k}\rangle
\end{pmatrix},
\end{aligned}
\end{equation}
and the quantized charge conductance can be obtained by $\sigma_{xy}=e^{2}/(2h)$ in the Landauer-B\"{u}ttiker framework,
which is just like the half-Hall conductance in the QAH phase for 3D TI\cite{Wakatsuki R}.

\subsection{Inhomogeneous electromagnetic wave}
Since the electromagnetic wave was applied aslant to the silicene sheet with a angle $\theta$,
the electromagnetic action of Dirac-semimetal in quantum field theory is\cite{Wilczek F,Sekine A,Fujikawa K}
\begin{equation} 
\begin{aligned}
S_{\theta}=\frac{e^{2}}{4\pi^{2}\hbar c}\int dt\ d^{3}r\ {\rm Tr}(F_{\mu v}\tilde{F}_{\mu v}),
\end{aligned}
\end{equation}
with ${\rm Tr}(F_{\mu v}\tilde{F}_{\mu v})\approx {\bf E}\cdot{\bf B}$
where ${\bf E}$ is the electric field vector (with unit vector $({\rm sin}\ \theta,0,{\rm cos}\ \theta)$ in the $x-z$ plane)
and ${\bf B}$ is the magnetic field vector which is in $y$-direction (with unit vector $(0,\epsilon_{1}{\rm cos}\theta-i\epsilon_{2}{\rm sin}\theta,0)$),
$F_{\mu v}=\partial_{\mu}{\bf A}_{v}-\partial_{v}{\bf A}_{\mu}+[{\bf A}_{\mu},{\bf A}_{v}]$,
and a charge current is induced by the $\theta$ term which 
induce the net flux and make it move from the $\theta=0$ boundary into the cell and then obtains the charges\cite{Pretko M,Sekine A},
the resulting charge current density in a perturbatively (variational) form is
${\bf J}({\bf r})=\delta S_{\theta}/\delta {\bf A}=\frac{e^{2}}{4\pi^{2}\hbar c}[\nabla\theta({\bf r}){\bf E}+\dot{\theta}({\bf r}){\bf B}]$,
which with the Hall current due to the QAHE in the first term and the chiral magnetic effect 
in second term, and here the term $\nabla\theta({\bf r}){\bf E}$ can be viewed as a electrical conductivity tensor.
Note that the Hall current by the QAHE here is not depends on the external magnetic field like the QHE, but depends on the spin magnetization.
The electromagnetic potential has ${\bf A}(\theta)=A({\rm sin}\ \theta,0,{\rm cos}\ \theta)$,
where $A$ is the amplitude of the inhomogeneous transverse wave
(which does not obeys the Coulomb gauge (zero-divergence constraint) 
$\nabla\cdot {\bf A}=0$ but obeys the axis current conservation $\partial_{\mu}{\bf J}^{\mu}=0$
with ${\bf J}=\frac{e}{\hbar}|{\bf J}_{L}-{\bf J}_{R}|$ (left and right chirality $L$ and $R$),
in the case of even dimension and with the chiral edge current which with opposite spin polarization by the chiral polarization field formed by the 
inhomogeneous electromagnetic wave or circular polarized light,
but it's not conserved anymore for the odd (space+time) dimension case\cite{Sekine A}),
and it has a dimensionless form as $A=\frac{e}{2\pi^{2}\epsilon_{0}\epsilon_{s}\omega_{p}c}$.
And the momentum $\hbar{\bf k}\ ({\bf k}=-i\nabla)$ can be replaced by the covariant momentum $\hbar{\bf k}+\frac{e}{c}{\bf A}$ (or canonical momentum\cite{Tahir M})
with first term the standard momentum of the Dirac quasiparticles and the second term describes the interaction with electromagnetic potential,
under the Landau gauge (U(1) gauge potential) ${\bf A}=Ax(0,-{\rm cos}\ \theta,0)$,
which also called Peierls substitution\cite{Tabert C J} or minimal substitution\cite{Ezawa M5} or minimal coupling\cite{Sharapov S G}. 
The QAHE can be induced by the internal magnetization and the SOC and Rashba-coupling with zero $E_{\perp}$ and
in the case that the TRI was broken by the internal magnetization,
and the electric displacement tensor can be obtained\cite{Ukhtary M S} by the frequency-dependent complex dielectric function $\epsilon(\Omega)$
which make it still works even in the thick monolayer limit (the 2D sheet)
\begin{equation} 
\begin{aligned}
\begin{pmatrix} D_{x}\\ D_{y}\\D_{z} \end{pmatrix}
&=\begin{pmatrix} \epsilon_{1}E_{x}+i\epsilon_{2}E_{z}\\ 0\\-i\epsilon_{2}E_{z}+\epsilon_{1}E_{z} \end{pmatrix},\\
\epsilon_{1}&=\epsilon_{0}\epsilon_{s}(1-\frac{\omega_{p}^{2}}{\omega^{2}}),\\
\epsilon_{2}&=\frac{e^{2}a}{2\pi^{2}\omega\hbar c}.
\end{aligned}
\end{equation}
Under the effect of both the QAHE and the magneto-optical effect with the dc-driven charge carriers 
and a dc-driven Hall conductivity $\sigma_{xy}^{dc}$ proportional to the filling factor $v_{f}$ (and implement a coherently propagate in low-frequency) 
in the insulation medium
which corresponds the nonzero and zero components of
angular momentum, respectively, and leading to the symmetric off-diagonal dielectric
tensor matrix\cite{Toyosaki H}
whose direction is depends on the sign of ${\bf q}/|{\bf q}|$.
And the frequency-dependent conductivity is\cite{Gajdo? M,Matthes L3}
\begin{equation} 
\begin{aligned}
{\rm Re}\ [\sigma(\Omega)]=\frac{2\pi e^{2}}{V}\lim_{{\bf q}\rightarrow 0}\frac{1}{|{\bf q}|^{2}}\sum_{c,v,{\bf k}}[\delta(\epsilon_{c,{\bf k}+{\bf q}}
-\epsilon_{c,{\bf k}}+\hbar w)-\delta(\epsilon_{c,{\bf k}+{\bf q}}-\epsilon_{c,{\bf k}}-\hbar w)],
\ V=\epsilon_{0}^{2}\epsilon_{s}^{2}\omega A.
\end{aligned}
\end{equation}
It's related to the in-plane-conductivity $\sigma_{{\rm PL}}$ as $\sigma(\Omega)=a\sigma_{{\rm PL}}(\Omega)$,
and the in-plane dielectric has $\epsilon_{{\rm PL}}(\Omega)=1+\frac{i\sigma_{{\rm PL}}(\Omega)}{\epsilon_{0}\omega \overline{\Delta}}$\cite{Sorianello V,Falkovsky L A}
where $\epsilon_{0}$ is the dielectric constant of vacuum.
In fact, the magneto-optical effect in the monolayer silicene is not obviously compare to the other bulk TI since its low-buckled distance
unless the frequency of light is large enough which is theoretical about $6\times 10^{6}$ THz to realize $\overline{\Delta}\simeq \lambda$. 
Since we suppose the ihomogeneous transverse wave, the $k_{y}$ is zero and the $y$-component of electric field $E_{y}=0$
due to the zero $\epsilon_{xy}$ and $\epsilon_{zy}$, and inhomogeneous in the $x$-direction.

The nearest-neighbor-site amplitudes satisfy $\Psi_{A}=i\pm\Psi_{B}$ with $A,B$ the two sublattices,
which is also the zero-energy solution with particle-hole symmetry,
and the sign $+$ here with positive imaginary part is correspond to the bounded state with negative energy 
while the negative part is corresponds to the zero or positive one.
And the zero models (the gapless edge state with particle-hole symmetry) is appear along the $m_{D}(x)=0$ and it's angle $\theta$-independent in the $y$-diraction
where the exchange coupling along the $z$-direction is pseudospin-independent,
Here we introduce a fixed $k_{z}$ which is a good quantum number with the transitional invariance althought the electromagnetic wave is along the $z$ direction,
and 
the envelope function spinor which in the four component basis
is $\Psi(x,z)=e^{ik_{z}z}\Phi(x)$ which is in a form silimar to the solitary wave solutions.
Then the eigenvalue problem yields eigenfunction 
\begin{equation} 
\begin{aligned}
H(\partial_{x},k_{z})\Psi(x,z)=E(k_{z})\Psi(x,z),\\ 
\begin{pmatrix}[1.5] m_{D}(x)&-\hbar v_{F}\partial_{x}\\
\hbar v_{F}\partial_{x}&-m_{D}(x)
\end{pmatrix}
\begin{pmatrix}[1.5] \Phi_{A}(x)\\
\Phi_{B}(x)
\end{pmatrix}
=0,
\end{aligned}
\end{equation}
with $E(k_{z})=\pm\eta\hbar v_{F}k_{z}$ the linear relation in the charge-neutral point and $\theta$-independent for zero model,
and perturbatively, the motion of $\Psi(x)$ can be represented by the Dirac mass term $m_{D}$ and the wave-induced inhomogenate Dirac-mass $m_{{\rm w}}(x)$,
\begin{equation} 
\begin{aligned}
&\eta\hbar v_{F}\partial_{x}\Psi(x)=-m_{D}\Psi(x),\\
&m_{D}(x)=(-\eta\hbar m_{{\rm w}}(x)+\eta\lambda_{{\rm SOC}}s_{z}-\frac{\overline{\Delta}}{2}E_{\perp}+Ms_{z}),\\
&m_{{\rm w}}=\frac{v_{F}^{2}e^{4}a^{2}}{4\pi^{4}\epsilon^{2}_{0}\epsilon^{2}_{s}\hbar^{2}c^{2} \omega(x) \omega_{p}^{2}}
\end{aligned}
\end{equation}
and can be solved by using the good quantum number $s_{z}$ (since here the rashba-coupling are ignored) as
\begin{equation} 
\begin{aligned}
\Psi(x)=C\ {\rm exp}\left\{{\rm exp}\left[\frac{-1}{\eta\hbar v_{F}}\int m_{D}(x')dx'\right]\right\},
\end{aligned}
\end{equation}
where $C$ is the normalization constant.
And the energy spectrum is
\begin{equation} 
\begin{aligned}
\varepsilon_{\eta}({\bf k})=Ms_{z}\pm\sqrt{\hbar^{2}v_{F}^{2}{\bf k}^{2}+(\frac{\overline{\Delta}}{2}E_{z}+\eta\hbar m_{{\rm w}}-\eta s_{z}\lambda_{{\rm SOC}})^{2}}.
\end{aligned}
\end{equation}
For the case of particle-hole symmetry, 

For a disturbed rotation angle $\theta$ from the electromagnetic wave, the valley-dependent interband transition (i.e., the pseudospin texture) can be described by
\begin{equation} 
\begin{aligned}
k_{x}+i\eta k_{y}=\frac{{\bf k}}{\epsilon_{\eta}}e^{i\eta\theta}，
\end{aligned}
\end{equation}
and the $n=0$ Landau level (see Sect.4.3) which is splited to $\pm|\lambda_{{\rm SOC}}-\frac{\overline{\Delta}}{2}E_{\perp}|$
is
\begin{equation} 
\begin{aligned}
H_{n=0}=&\pm\hbar v_{F}e^{-i\theta\sigma_{y}/2}(k_{z}{\rm cos}\theta+k_{y}{\rm sin}\theta),\\
\theta=&\pm{\rm arctan}\frac{A_{y}/({v_{y}})}{A_{z}/({ v_{z}})},
\end{aligned}
\end{equation}
where $v_{y/z}$ is the velocity operator as mentioned above, and the sign $\pm$ depends on the right and left polarization in the $x$-direction.
While for the undisturbed case, 
\begin{equation} 
\begin{aligned}
H_{n=0}=\pm\hbar v_{F}k_{z},
\end{aligned}
\end{equation}
where $k_{z}$ can be viewed as a mass term, i.e., the $z$-component of vector $\bf K$.
The effect of $m_{{\rm w}}$ with chirality (which with the $s_{z}$ different to the helical state) 
can be viewed as a purely imaginary next-nearest-neighbor hopping term with broken TRI but preserved particle-hole symmetry,
\begin{equation} 
\begin{aligned}
-i\frac{\hbar m_{{\rm w}}}{3\sqrt{3}}\sum_{\langle\langle i,j\rangle\rangle ;\sigma\sigma'}\upsilon_{ij}c^{\dag}_{i\sigma}c_{j\sigma'}.
\end{aligned}
\end{equation}
We can also easily find the four energy levels in the energy spectrum (band structrue) as:
\begin{equation} 
\begin{aligned}
(\hbar m_{{\rm w}}(x)+\lambda_{{\rm SOC}}+\frac{\overline{\Delta}}{2}E_{\perp}-M)>\\
(-\hbar m_{{\rm w}}(x)+\lambda_{{\rm SOC}}-\frac{\overline{\Delta}}{2}E_{\perp}+M)>\\
(\hbar m_{{\rm w}}(x)-\lambda_{{\rm SOC}}+\frac{\overline{\Delta}}{2}E_{\perp}+M)>\\
(-\hbar m_{{\rm w}}(x)-\lambda_{{\rm SOC}}-\frac{\overline{\Delta}}{2}E_{\perp}-M)
\end{aligned}
\end{equation}
for the TI phase,
and 
\begin{equation} 
\begin{aligned}
(\hbar m_{{\rm w}}(x)+\lambda_{{\rm SOC}}+\frac{\overline{\Delta}}{2}E_{\perp}-M)>\\
(\hbar m_{{\rm w}}(x)-\lambda_{{\rm SOC}}+\frac{\overline{\Delta}}{2}E_{\perp}+M)>\\
(-\hbar m_{{\rm w}}(x)+\lambda_{{\rm SOC}}-\frac{\overline{\Delta}}{2}E_{\perp}+M)>\\
(-\hbar m_{{\rm w}}(x)-\lambda_{{\rm SOC}}-\frac{\overline{\Delta}}{2}E_{\perp}-M)
\end{aligned}
\end{equation}
for the trivial band insulator phase,
which are may corresponds to the four electron states $|\pm\frac{1}{2}\rangle,|\pm\frac{3}{2}\rangle$
(orbital characters of the electron wave functions) near the FS
in the case with particle-hole symmetry and spectrum symmetry and have not electron interaction.

\subsection{circular polarized light}
While for the circular polorized light 
under a Chern-Simons action which in a similar form with Eq.(65):
$S(t,{\bf r})=(e^{2}/4\pi^{2}\hbar c) \int dtd^{3}r\partial_{\mu}(\omega_{l} t)\varepsilon^{\mu\nu\rho\sigma}A_{\nu}\partial_{\rho}A_{\sigma}$
where $\varepsilon_{\mu\nu\rho\sigma}$ is the Levi-Civita symbol,
and here the term $(\omega_{l} t)$ won't restrict the photon but corresponds a free photon field.
The gauge electromagnetic potential ${\bf A}(t)=A(\pm{\rm sin}\ \omega_{l} t,{\rm cos}\ \omega_{l} t)$\cite{Kitagawa T}
where $+(-)$ denotes the right (left) polarization with the time periodicity $T=2\pi/\omega$ and with $\nabla\times {\bf A}={\bf K}$,
thus this circular polorized light-induced Dirac mass is spin-valley-dependent,
while the linear polorized light can't brings the Dirac mass.
And the above-mentioned Hall current will change sign when the circular polorization of light reverses.
The effect of off-resonant circular polorized light vector potential around the two neighbor-valley is described by the effective Hamiltonian
\begin{equation} 
\begin{aligned}
H_{{\rm eff}}^{+}=
\begin{pmatrix}[1.5] iaR(A_{x}+iA_{y})&\hbar v_{F}(A_{x}-iA_{y})\\
\hbar v_{F}(A_{x}+iA_{y})&iaR(A_{x}-iA_{y})
\end{pmatrix},\\
H_{{\rm eff}}^{-}=
\begin{pmatrix}[1.5] -iaR(A_{x}+iA_{y})&-\hbar v_{F}(A_{x}+iA_{y})\\
-\hbar v_{F}(A_{x}-iA_{y})&-iaR(A_{x}-iA_{y})
\end{pmatrix},
\end{aligned}
\end{equation}
and it resulting a quantized Hall conductance with the possible QAHE and a gap opened by the virtual photon process.
In the case of TRI, $H_{{\rm eff}}^{+}=H_{{\rm eff}}^{-*}$.
The effective Hamiltonian given by the time-dependent vector potential is 
$H_{{\rm eff}}=\frac{i\hbar}{T}{\rm log}[\mathcal{T}{\rm exp}(-it\int^{T}_{0}e^{i{\bf A}(t)dt})]$
where $\mathcal{T}$ is the Feynman-Dyson time-ordering operator, as shown in the Fig.18 with the frequency of light (laser)
3000 THz, 1000 THz, and 500 THz, which with the harmonic oscillator with nature frequency $\omega$ according to Floquet theory.

The optical absorbtion $A_{op}(\omega)=(1/\epsilon_{0}c){\rm Re}[\sigma(\omega_{l})]$\cite{Matthes L}
as a first-order-process which is shown in the Fig.19(a) would happen when the frequency is lowerd to below the $3t=4.8 $ eV $=1000$ THz\cite{Ezawa M3}
(and becomes on-resonant light; while for graphene, this critical frequency is 6t$\approx 14.4$ eV$=3000$ THz),
and the second-order-process, like the photocoupling and the reflection, vanish in this region.
In this case, the time-dependent damped oscilation would happen due to the interband electronic transition 
(between the empty conduction band and the filled valence band) with the optical absorbtion,
and results in a unquantized (anomalous) hall effect
due to the subsequent relaxation in the nonequilibrium dynamics after quenched way from the steady state.
And there are massive interband-scattering (especially for the inter-flat-band scattering),
which are mainly formed by the $\pi$ and $\pi^{*}$ bands or $\sigma$ and $\sigma^{*}$ bands,
and even the $\pi$ and $\sigma$ bands which is similar to the graphene.
The spin-valley-dependent optical absorbation and the selection rules also support the tuneable linear bands by the circular polarized light
together with the vertical electric field.
We present some of the optical propertices of silicene in Fig.19.
We can see that the main peaks of optical absorbtion(a), real part of the optical conductivity(b), and the energy loss spectra(c)
are roughly at the same energy region ($5$ eV$\sim 7.5$ eV).
That suggest they have similar spectral behavior and such phenomenons also found in the graphene which has a real optical conductivity distribute similar to 
the plasmon peaks\cite{Eberlein T}.
The dielectric function (Fig.19(d)) in the ($k,\omega$)-space has $\epsilon^{-1}(k,\omega)=1-\frac{8\pi^{2}c_{s}}{k}\chi(k,\omega)$ 
where $\chi(k,\omega)$ is the complex polariztion (or susceptibility) in RPA.
The transmission peak of real dielectric function in the range below 1 eV is
also related to both the inter-band transitions and intra-band transitions which characterized by the Drude factor
which is zero in the thermodynamic limit $N\rightarrow \infty$.
And the possible excitonic transitions in the low-energy range has been suggested\cite{Mohan B2}. 
While the reflection valley is nearly 5 eV.
The energy loss function is well fits with the dielectric function by ${\rm  Im}\epsilon^{-1}(k,\omega)$ as shown in the Fig.19(c).
The $\omega$-dependent kinetic energy is obtained by sum up the real part of the frequency-dependent planar optical conductivity 
and are approximate tothe single-time optical conductivity,
$-E_{{\rm kin}}(\omega)=\frac{1}{\pi}\int d\omega{\rm Re}[\sigma]
(\omega)\sim \sigma(t)$ and calculated as $E_{{\rm kin}}$=-6.65$/\pi$.

The complex dielectric function $\epsilon(\omega)$ can be expressed by
$\epsilon(\omega)=(n_{t}(\omega)-i\kappa(\omega))^{2}$\cite{Matthes L} as shown in the (d).
Indeed, the resonance frequency rises the semiclassic features and its linear with the increse of electric field and circular polarized light.
The Sommerfeld vacuum fine structure constant
$c_{s}=\frac{e^{2}}{2\epsilon_{0}h c}=1/137.036$\cite{Nair R R}.
is related to the zero-$\omega$ optical absorption in the limit of vanishing SOC by $A_{op}(0)=\pi c_{s}$\cite{Matthes L2},
which is applicablr for all the group IV atoms.

The circular polarized light-indeuced periodically driven nonequilibrium system has results a dc-driven charge current,
which disobey the current continuity $\nabla \cdot{\bf J}+\frac{\partial \rho}{\partial t}=0$ and 
the probability current conserved $-\frac{i\hbar}{2m}(\phi^{*}\nabla\phi-\phi\nabla\phi^{*})$ and
the Gaussian distribution (wigner-Dyson type), and can be represented by a variant reservior response form 
in a determined Landau level with the frequency $\omega$ before the optical coupling:\cite{Kitagawa T}
\begin{equation} 
\begin{aligned}
{\bf J}_{{\rm res}}=&\sum_{n,b}\int 
\frac{d\omega}{2\pi}t_{a}^{2}\rho_{a}(\omega+n \hbar\omega_{l})t_{b}^{2}\rho_{b}(\omega)G_{ij}(n,\omega)
(f_{b}(\omega)-f_{a}(\omega+n \hbar\omega_{l})),\\
G_{ij}(n,\omega)=&\int dt e^{in\hbar\omega_{l}t}\int dt'G_{ij}^{{\rm ret}}(t,t')e^{i(\omega+i0^{+})(t-t')},\\
\rho_{a}(\omega+n \hbar\omega_{l})=&\frac{1}{N}\sum_{{\bf k}}\delta(\omega+n \hbar\omega_{l}-t|\epsilon_{{\bf k}}|),\\
\rho_{b}(\omega)=&\frac{1}{N}\sum_{{\bf k}}\delta(\omega-t|\epsilon_{{\bf k}}|),\ t=\frac{2\sqrt{3}\hbar v_{F}}{3a}\approx 1.6 eV,\ v_{F}\approx 5.5\times 10^{5}\ {\rm m/s}\\
f_{b}(\omega)=&\frac{1}{1+{\rm exp}[\beta_{b}(\omega-\mu_{b})]},\\
f_{a}(\omega+n \hbar\omega_{l})=&\frac{1}{1+{\rm exp}[\beta_{a}(\omega+n \hbar\omega_{l}-\mu_{a})]},\\
\end{aligned}
\end{equation}
where the $a$ corresponds the channel which coupling with the photons (absorbs (or emits) $n$ photons),
while the $b$ is the one which not couple with the photons
(i.e., describe the transport between two leads $a$ and $b$).
$f_{a/b}$ is the Fermi-Dirac distribusion function, 
$\omega_{l}$ denotes the frequency of the light,
$\rho_{a/b}$ is DOS per unit cell of each channel,
and $G_{ij}^{{\rm ret}}(t,t')$ is the retarded Green's function
$G_{ij}^{{\rm ret}}(t,t')=-i\theta(t-t')\langle\{c_{i}(t),c_{j}^{\dag}(t)\}\rangle$ ($\theta(t-t')$ is the Heaviside step function),
and th term $G_{ij}^{{\rm ret}}(t,t')e^{i(\omega+i0^{+})(t-t')}$ 
can be replaced by the advanced Green's function as $G_{ij}^{{\rm adv}}(t,t')e^{i(\omega+i0^{-})(t'-t)}$ 
with $G_{ij}^{{\rm adv}}(t,t')=i\theta(t'-t)\langle\{c_{i}(t),c_{j}^{\dag}(t)\}\rangle$ in the above equation
since the relation $G_{ij}^{{\rm ret}}(t,t')=(G_{ij}^{{\rm adv}}(t,t'))^{*}$.
While for the Matsubara frequency $\omega_{M}$, wejust need to replace the real time by the imaginary time $\tau$.
The reservior variables can be well described by the master equation in the Liouville space (see Ref.\cite{Wu C H}) 
with the unperturbed density operator $\mathcal{J}$
\begin{equation}   
\begin{aligned}
\partial_{t}\mathcal{J}=-i[H,\mathcal{J}]+\mathcal{K}\sum_{i}[O_{i}\mathcal{J}O_{i}^{\dag}-\frac{1}{2}(O_{i}^{\dag}O_{i}\mathcal{J}+\mathcal{J}O_{i}^{\dag}O_{i})]\equiv\mathcal{L}\mathcal{J},
\end{aligned}
\end{equation}
where $\mathcal{J}$ corresponds to the pure state or mixed state and $O_{i}$ is the Lindblad operator describing the bath coupling.
For a detail discussing about the mattter+dielectric mediated+reservoir system, see Ref.\cite{Behunin R O}.
Then the Hall conductivity can be obtained by the linear response Kubo formula.
First the Hall currents correlation\cite{Lykken J D,Hsu Y F,Gusynin V P}
\begin{equation}
\begin{aligned}
\Pi_{xy}(i\hbar\omega_{l})=&-\frac{ie^{2}\hbar^{2}v_{F}^{2}}{2\pi\hbar\ell_{B}\beta}\sum_{n=0}^{\infty}\int\frac{d\omega}{2\pi}\frac{d^{2}k}{(2\pi)^{2}}{\rm Tr}
\left[\gamma^{x}S_{\beta}(i\omega)\gamma^{y}S_{\beta}(i\omega+i\omega_{l})\right]\\
=&-\frac{ie^{2}\hbar^{2}v_{F}^{2}}{2\pi\hbar\ell_{B}\beta}\sum_{n=0}^{\infty}\int\frac{d\omega}{2\pi}\frac{d^{2}k}{(2\pi)^{2}}
\left[(i\omega+\mu+i\eta_{s})\left(\frac{1}{(i\omega+\mu+i\eta_{s})-\varepsilon^{2}_{\eta N}}+\frac{1}{(i\omega+\mu+i\eta)-\varepsilon^{2}_{\eta (N+1)}}\right)\right.\\
&\left.+im_{D}\left(\frac{1}{(i\omega+\mu+i\eta_{s})-\varepsilon^{2}_{\eta (N+1)}}-\frac{1}{(i\omega+\mu+i\eta)-\varepsilon^{2}_{\eta N}}\right)\right]\\
&\times\left[(i\omega+i\omega_{l}+\mu+i\eta_{s})\left(\frac{1}{(i\omega+i\omega_{l}+\mu+i\eta_{s})-\varepsilon^{2}_{\eta N}}
+\frac{1}{(i\omega+i\omega_{l}+\mu+i\eta_{s})-\varepsilon^{2}_{\eta (N+1)}}\right)\right.\\
&\left.+im_{D}\left(\frac{1}{(i\omega+i\omega_{l}+\mu+i\eta_{s})-\varepsilon^{2}_{\eta (N+1)}}-\frac{1}{(i\omega+i\omega_{l}+\mu+i\eta_{s})-\varepsilon^{2}_{\eta N}}\right)\right],
\end{aligned}
\end{equation}
which can be written in a retarded form by the
analytical continuation as $i\hbar\omega_{l}\rightarrow \hbar(\omega_{l}+i0^{+})$
and then the dc Hall conductivity can be obtained as 
$\sigma_{xy}={\rm lim}_{\omega_{l}\rightarrow 0}{\rm Im}[\Pi_{xy}(\hbar(\omega_{l}+i0^{+}))]/(\hbar\omega_{l})$.
We comment that the dc-Hall charge conductivity is always zero since it sum over both the spin and valley index as we show in above.
The $S_{\beta}$ in above function is the temperature-related partial function 
\begin{equation}   
\begin{aligned}
S_{\beta}(i\omega)=\int d\omega\frac{A(\omega)}{(i\omega+\mu+i\eta_{s})},
\end{aligned}
\end{equation}
with the momentum-space Green's function $G(i\omega+\mu+i\eta_{s},k)$ has the pole in the eigenfrequency $\tilde{\omega}$ as
$G(i\omega+\mu+i\eta_{s},k)\propto 1/(\omega-\tilde{\omega})$.
And the frequency (or dispersion)-dependent spectral weigth function which can be obtained directly by the Angle-resolved photoemission spectroscopy (ARPES) is
\begin{equation}   
\begin{aligned}
A({\bf k},\omega)=\frac{-{\rm Tr}G(i\omega+\mu+i\eta_{s},k)}{\pi},
\end{aligned}
\end{equation}
whose time-evolution can be obtained by
the above Floquet retarded Green's function as
\begin{equation}   
\begin{aligned}
A(t,\omega)=\frac{-{\rm Im}\int_{t_{i}}^{t_{f}}dt'e^{i(\omega+i0^{+})(t-t')}G^{{\rm ret}}(t,t')}{\pi}.
\end{aligned}
\end{equation}
Note that this time-dependent spectral weigth function is different from the $A({\bf k},\omega)$.
The retarded Green's function here contains a single-time different-site term as explained above,
and we can use the spin-correlation (magnetic order or the local magnetization magnetization) to carry out the simulations,
as we present in the Fig.20.
The two-site time-dependent correlation $-i\langle S_{i}(t)S_{j}(t)\rangle$ from the Mott insulating initial state is taken into accout,
and the transition of the kinetic energy during the relaxation process is on-site interaction-dependent as shown in the lower-left inset in Fig.20.
Note that the scattering factor is enhanced during the relaxation of the above spectral weigth function as shown in the Fig.20.
The oscillations during the relaxation (especially for low-temperature) 
is related to the coupling between the spin and charge dynamic in the 3D momentum space (unlike the 1d chain which the spin and charge dynamics are independent)
just like what happen in the lowest Landau-level
as mention below.
The eigenenergy $\tilde{E}$ corresponds to this eigenfrequency $\tilde{\omega}$ has the following relation under the magnetic field
(we still assume only the $x$-direction be a non-good quantum number)
\begin{equation} 
\begin{aligned}
\begin{pmatrix}[1.5] 
m_{D}(x)&-i\frac{\hbar v_{F}}{\ell_{B}}(\frac{x}{\ell_{B}}+\ell_{B}\partial_{x})\\
i\frac{\hbar v_{F}}{\ell_{B}}(\frac{x}{\ell_{B}}-\ell_{B}\partial_{x})&-m_{D}(x)
\end{pmatrix}
\begin{pmatrix}[1.5] \Phi_{A}(x)\\
\Phi_{B}(x)
\end{pmatrix}
=\tilde{E}
\begin{pmatrix}[1.5] \Phi_{A}(x)\\
\Phi_{B}(x)
\end{pmatrix}
\end{aligned}
\end{equation}

For a non-hermiticity topological system under the effect of the light, due to the complex next-nearest-neighbor hopping, the particle-hole symmetry don't exist,
then for the Haldane model in such system, the motion of amplitude of the particle $A^{e}$ and hole $A^{h}$ in the sublattices $A$ and $B$, respectively,
\begin{equation}   
\begin{aligned}
i\partial_{t}A^{e}_{ij}=\omega A^{e}_{ij}+t\sum_{\langle i,j\rangle}A^{h}_{ij}+t'\sum_{\langle\langle i,j\rangle\rangle}A^{e}_{ij},\\
i\partial_{t}A^{h}_{ij}=\omega A^{h}_{ij}+t\sum_{\langle i,j\rangle}A^{e}_{ij}+t'\sum_{\langle\langle i,j\rangle\rangle}A^{h}_{ij}e^{-2i\phi_{ij}},\\
\end{aligned}
\end{equation}
where $\omega$ is the on-site resonance frequency, $\phi_{ij}$ is the Haldane flux.
And here $|A^{e}|\neq|A^{h}|$ since the particle-hole symmetry is broken.
The graphs are shown in the Fig.21.
For the inhomogeneous vector field and according to the Eq.(71), the above equations can be represented by the 2D nonlinear Soler model by the
polar coordinates as\cite{Cuevas-Maraver J,Merle F}
\begin{equation}   
\begin{aligned}
i\partial_{t}A^{e}_{ij}=\omega A^{e}_{ij}+e^{-i\theta}(i\partial_{r}+\frac{1}{r}\partial_{\theta})A^{h}_{ij}+m_{D}\left||A^{e}_{ij}|^{2}-|A^{h}_{ij}|^{2}\right|^{k}A^{e}_{ij},\\
i\partial_{t}A^{h}_{ij}=\omega A^{h}_{ij}+e^{i\theta}(i\partial_{r}+\frac{1}{r}\partial_{\theta})A^{e}_{ij}-m_{D}\left||A^{e}_{ij}|^{2}-|A^{h}_{ij}|^{2}\right|^{k}A^{h}_{ij},\\
\end{aligned}
\end{equation}
where $k\in (0,1)$ is the nonlinear factor, and $m_{D}$ is the Dirac mass which is given in the Eq.(72).

As we mentioned in the Sect.3, the insulator-like transport properties in the Kondo dot with the Coulomb interactions,
the single-particle action in the Nonequilibrium dynamical mean-field theory (DMFT) 
which is useful to dealing with the non-equilibrium impurity problem can be described as\cite{Eckstein M3}
\begin{equation} 
\begin{aligned}
S=\int_{\mathcal{C}}dtdt'c^{\dag}(t)\Lambda(t,t')c(t')+\int_{\mathcal{C}}dtV(t),
\end{aligned}
\end{equation}
where $\mathcal{C}$ is the Keldysh contour and $\Lambda(t,t')$ is the self-consistently hybridization function,
$V(t)$ is the local interaction term which is $V(t)=U(t)(n_{\uparrow}-1/2)(n_{\downarrow}-1/2)$ for the common Hubbard model.
Next we simulate the nonequilibrium dynamics of the silicene in the $1/r$ (long-range) Hubbard model at half filling $v_{f}=1/2$ ($\mu=0$).
and detect the double occupation as a function of the disturbed on-site interaction $U$ and with a variational band gap 
(which can be treated as index of the flatness ratio\cite{Yang S} for the topological flat band which with nonzero Chern number when taking the 
third-nearest-neighbor hopping into account just like the treating for the bilayer or multilayer silicene).
Fig.22 shows the relusts of the mean-field variational simulation, the double occupation $d_{hf}=\sum_{i}n_{i\uparrow}n_{i\downarrow}/N$ for the on-site interaction
quench from 0 to U with different bandgaps in nonequilibrium DMFT.
Note that the large diamagnetic moment (compared to the paramagnetic one) makes sure a lower Coulomb repulsive potential (with large ${\bf r}$)
and thus allows the double occupation to exist (see, e.g., Ref.\cite{Thouless D J2}).
The more large bandwidth can be achieved by considering the third-nearest-neighbor hopping in silicene.

For linear-polarized light incident in $z$-direction with a angle $\theta$, the surface plasma can be observed through the nonelastic scattering, 
the time-dependent wave packet amplitude for the plane one and the $z$-direction polarizaed one are $\phi_{1}\propto e^{i(ky-\omega t)}$
and $\phi_{2}\propto e^{i(ky-\omega t)}+v_{g}pz$ with $p=\sqrt{k^{2}-k_{0}^{2}}$ where $k_{0}$ here is smaller than the one for plasma frequency $k_{p}=\omega_{p}/c$,
respectively,
and with the light intensity $I=\kappa_{n}e^{i{\bf k}\cdot{\bf r}}[{\rm cos}\theta\hat{z}+{\rm sin}\theta(\hat{z}\times{\bf k})]$
where $n$ denotes the number of the absorbtion/emition photon.
The $z$-direction polarizaed frequency is affected by the Zeeman splitting of the ground state\cite{Price H M}.
The frequency of linear-polarized light in plane is in the spin basis of $\frac{1}{\sqrt{2}}(\sigma_{x}+\sigma_{y})$ while the $\sigma_{z}$-polarized one is 
in the basis of $\sigma_{z}$,
which can be transfer through the diagonalization procedure by the unitary operator $U=\frac{1}{\sqrt{2}}(\frac{i}{\sigma_{z}}+1)$
as $U^{\dag}(\frac{1}{\sqrt{2}}(\sigma_{x}+\sigma_{y}))U=\sigma_{z}$.

\subsection{Landau gauge}
Firstly we note that both the ac transverse (Hall) and longitudinal conductivity are proportional to the Fermi species $N_{f}$ which in our model
can be view as 4 for the usual case.
For the QHE under the magnetic field ${\bf B}=\nabla\times{\bf A}$, 
since the Hall conductivity $\sigma_{xy}=2e^{2}/h$ (2 denotes the spin degrees of freedom)
which is antisymmetry in the Fermi energy spectrum
is odd under the time reversal, i.e., with the odd filling factor and 
takes the integer as $\sigma_{xy}=2(n+1)e^{2}/h,\ n\in N$.
This Hall conductivity dominate at low-temperture expecially in the zero-temperature-limit and the zero field-limit
which will be discussed in the below.
A nontrivial integer QHE also been found as $\sigma_{xy}=(n+1/2)(4e^{2}/h)=2(2n+1)e^{2}/h,\ n\in N$ (4 denotes the spin
and valley degeneracy) in the 2D electron gases\cite{Zhang Y,Hsu Y F}
and in the zero field-limit($\Delta_{\perp}=0$),
and it's also coincide with the quantized Hall conductivity of the monolayer graphene with the massless Dirac dispersion in the low-energy-limit\cite{Cao Y}.
For silicene in the Hall device, the quantized Hall conductivity becomes $\sigma_{xy}=0$ for the $n=0$ Landau level and $\sigma_{xy}=2(2n+1)e^{2}/h$ for $n\neq 0$
level due to the strong SOC.
While the integer Hall conductivity $\sigma_{xy}=(2n+1)e^{2}/h,\ n\in N$ is emerge in the SVPSM region ($\Delta_{\perp}=\Delta_{SOC}$).
The vanishing of the $\sigma_{xy}=0$ plateau in this region is due to the opened band gap.
Thus the range of the $\sigma_{xy}=0$ plateau 
for our QSH quantized Hall conductivity around the Fermi energy $E_{F}=0$ is very close to the value of $|2m_{D}|$,
while that for the QAHE anomalous Hall conductivity large as 9 meV\cite{Zhang J},
and such a zero conductivity plateau in the charge-neutral point 
supports a trivial insulator.
The plateau $\sigma_{xy}=e^{2}/h$ also emerges as a lowest integral nonzero conductivity in the QSH devices,
by gap out one helical edge states in the top edge or bottom edge of two-terminal silicene
when through a non-spin-polarizaed current,
due to the broken TRI by the in-plane magnatization\cite{Rachel S,An X T}
or the Rashba coupling $R$ and $R_{2}(E_{\perp})$ with strehgth that enable to open up the edge gap.
and keeps the another edge state gapless.
Then it's possible for one edge to realize the spin accumulation
(here we don't consider the limitation
from the spin-flip relaxation generated by the spin polarized
current as well as the magnetization damping since they are negligible in the monolayer silicene)
and the spin-polarization $P_{s}=\sum_{\eta}\frac{\sigma_{\eta\uparrow}-\sigma_{\eta\downarrow}}{\sigma_{\eta\uparrow}+\sigma_{\eta\downarrow}}=\sigma^{s}/\sigma^{v}$
is rised due to the spin-momentum locking effect of TI as a spin filter.
Both the spin-polarization and the induced conductance is robust against the disorder\cite{An X T,Sheng D N} and
The full-spin-polarized current emerge under the strong in-plane-magnetization protected by the TRI and give rises ferromagnetism.
A fully spin-polarized current in a QSH device of silicene with conductivity plateau $e^{2}/h$ emerges when the strength of 
applied in-plane FM or AFM exchange field 
is $\le$0.3t\cite{An X T}.
The $n=0$ Landau level has a two-fold (Kramers or valley) degenerate (it will be four-fold degenerate if don't consider the Zeeman splitting)
while other LL have only the two-fold degenerate.
The Zeeman splitting induce a gap as $\Delta_{z}=\mu-\sigma_{z}\eta\mu_{B}Bg/2$ 
where $g\sim 2$ is the electron Lande $g$-factor, $\eta$ is the valleys index, and
the chemical potential here can be viewed
as the energy of a N-level $\varepsilon_{N}=\varepsilon_{\eta N}$ and it has $\mu=0$ in the charge-neutral point at low-temperature
and $\mu=1$ and -1 in the conduction band and valence band, respectively.
And the intraband scattering factor is
\begin{equation} 
\begin{aligned}
\eta_{s}=\frac{\hbar^{2}v_{F}^{2}|eB|}{\hbar c\varepsilon_{\eta N}},
\end{aligned}
\end{equation}
which is identical to the cyclotron resonance frequency $\omega_{c}=\frac{\sqrt{2}\hbar v_{F}}{\ell_{B}}=\frac{|eB|}{cm^{*}}$
where $\ell_{B}=\sqrt{\hbar c/|eB|}$ is the magnetic length which play the role of quantized cyclotron orbit radius here\cite{Moon K}
and estimated nearly 33.17 nm for $B=1$ T in this work.
The effective mass (or cyclotron mass) 
$1/m^{*}=\sqrt{\frac{2c}{\hbar |eB|}}\hbar v_{F}=v_{F}^{2}/\varepsilon_{N}$ which scale as $V_{F}^{2}$
and the $m^{*}$ it's proportional to the bulk band gap.
For the Landau level spectrum with $\omega_{c}=2m_{D}$, and take the center of the Landau level spectrum as $\mu=0$ for the 
band structure,
then the energies $E=m_{D}$ and $E'=-m_{D}$ corresponds to the $n=0$ Landau level for the K valley and K' valley, respectively,
and in the range of $\mu>E>-\mu$, the upshift Landau levels equals the downshift one\cite{Hsu Y F}.
Through the expression of the DOS in the following section, we can see that the DOS is nonzero even in the place of $\mu=0$ when the $|\varepsilon|>|m_{D}|$.

Under the optical vector potential, the transverse off-diagonal Hall conductivity in the linear response Kubo formula becomes
\begin{equation} 
\begin{aligned}
\sigma_{xy}=\frac{i\hbar e^{2}}{N}\sum_{m\neq n}\frac{f_{m}-f_{n}}{(E_{n}-E_{m}+n\hbar\omega_{l}+i\eta)(E_{m}-E_{n})}\langle m|v_{x}|n\rangle \langle n|v_{y}|m\rangle,
\end{aligned}
\end{equation}
where $f_{m}=1/(e^{\beta(E_{m}-\mu)}+1)$ is the Fermi-Dirac distribusion function, 
which can be replaced by the Heaviside step function $\theta$ in the zero-temperature limit\cite{Tabert C J},
e.g., $f=1$ for the electron-like (occupied) Landau level, and $f=0$ for the hole-like (unoccupied) Landau level\cite{Hsu Y F}
and with the Fermi level inside them.
$E_{m}$ is the energy of $m$-th electron state,
and $\eta_{s}=1/(2\tau_{e/h})$ is a small quantity of the transport scattering factor 
(or a broaden-factor (band width) in the spectrum which will be discussed below) with the mean 
quasiparticle lifetime $\tau_{e/h}$
in the electron-like and hole-like Landau level\cite{Tse W K2}.
The vector potential in Landau gauge has been mentioned above.
For the CDW order which can change the QAH into the trivial band insulator,
can be represented by 
For the low-energy case with both the SDW and CDW, consider the right and left polarization in $x$-direction (while the $k_{y}$ and $k_{z}$ are
the good quantum number), and the Landau gauge ${\bf A}=(A_{x},A_{y},A_{z})=(0,-A_{z}x,A_{z}x)=Ax(0,-{\rm cos}\ \theta,{\rm sin}\ \theta)$,
the effective Hamiltonian is
\begin{equation} 
\begin{aligned}
H=\hbar v_{F}[(\psi^{\dag}_{L}(-i\partial_{x}+{\bf k}+\frac{e}{\hbar c}{\bf A}){\pmb \sigma}(\sigma_{x})\psi_{L})
+(\psi^{\dag}_{R}(-i\partial_{x}-{\bf k}+\frac{e}{\hbar c}{\bf A}){\pmb \sigma}(-\sigma_{x})\psi_{R})],
\end{aligned}
\end{equation}
or in the Landau band form\cite{Yang K Y}
\begin{equation} 
\begin{aligned}
H_{i}=\sum_{{\bf q},{\bf k},{\bf k}',i_{1,2,3,4}}U_{i_{1}i_{2},i_{3}i_{4}}c^{\dag}_{{\bf k},i_{1},\uparrow}c_{{\bf k}+{\bf q},i_{2},\downarrow}
c_{{\bf k}',i_{3},\uparrow}c^{\dag}_{{\bf k}'+{\bf q},i_{4},\downarrow},
\end{aligned}
\end{equation}
where $i$ labels the Landau bands.
The above energy spectrum can be rewritten as 
\begin{equation} 
\begin{aligned}
\varepsilon_{\eta n}=
Ms_{z}\pm\sqrt{2|n|\hbar^{2}v_{F}^{2}\frac{eB}{\hbar c}+(\frac{\overline{\Delta}}{2}E_{z}+\eta\hbar m_{{\rm w}}-\eta s_{z}\lambda_{{\rm SOC}})^{2}},
\end{aligned}
\end{equation}
and the Dirac-like quasiparticle excitation dispersion can be written as $-\mu+\epsilon_{\eta n}$.
And it's found that $\varepsilon_{\eta n}=\frac{2n+1}{2}\omega_{c}$ in the nonrelativistic case\cite{Schakel A M J}.
The Dirac-like effective Lagrangian density in the (3+1)-QED is\cite{Sharapov S G,Gorbar E V,Gusynin V P}
\begin{equation} 
\begin{aligned}
\mathcal{L}_{{\rm eff}}({\bf r},t)
=\sum_{\sigma=\uparrow,\downarrow}
&\left[\gamma^{0}(i\hbar\partial_{t}+\mu)
+v_{F}\gamma^{1}(i\hbar \partial_{y}+\frac{e{\bf A}_{y}}{\hbar c})\right.\\
&\left.+v_{F}\gamma^{2}(i\hbar \partial_{z}-\frac{e{\bf A}_{z}}{\hbar c})
+(\gamma^{3}\hbar v_{F}\partial_{x})-m_{D}-v_{F}\Delta_{z}
\right]
\Psi_{\sigma}({\bf r},t),
\end{aligned}
\end{equation}
where $\tilde{\Psi}_{\sigma}=\Psi_{\sigma} \gamma^{0}$ is the Dirac conjugated spinor 
with $\Psi_{\sigma}=[\psi_{\sigma}^{A},\psi_{\sigma}^{B}]^{T}$
and the $4\times 4$ Gamma matrices: $\gamma^{0}=\sigma_{z}\otimes\sigma_{z}$, $\gamma^{1}=\sigma_{z}\otimes i\sigma_{y}$, 
$\gamma^{2}=\sigma_{z}\otimes i\sigma_{x}$, $\gamma^{3}=i\sigma_{z}\otimes i\sigma_{z}$,
and satisfy the anticommutative Clifford algebra $\{\gamma^{\mu},\gamma^{\nu}\}=2\eta^{\mu\nu}{\bf I}_{4\times 4}=-2\delta_{\mu\nu}$ with the
Minkowski metric $\eta^{\mu v}={\rm diag}(1,-1,-1,-1)$.
If the CSB is not broken by the Higgs-Yukawa coupling and keeps the isotropic Fermi velocityand massless Dirac fermions (not in the (3+1)-QED)
it has the Lorentz-invariant under the gauge transformation: $\psi\rightarrow e^{i\gamma^{5}\phi}\psi$
where Hermitian $\gamma^{5}=i\gamma^{0}\gamma^{1}\gamma^{2}\gamma^{3}=-\gamma^{0}$ is anticommutates with other Gamma matrices,
and the group velocity is invariant in this case and thus the effective mass is zero.
The above effective Lagrangian densitycontains the Dirac-like kinetic energy, Dirac-mass, and Zeeman-term in the low-energy region of silicene,
whose Hamiltonian is $H=\mathcal{L}_{{\rm eff}}+g_{c}\sum_{\sigma\sigma'}\int d{\bf r}_{1}d{\bf r}_{2}
\frac{\tilde{\Psi}_{\sigma}({\bf r}_{1})\gamma^{0}\Psi_{\sigma}({\bf r}_{1})\tilde{\Psi}_{\sigma'}({\bf r}_{2})\gamma^{0}\Psi_{\sigma'}({\bf r}_{2})}
{4\pi|{\bf r}_{1}-{\bf r}_{2}|}$.
In this case, the state $\langle\overline{\Psi}\Psi\rangle$ also be a chiral pairing state.
Under the gauge vector potential ${\bf A}^{\mu}$ where the index $v=0,1,2,3$, there exist the following relation
\begin{equation} 
\begin{aligned}
\frac{{\bf J}}{ev_{F}}=
\sum_{\sigma=\uparrow,\downarrow}\tilde{\Psi}_{\sigma}({\bf r},t)\gamma^{\mu}\Psi_{\sigma}({\bf r},t)\xlongequal{{\bf A}^{\mu}}
\frac{e}{4\pi^{2}\hbar v_{F}c}\int dtd^{3}r\partial_{\mu}(\omega_{l} t)\varepsilon^{\mu\nu\rho\sigma}\partial_{\rho}A_{\sigma},
\end{aligned}
\end{equation}
Uner the Feynman gauge with Lorentz-invariant, the current has $\partial_{\mu}{\bf J}^{\mu}=0$
and thus satisfy the continuity equation $\frac{\partial\rho}{\partial t}+\nabla\cdot {\bf J}$ where $\rho$ is the probability density.
The electromagnetic potential also has $\partial_{\mu}{\bf A}^{\mu}=0$,
and the electromagnetic coupling-related causal retarded propagator (especially in the nonrelativistic limit which with $c\rightarrow\infty$ and $m_{D}\rightarrow 0$)
with lorenz invariant by the time ordered product as 
\begin{equation} 
\begin{aligned}
G({\bf r}'-{\bf r},t'-t)=-i\langle\mathcal{T}{\bf A}^{\mu}({\bf r},t){\bf A}^{\mu}({\bf r}',t')\rangle\\
=\int\frac{d^{3}k}{(2\pi)^{3}}\frac{d\omega}{2\pi}\frac{e^{i[k({\bf r}'-{\bf r})-\omega(t'-t)]}}{\omega^{2}-\varepsilon^{2}+i\eta_{s}}.
\end{aligned}
\end{equation}

\section{Long- and short-range correlation with Coulomb effect and the strong SOC in silicene}

Since the kinetic energy in the lowest Landau level (LLL) in silicene with QHE is the one been quenched to the zero-kinetic-energy state
in the non-Abelian generalized Berry phase (with smooth Berry curvature).  
In the inhomogenerate case mention above, the Fermi velocity in Diac-cone is anisotropic in generate,
and can be written as $v_{F}=(v_{x},\frac{x}{y}v_{y},\frac{x}{z}v_{z})$ in the Landau gauge, thus the Diac-cone is anisotropic too,
and that's a different to the isotropic Dirac cone case within the (2+1)-QED which has been discussed in the, e.g., Refs.\cite{Sharapov S G,Khveshchenko D V}
which with a larger $v_{F}$ and larger $t$ (about twice of ours: $9.7\times 10^{5}$ m/s and 3 eV) and thus have a more obvious relativistic effect
(compare to the Zeeman effect),
and this means that the splitting effect of magnetic field is dominantly that the SOC, 
expecially for the LLL which has zero kinetic energy given by SOC and has only the energy given by the Coulomb interaction
(which may give rise the fractional QHE especially under strong magnetic field like the Halperin non-Abelian singlet state in the SU(2) symmetry-limit
\cite{Kou A,Sterdyniak A})
and the charges are carried by the strong spin textute of silicene $\langle u_{n{\bf k}}|{\pmb \sigma}|u_{n{\bf k}}\rangle$ in the LLL
which can be observed in the quantum Hall device even without any external magnetic fields.
For silicene, the strong SOC make the interaction strength related to the Zeeman coupling even with a nonzero filling factor.
With the intersite Coulomb interaction, since the Hund's rule turns to minimized the system energy, the size of the Skyrmion $A_{s}$ is become larger 
($A_{s}\gg \ell_{B}\gg a$)
and with lower energy $E_{s}=4\pi \rho_{s}$\cite{Moon K} due to the reduced spin stiffnes with the increasing current-current correlation,
and give rise the maximum spin-polarized CDW ground state contributed by the charges on LLL.
For silicene in the half-filling Hubbard model (like the KMH model), this minimized-energy system (with minimized-Coulomb-interaction) 
corresponds to the band structure 
in low-energy region that the upper band
is completely empty while the lower band is fully filled,
since in this case the total spin is becomes maximum as $\hbar N/2$ ideally in the ferromagnetic ground state where $N$ is the number of electrons
which equal to four times of the cell number in the four (doubly degenerate-) band model.
In our inhomogenerate case, the anisotropic interaction quasiparticles breaks both the global spin SU(2) symmetry 
(i.e., chiral symmetry breaking (CSB) with the linear damping of Dirac fermion,
which can be achieved even in the isotropic interaction caused by the Higgs Yukawa couplings
with the massive Dirac fermions by a mass term $\sum_{\sigma}\tilde{\Psi}_{\sigma}m_{\sigma}\Psi_{\sigma}$
and the Coulomb term $\sum_{\sigma}\tilde{\Psi}_{\sigma}\frac{g_{c}\gamma^{0}}{|{\bf r}|}\Psi_{\sigma}$
where $g_{c}=2\pi e^{2}/(\epsilon_{0} v_{F})=4\pi c_{s}hc/v_{F}$ is the long-range Coulomb coupling;
Note that for more generate case, this dimensionless (bare and nonlocal) 
Coulomb coupling can be represented by the ratio between the Coulomb potential and the thermodynamic potential:
$g_{c}=V_{c}({\bf r})/V_{\beta}=V({\bf r})/(-\frac{4}{2\pi\hbar^{2}v_{F}^{2}}\sum_{\mu}[\int\frac{1}{1+e^{\beta(E-\mu)}}d\mu-\mu])=
(e^{2}/(4\pi\epsilon|{\bf r}|))/(-\frac{4}{2\pi\hbar^{2}v_{F}^{2}\beta}\sum_{\mu}{\rm ln}(1+e^{\beta(E-\mu)})$ 
where the factor 4 denotes the spin and valley degrees fo freedom,
while the thermodynamic potential under a magnetic field with the Landau levels has been presented in the Eq.(13) of Ref.\cite{Koshino M})
and the U(1) rotation symmetry (in pseudospin space) by the decoupling.
Since the terms $\partial_{x,y,z}$ in Eq.(94) are relted to the kinetic energies of the tight-binding Hamiltonian,
the (${\bf r}t$)-dependent kinetic energies are important which been detected in Ref.\cite{Wu C H}.
As we mention above, the commensurate filling at the point that photon frequency $\hbar \omega=U$ where the on-site interaction $U\approx$ 5 eV here
and corresponds the a peak in the absorption spectrum.
The quantum transition near this point is also related to the entanglement between the long-wavelength SDW and CDW expecially in the LLL.
The presence of the Coulomb interaction also leads to the dampling of self-energy of the Dirac-quasiparticle 
(due to the electron-phonon coupling, electron-collision, or the acoustic phonon scattering)
in the low-energy limit 
and follows the Kramers-Kronig relation which also associated with the longitudinal conductivity which as a function of the charge carrier density,
$\varepsilon_{Dq}=\sqrt{(\hbar v_{F}{\bf k}+m_{D})}+\Sigma,
\ {\rm Re}\Sigma \sim g_{c}e^{1/g_{c}},\ {\rm Im}\Sigma \sim g^{2}_{c}e^{1/g_{c}}$\cite{González J,Khveshchenko D V},
with first term the linear dispersion term and the second term $\Sigma$ the nonlinear dispersion term due to the screening effect.
In the symmetry unbroken DMFT, 
the self-energy $\Sigma$ of the silicene cluster 
which is vanish for the full gapped band insulator,
can be represented by the lattice Green's function in helicity basis (gapless) which mension above
$G_{k}=(i\partial_{t}+\mu-\epsilon_{k}-\Sigma)^{-1}$ as the Green function of the reference system (not the target one)
with $i\partial_{t}=i\omega\langle u_{n{\bf k}}|\Theta|u_{n{\bf k}}\rangle$ by the Fourier transformation in the momentum space
where the analytical continuation can be used as $i\omega\rightarrow\omega+i0^{+}$ where the small real quantity $0^{+}$
comes from the quasiparticle scattering factor or a small kinetic energy induced by the Fermion loop,
i.e., replace the imaginary frequency in the one-particle Green’s function by the real frequency.
$\Sigma^{ab}_{cd}=[G^{-1}_{w}-G_{k}^{-1}]^{ab}_{cd},\ a,b,c,d=1,2,3,4$ is the orbits index for the four sites. 
and note that the term $(i\partial_{t}+\mu-\Sigma)$ in $G_{k}^{-1}$ is nonzero only for the single-site case, i.e., $a=b=c=d$.
The 4$\times$4 target Weiss field matrix (four unit cell/cluster) which in a inverse Green function form which brings a inverse eigenvalue 
\begin{equation} 
\begin{aligned}
G^{-1}_{w}=\int d\varepsilon\rho_{k}(\varepsilon,\omega)G^{-1}_{k}=\frac{1}{i\partial_{t}+\mu-t^{2}G^{-1}_{k}-\Sigma}
\end{aligned}
\end{equation}
where the nearest neighbor hopping $t$ can be treated as the matrix element of the self-consistent hybridization function 
$\Lambda^{ab}_{cd}=[t^{*}t/G^{-1}_{w}]^{ab}_{cd}$ and the positive-defined electron spectral DOS
as shown in the left panel of Fig.23\cite{Ludwig A W W,Kotliar G,Go A}
\begin{equation} 
\begin{aligned}
\rho_{k}(\varepsilon,\omega)&=\frac{1}{2\pi i}{\rm Tr}[G^{A}(\varepsilon,\omega)-G^{R}(\varepsilon,\omega)]=
\frac{1}{2\pi i}{\rm Tr}[G^{ab}_{cd}(\omega-i\eta_{s})-G^{ab}_{cd}(\omega+i\eta_{s})],\\
\rho_{k}(\varepsilon,\omega)_{T\rightarrow 0}&=\frac{4|\varepsilon|}{2\pi\hbar^{2}v_{F}^{2}}\theta(|\varepsilon|-|\varepsilon_{k}|)
\end{aligned}
\end{equation}
where $G^{ab}_{cd}$ is the 16$\times$16 causal local bath matrix in the four band model just like the ${\bf U}$ in the Eq.(11),
and the factor 4 denote the fourfold degrees of freedom (spin and valley)
and it's important for the redefinition of the fine structure constant\cite{Son D T}.
From the plot we can see that the zero DOS region is determined by the band gap.
In fact, the interaction between the metallic subtrate may give rise the nonlinear damping and breaks the long-range Coulomb interaction with large $g_{c}$
and short-range strong correlated.

The vertex function $\Gamma$ which is important in the variant cluster approximation is associated with the jump of the self-energy
in the spectral weight in momentum space
of the first BZ in the MFT, and it's piecewise and in the direction which normal to the boundary between different pieces with different self-energies,
thus the DOS is also piecewise as
\begin{equation} 
\begin{aligned}
\rho_{k}(\varepsilon,\omega)=
\frac{4|\varepsilon|}{2\pi\hbar^{2}v_{F}^{2}}\frac{1}{2}\sum_{\eta=\pm 1}\left[\theta(|2\varepsilon|-2|m_{D}|_{\eta})\right],
\end{aligned}
\end{equation}
where $\epsilon$ is the energy spectrum in Eq.(74),
and the intraband DOS under the magnetic field is
\begin{equation} 
\begin{aligned}
\rho_{k}(\varepsilon,\omega)=
\frac{4\varepsilon_{\eta n}}
{2\pi\hbar^{2}v_{F}^{2}}\sum_{s_{z}=\pm 1,\eta=\pm 1}\sum^{\infty}_{N=-\infty}\theta(\varepsilon_{\eta n}-\varepsilon_{N}),
\end{aligned}
\end{equation}
In the low temperature-limit,
the inverse effective mass is $1/m^{*}=2\pi N_{f}/(h^{2}\rho_{k})$
where the DOS $\rho_{B}$ under the magnetic field can be obtained through the Shubnikov-de Haas osillation
as $\rho_{B}=4B/\Phi_{0}$\cite{Cao Y,Novoselov K S},
and there is dominated by the electrons elastic scattering 
due to the randomly charged Coulomb impurities with the $g_{c}\ll 1$ and $\omega\approx \hbar v_{F}k\gg 1/\beta$
(or in the weak magnetic field-limit $\frac{2eB}{c\hbar}\hbar v_{F}\gg 1/\beta$),
and the momentum-dependence is weaker but more dependents on the temperature and frequency $\omega$.
In fact, it's independent of any logarithmic renormalization which in higher energies\cite{Aleiner I L} like the ${\rm ln}(\hbar v_{F}k\beta)$.
In the zero-temperature limit with weak coupling where the scattering factor is not more a frequency-dpendent quantity but with $\omega=0$ and 
the impurity vertex function can be ignored.
The downward renormalized Femion velocity can't be observed in this AFM Mott insulator regime which with the BCS-like weak interaction pairing
(with positive chemical potential).
The scattering factor becomes
\begin{equation} 
\begin{aligned}
\eta_{s}=-\left[ \frac{1}{N}\int\frac{d^{2}k}{(2\pi)^{2}}
\ {\rm Im}(U_{{\bf q}}(\omega))^{2}\times {\rm Im}\frac{1}{\chi(\omega,{\bf k})}\right],
\end{aligned}
\end{equation}
where $U_{{\bf q}}(\omega)=\frac{{\bf q}}{\sqrt{{\bf q}^{2}+\hbar^{2}v_{F}^{2}}}$, ${\bf q}=\omega+\epsilon+i\Sigma(\epsilon+\omega)$.
In this case, the orbital susceptibility which is the is\cite{Koshino M}
\begin{equation} 
\begin{aligned}
\chi(m_{D})=\frac{-4e^{2}\hbar^{2}v_{F}^{2}}{6\pi c^{2}}\frac{1}{2|m_{D}|}\Theta(|m_{D}|-|\varepsilon|).
\end{aligned}
\end{equation}
and the diagonal ac conductivity under the elastic scattering is
\begin{equation} 
\begin{aligned}
\sigma_{xx}=&\sigma_{yy}=\frac{\beta e^{2}}{N}\sum_{m}f_{m}(1-f_{m})\frac{\left|\langle m|v_{x}|m\rangle \langle m|v_{y}|m\rangle\right|}{i\omega+\mu+i\eta_{s}},\\
\eta_{s}=&\frac{2\pi\rho^{{\rm imp}}}{\hbar}\sum_{{\bf q}}|U^{{\rm imp}}_{{\bf q}}|^{2}(1-\frac{m_{D}}{\varepsilon_{\eta}}),
\end{aligned}
\end{equation}
where the screened impurity potential $U^{{\rm imp}}_{{\bf q}}=\frac{e^{2}}{2\epsilon_{0}\epsilon_{s}\sqrt{{\bf q}^{2}+{\bf k}_{0}^{2}}}{\bf q}$\cite{Vargiamidis V}
with ${\bf k}_{0}=2\pi e^{2}\rho_{k}/\epsilon_{s}$, and ${\bf q}=2k\sqrt{\frac{\varepsilon_{\eta}-m_{D}}{2\varepsilon_{\eta}}}$
and
thus it exhibits a linear-temperature-dependence under the not-too-strong magnetic field
and such a linear-temperature-dependence behavior suggest the exist of the extended state
while for the localized state it will exhibits a exponential behavior\cite{Thouless D J2,Ma R}.
The non-diagonal Hall conductivity Eq.(85) in the zero-temperature-limit can be reduced to\cite{Schakel A M J,Gusynin V P2,Hsu Y F}
\begin{equation} 
\begin{aligned}
&\sigma_{xy}=\frac{2e^{2}}{h}{\rm sgn}(\mu){\rm sgn}(eB)[\theta(\mu+|m_{D}|)\theta(\mu-|m_{D}|)+\sum_{n=1}^{\infty}
\sum_{\eta=\pm 1}\theta(\mu+\varepsilon_{\eta n})\theta(\mu-\varepsilon_{\eta n})],\\
&\sum_{n=1}^{\infty}
\sum_{\eta=\pm 1}\theta(\mu+\varepsilon_{\eta n})\theta(\mu-\varepsilon_{\eta n})=
{\rm Int}\left[\frac{\hbar c}{2\hbar^{2}v_{F}^{2}|eB|}(\mu+|m_{D}|)^{2}\right]
(\theta(\mu-\varepsilon_{\eta=1,n})+1)+\\
&{\rm Int}\left[\frac{\hbar c}{2\hbar^{2}v_{F}^{2}|eB|}(\mu-|m_{D}|)^{2}\right]
(\theta(\mu+\varepsilon_{\eta=-1,n})+1).
\end{aligned}
\end{equation}
where the sign of $\mu$ is presented in the Eq.(6) of the Ref.\cite{Gusynin V P2}
and ${\rm Int}[\cdot]$ denotes the integer pair of $\cdot$.
The term $\theta(\mu+|m_{D}|)\theta(\mu-|m_{D}|)$ describes the $N=0$ anomaly level while the other term describes the $n\neq 0$ levels.
The dc-conductivity by the above dc-driven charge current (Eq.(83)) is in a semiclassical Drude formular.

For strong Coulomb coupling $g_{c}\gg 1$ with the non-BCS type pairing
which can be obviously observed in by the metallic substrate with the strong screening effect due to the violent change of the electronic band structure
by the massive polarized charge,
and the strongly momentum-dependent self-energy, 
the renormalization of Fermi velocity is effective as $v_{F}^{*}/v_{F}=(\Lambda/{\bf k})^{8/(\pi^{2}N_{f})}<1$
\cite{Jafari S A,Khveshchenko D V2,Khveshchenko D V},
where the scale $\Lambda$ is in the region of the quantum critical point which with the gapless marginal Dirac fermion
and the fermion species $N_{f}$ is can be reduced by increasing the temperature or by using the 
substrate with large dielectric constant and thus with a large screening effect\cite{Son D T}.
And note that the $v_{F}^{*}$ here is in a scale of lattice constant initially due to the Coulomb coupling in the nonrelativistic approximation,
A singular peak of the orbital susceptibility (negative diagmagnetic susceptibility) well appear in the in the Dirac-liquid\cite{Koshino M,Jafari S A} region 
and with the zero DOS
due to the strongly momentum-dependence, and it's increse with the strength of SOC and exchange field,
and the temperature-dependent orbital susceptibility in this case is
\begin{equation} 
\begin{aligned}
\chi(\beta,m_{D})=\frac{-4e^{2}\hbar^{2}v_{F}^{2}}{6\pi c^{2}}\frac{1}{2|m_{D}|}{\rm tanh}(2m_{D}\beta).
\end{aligned}
\end{equation}
As we shown in the right panel of Fig.23, the absolute value of the susceptibility (peak) is reach when the $m_{D}\rightarrow 0$,
and are increase with the decrese of the temperature as a whole,
which is also well agree with the results presented in other literatures\cite{Ezawa M5,Yarmohammadi M,Fukuyama H}.
In fact, both the diamagnetic and paramagnetic response which with opposite magnetic moment (i.e., 
diamagnetic moment and paramagnetic moment with the spin carriers along the edge direction carriers the up- and down- spin, respectively)
are coexist in the silicene due to the interactions between the magnetic field and the charge
carriers with spin-up and spin-down, respectively, and they are both increse with the temperature.
In this large $g_{c}$ region, the $V_{c}({\bf r})$ contains no kinetic term
and the interactions within the Fermi liquid may retain weak in the large-species case\cite{Son D T}.
And the $\sigma_{xx}$ may disapper under the strong disorder from the strong magnetic field
even with the contributions from the particle hopping
(i.e., the $\sigma_{xx}$ will becomes insulating due to the strong disorder like in the high-frequency off-resonance case
which the insulativity is in a order of $\frac{2v_{F}^{2}}{\omega(x)}(e^{4}a^{2}4\pi^{4}\epsilon^{2}_{0}\epsilon^{2}_{s}\hbar^{2}c^{2} \omega_{p}^{2})^{2n}$
and the quantized $\sigma_{xy}$ will exponentially tend to a saturate value\cite{Ma R,Thouless D J2}.
and the frequency-independent scattering factor (scattering rate) which is mentioned above $\eta_{s}=1/(2\tau_{e/h})$
can be represented as the imaginary part of the fermion self-energy\cite{Altherr T} which can be obtained by the Schwinger-Dyson function
\begin{equation} 
\begin{aligned}
\eta_{s}(\omega=0)=-\left[ \frac{1}{N}\int\frac{d^{2}k}{(2\pi)^{2}}{\rm tanh}\frac{\varepsilon\beta}{2}
\ {\rm Im}(U^{i}_{{\bf q}})^{2}\times {\rm Im}\frac{1}{\chi(0,{\bf k})}\right]\ge 0,
\end{aligned}
\end{equation}
where $1/N$ here denotes the single-band density.

\section{Discussion}

Through the analysis, we also find that the SDC state which has the fully spin-polarized and dissipationless edge models and the spin-valley TI which 
has also a fully spin-polarized dissipationless valley current but both has not the spin helical,
are favorable to make the ideal wire that with $100\%$ transmission rate.
For the fully spin-polarized gapless edge states in above phases, 
the large polarized-spin degenerate with the total spin (almost equivalent),
and act like the 1D Luttinger liquid but with zero energy-cutoff (bandwidth).
Evenmore,
the dipole coupling between two high spin-polarized edge states are possible through the optical absorption for the on-resonance light.

As we mention at the begining, for perfect normal retroreflection, the edge state is flat and there haven't chiral edge and the spin or charge edge current.
In fact, even for the Andreev retroreflection (which associate with the conserved group velocity
between the particle-like to hole-like excitations) happen in the topological-protected-interface between the $M_{c}$-dominated region (normal silicene)
and the $M_{s}^{AFM}$-dominated region (s-wave superconductor based) when the chemical potential in the SC region is $\mu_{sc}\gg M_{c}$
(by the doping), the spin edge state is also vanish due to the fully spin-polarized as mentioned above for the device of three-terminate spin-valley TI,
nomatter the electric field is exist or not, and the scattering angle ($\theta_{s}\in[-\pi/2,\pi/2]$) is nearly zero in this case.
The approximated one-component spin current flow in the 1D helical edge is $\sum_{i}{\bf e_{x}}{\rm cos}\theta_{s}
\langle n_{iA\uparrow}-n_{iB\downarrow}\rangle$.

\section{Appendix A : Spin fluctuation and susceptibility in harmonic and anharmonic potential}

Firstly we imagine a system, which with long-enough coherence time, have a large number of degrees of freedom,
and both the thermodynamic limit and the long-time limit can be taken.
Then the unitary dynamics of such system can be investigate effectively.
For the probability distribution after the local or global quenching,
it often show a edge singularity\cite{Silva A}
near the critical point for phase transition.
For eaxmple,
the perturbations of long-range
spin-spin interaction will breaking the integrability of integrable system and force it exhibit a effectively asymptotic thermal behavior\cite{Wu C H} (like the 
transverse field Ising model). The probability distribution will evolve to non-diagonal Gaussian distribution from the previous Possion one.
We already know that the strong interaction emerge after the interaction quenching for nonintegrable Hubbard model
and lead to the stationary prediction by the generalized Gibbs ensemble (GGE)\cite{Wu C H}.
In fact, for such a nonintegrable model, the strong on-site interaction (repulsion) also leads to the incommensurate spin order which appear in
the topological insulating phase or the topological superconducting (SC) phase\cite{Farrell A} and reflected in
a non-diagonal and anisotropic spin Hamiltonian, and the ferromagnetism is also exist in such nonintegrable half-filling Hubbard model.
In fact, it's been confirmed that the spin-orbit coupling (SOC) is pivotal to leads to the topological SC\cite{Farrell A},
and may give rise to many non-trivial effects even for the noninteraction Ferminic system, like the noninteracting Bogoliubov quasiparticles.

In the nonequilibrium protocol for above Hubbard model, the damping effect on amplitude of oscillations is exist for the regime that the on-site 
interaction is small than the critical value which is as a phase transition point, and in fact such a damping is exist in most the dissipative model
which tends to steady state and
with a observable damping spectrum\cite{Diehl S}.
Since in damping system, the amplitude fluctuation 
in a Gaussian probability distribution which can be expressed in the form\cite{Wu C H}
\begin{equation}   
\begin{aligned}
P=w\sum_{{\rm Gaussians}}{\rm exp}(-\frac{[\mathcal{Z}(x)-\mathcal{Z}(x_{G})]^{2}}{2(\delta\mathcal{Z})^{2}}),
\end{aligned}
\end{equation}
which is guided by the differece of free energy $E(\mathcal{Z})-E_{G}(\mathcal{Z},\tau)$,
and here $\mathcal{Z}(x)$ is a coordinate-dependent variable and $\delta\mathcal{Z}$ is the width of the Gaussians,
while the $w$ is the width-dependent model amplitude of the Gaussians.
In out-of-equilibrium protocol, the width $\delta\mathcal{Z}$ describe the fluctuation of the noise from perturbation of spin interaction,
and the difference between $\mathcal{Z}(x_{G})$ and the initial one $\mathcal{Z}_{0}$ is a good estimator of the amplitude of the quenching.
The $x_{G}$ is the final positon of Gaussians, i.e., the position of the center of mass after a long time (relaxation),
in other words,
the final part of the trajectorys which Gaussians centered along.
Such amplitude for damping system will suppressed by the global phase correlation when the frequency is large which corresponds to the weak-interaction,
The strong interaction will lead to the detuning and dephasing as well as the vanishing of nondiagonal contribution.
The detuning cause a anharmonic potential which contain the three-order (asymmetry) term and quartic (symmetry) nonlinear term\cite{Wu C H,Hou J X}
and in fact such three-order or quartic nonlinear term will exist in the interaction potential term of the
quantum systems with anharmonic trap (include the Bose–Einstein condensate, quantum gas, or the lattice models),
which is originate from nonlinear Hamiltonian and the dispersion.

For such anharmonicity case, we nextly introduce the bare action of quantum system with $N$-component continuous quantum field 
$\phi_{\alpha}\ (\alpha=1\cdot\cdot\cdot N;{\rm with\ O(N)-symmetry})$ 
in $\phi^{4}$ field theory with SI units\cite{Wu C H}
\begin{equation} 
\begin{aligned}
S=\int d^{d}r d\tau \frac{1}{2}[(\nabla_{r}\phi_{\alpha})^{2}+\frac{(\partial_{\tau}\phi_{\alpha})^{2}}{c^{2}}-(r_{c}+r)\phi_{\alpha}^{2}+\frac{\lambda x^{4}}{N}\phi_{\alpha}^{4}],
\end{aligned}
\end{equation}
where $c$ is the velocity, $\lambda x^{4}$ is the quartic nonlinear local potential term,
and the critical bare (unphysical) coupling $r_{c}$ is reach in the $r=0$.
The first two terms within the bracket are the fluctuation of $\phi_{\alpha}$ under the control of action $S$ and with determined amplitude,
while the $(r_{c}+r)\phi_{\alpha}^{2}$ is the term describing the bare coupling to the amplitude model and play a key role in the process of phase transition near the 
phase transition point. The $r_{c}$ is\cite{Polkovnikov A}
\begin{equation} 
\begin{aligned}
r_{c}=\frac{2c\lambda x^{4}(N+2)}{N}\int\frac{d^{d}k}{(2\pi)^{d}}\int\frac{d\omega}{2\pi}\frac{1}{p^{2}}
\end{aligned}
\end{equation}
where $p^{2}=k^{2}+(-i\omega/c)^{2}$ is the $(d+1)$-dimensional Euclidean momentum where $k$ is a wave vector.
The strong dampling is appear when the $p^{2}$ is close to zero.
Thus the trapping frequency $\omega$ (include the transverse one and the longitudinal one) and $r$ are decrease with the increasing of damping,
and cause the overdamp when the $r$ close to zero, and reach the phase transition point when $r=0$.
So, the strong-interaction which with low-frequency and low-momentum propagator has the suppressed amplitude and the possible strong dampling if the frequency is low
enough,
while for the weak-interaction, it has the opposite result.
The characteristics of spin-wave stem from the spin-wave model with the transverse susceptibility or longitudinal susceptibility
as presented in the main text.

In fact, since the SC order parameter which couples the electron and hole excitations is associate with the center-of-mass coordinates of the pairs\cite{Hu C R},
the Gaussian profile wave packet can be represented in the 3D model with the electron-phonon-interaction-indeuced nonlinear term.
And this nonlinear term may makes the wave packet collapese into the discrete solitons within the band gap
and with negative effective mass\cite{Trombettoni A}.
%
Through the analysis of Gaussian wave package, 
we can know that the phase $\phi$ which govern the vortex-phase only take effect on the next-nearest-neighbor hopping,
and through this we can theoretically realize all lattice direction by adjusting the phase in a artificial gauge field.
That also provide possible for the magnetic flux modulated single-particle chiral dynamics\cite{Tai M E}
and the Josephson tunneling junction-induced superconducting phase transition,
e.g., utilize the special Josephson junction arrays in the contact surfaces of the edges of silicene with a loop $s$-wave superconductor
and lead to a phase transition from singlet chiral $d_{1}+id_{2}$-pairing wave superconducting phase to the triplet $f$-wave superconducting phase
under a controllable magnetic flux\cite{Zhang L D}, although the annular FS suppress the spin-triplet SC\cite{Sato M}.

\section{Appendix B: Tight-binding Hamiltonian in the low-energy limit}
Considering the bulking distance in the freestanding silicene, the nearest-neighbor hopping vectors may become
\begin{equation} 
\begin{aligned}
r_{1}=&(\frac{\sqrt{3}k_{x}}{3},0,\frac{\sqrt{3}k_{z}}{3}{\rm tan}\theta),\\
r_{2}=&(\frac{-\sqrt{3}k_{x}}{6},\frac{k_{y}}{2},\frac{\sqrt{3}k_{z}}{3}{\rm tan}\theta),\\
r_{2}=&(\frac{-\sqrt{3}k_{x}}{6},-\frac{k_{y}}{2},\frac{\sqrt{3}k_{z}}{3}{\rm tan}\theta),
\end{aligned}
\end{equation}
where $\theta\approx 12^\text{o}55'$ is the angle between the Si-Si band with the $x-y$ plane,
and thus it has $\frac{\sqrt{3}k_{z}}{3}{\rm tan}\theta\approx-\frac{k_{z}}{2\sqrt{14}}$.
while the next-nearest-neighbor hopping vectors ${\bf r}'$ are not affected.
The resulting nearest-neighbor dispersion are\cite{Guzmán-Verri G G}
\begin{equation} 
\begin{aligned}
\epsilon_{1}=\begin{pmatrix}[1.5] \{1,1,1\}\\ \{1,\frac{1}{4},\frac{1}{4}\}\\
\{1,-\frac{1}{2},-\frac{1}{2}\}\\
\{0,\frac{\sqrt{3}}{2},-\frac{\sqrt{3}}{2}\}\\
\{0,\frac{3}{4},\frac{3}{4}\}\\
\{0,-\frac{\sqrt{3}}{4},\frac{\sqrt{3}}{4}\}\\
\{0,0,0\}\\
\end{pmatrix}
e^{ik{\bf r}}=
\begin{pmatrix}[1.5] \{1,1,1\}\\ \{1,\frac{1}{4},\frac{1}{4}\}\\
\{1,-\frac{1}{2},-\frac{1}{2}\}\\
\{0,\frac{\sqrt{3}}{2},-\frac{\sqrt{3}}{2}\}\\
\{0,\frac{3}{4},\frac{3}{4}\}\\
\{0,-\frac{\sqrt{3}}{4},\frac{\sqrt{3}}{4}\}\\
\{0,0,0\}\\
\end{pmatrix}
\begin{pmatrix}[1.5]
e^{ikr_{1}}\\
e^{ikr_{2}}\\
e^{ikr_{3}}
\end{pmatrix}
\end{aligned}
\end{equation}
which make up the $sp^{3}s^{*}$ model of silicene consider the $\sigma$-band
thus the valence (Kohn-Luttinger) band are more lower than the $sp^{3}$ one and the $\sigma$ band and $\pi$ band can't be crossing with each other in this case,
e.g., for the planar silicene the $\sigma$ band and $\pi$ band also can't be crossing with each other due to the orbital symmetry
unless there exist the intrinsic SOC.
The next-nearest-neighbor dispersion are\cite{Guzmán-Verri G G}
\begin{equation} 
\begin{aligned}
\epsilon_{2}=
\begin{pmatrix}[1.5] \{1,1,1,1,1,1\}\\ \{1,1,\frac{3}{4},\frac{3}{4},\frac{3}{4},\frac{3}{4}\}\\
\{1,1,\frac{1}{4},\frac{1}{4},\frac{1}{4},\frac{1}{4}\}\\
\{1,-1,-\frac{1}{2},\frac{1}{2},\frac{1}{2},-\frac{1}{2}\}\\
\{0,0,\frac{3}{4},\frac{3}{4},\frac{3}{4},\frac{3}{4}\}\\
\{0,0,-\frac{\sqrt{3}}{4},-\frac{\sqrt{3}}{4},\frac{\sqrt{3}}{4},\frac{\sqrt{3}}{4}\}\\
\{0,0,\frac{\sqrt{3}}{2},-\frac{\sqrt{3}}{2},\frac{\sqrt{3}}{2},-\frac{\sqrt{3}}{2}\}\\
\{0,0,0,0,0,0\}\\
\end{pmatrix}
e^{ik{\bf r}'}=
\begin{pmatrix}[1.5] \{1,1,1,1,1,1\}\\ \{1,1,\frac{3}{4},\frac{3}{4},\frac{3}{4},\frac{3}{4}\}\\
\{1,1,\frac{1}{4},\frac{1}{4},\frac{1}{4},\frac{1}{4}\}\\
\{1,-1,-\frac{1}{2},\frac{1}{2},\frac{1}{2},-\frac{1}{2}\}\\
\{0,0,\frac{3}{4},\frac{3}{4},\frac{3}{4},\frac{3}{4}\}\\
\{0,0,-\frac{\sqrt{3}}{4},-\frac{\sqrt{3}}{4},\frac{\sqrt{3}}{4},\frac{\sqrt{3}}{4}\}\\
\{0,0,\frac{\sqrt{3}}{2},-\frac{\sqrt{3}}{2},\frac{\sqrt{3}}{2},-\frac{\sqrt{3}}{2}\}\\
\{0,0,0,0,0,0\}\\
\end{pmatrix}
\begin{pmatrix}[1.5]
e^{ikr'_{1}}\\
e^{ikr'_{2}}\\
e^{ikr'_{3}}\\
e^{ikr'_{4}}\\
e^{ikr'_{5}}\\
e^{ikr'_{6}}
\end{pmatrix}
\end{aligned}
\end{equation}
as a $sp^{3}$ model which contains the effect of $\pi-\sigma$ rehybridization\cite{Guo Z X}.
It's obviously that the Eq.\{24\} consider the $\pi$-band (mainly contributed by the $p$-orbit) which contains both the nearest-neighbor hopping 
and the next-nearest-neighbor hopping, and it can be represented by\cite{Guzmán-Verri G G}
\begin{equation} 
\begin{aligned}
H_{\pi}=
\begin{pmatrix}[1.5]
E_{p}+V_{pp\pi}^{\{2\}}e^{ik{\bf r}'}&V_{pp\pi}^{\{1\}}e^{ik{\bf r}}\\
V_{pp\pi}^{\{1\}}(e^{ik{\bf r}})^{*}&\epsilon_{p}+V_{pp\pi}^{\{2\}}e^{ik{\bf r}'}
\end{pmatrix}
\end{aligned}
\end{equation}
where $V_{pp\pi}^{\{1\}}$ and $V_{pp\pi}^{\{2\}}$ are the first-order and second-order parameters of $\pi$ band made by the $p$ bands,
and $E_{p}$ is the $p$ band's energy,
and 
\begin{equation} 
\begin{aligned}
H_{\pi}=
\begin{pmatrix}[1.5]
0&0\\
0&0
\end{pmatrix}
\end{aligned}
\end{equation}
only in the point of $(k_{x}=0,k_{y}=0)$, i.e., the gapless Dirac-point (with the heavy particle/hole subband),
which have the zero effective mass $m^{*}=0$ for the charge carriers.
The total Hamiltonian is (we omitt the $e^{ik{\bf r}}$ and $e^{ik{\bf r}'}$ for simplicity in the following)
\begin{equation} 
\begin{aligned}
H_{\sigma/\pi}&=
\begin{pmatrix}[1.5]
H_{\pi}&N_{2\times 6}\\
N_{6\times 2}^{\dag}&H_{\sigma}
\end{pmatrix},\\
H_{\sigma}=&
\begin{pmatrix}[1.5]
L&T\\
T^{\dag}&L
\end{pmatrix},\\
T=&
\begin{pmatrix}[1.5]
V_{pp\sigma}^{\{1\}}\{1,\frac{1}{4},\frac{1}{4}\}+V_{pp\pi}^{\{1\}}\{0,\frac{3}{4},\frac{3}{4}\}&(V_{pp\sigma}^{\{1\}}-V_{pp\pi}^{\{1\}})\{0,-\frac{\sqrt{3}}{4},\frac{\sqrt{3}}{4}\}&-V_{sp\sigma}^{\{1\}}\{1,-\frac{1}{2},-\frac{1}{2}\}\\
(V_{pp\sigma}^{\{1\}}-V_{pp\pi}^{\{1\}})\{0,-\frac{\sqrt{3}}{4},\frac{\sqrt{3}}{4}\}&V_{pp\sigma}^{\{1\}}\{0,\frac{3}{4},\frac{3}{4}\}+V_{pp\pi}^{\{1\}}\{0,\frac{3}{4},\frac{3}{4}\}&-V_{sp\sigma}^{\{1\}}\{0,\frac{\sqrt{3}}{2},-\frac{\sqrt{3}}{2}\}\\
V_{sp\sigma}^{\{1\}}\{1,-\frac{1}{2},-\frac{1}{2}\}&V_{sp\sigma}^{\{1\}}\{0,\frac{\sqrt{3}}{2},-\frac{\sqrt{3}}{2}\}&V_{pp\pi}^{\{1\}}\{1,1,1\}
\end{pmatrix},\\
L=&
\begin{pmatrix}[1.5]
L_{1}&L_{3}&0\\
L_{3}^{\dag}&L_{2}&0\\
0&0&E_{p}+V_{pp\pi}^{\{2\}}\{1,1,1,1,1,1\}+\Delta_{sp}
\end{pmatrix},\\
L_{1}=&E_{p}+V_{pp\sigma}^{\{2\}}\{0,0,\frac{3}{4},\frac{3}{4},\frac{3}{4},\frac{3}{4}\}+V_{pp\pi}^{\{2\}}\{1,1,\frac{1}{4},\frac{1}{4},\frac{1}{4},\frac{1}{4}\},\\
L_{2}=&E_{p}+V_{pp\sigma}^{\{2\}}\{1,1,\frac{1}{4},\frac{1}{4},\frac{1}{4},\frac{1}{4}\}+V_{pp\pi}^{\{2\}}\{0,0,\frac{3}{4},\frac{3}{4},\frac{3}{4},\frac{3}{4}\},\\
L_{3}=&(V_{pp\sigma}^{\{2\}}-V_{pp\pi}^{\{2\}})\{0,0,-\frac{\sqrt{3}}{4},-\frac{\sqrt{3}}{4},\frac{\sqrt{3}}{4},\frac{\sqrt{3}}{4}\}，
\end{aligned}
\end{equation}
where $\Delta_{sp}$ is the energy difference between the $3s$ and $3p$ orbits,
which is corresponds to the Kana-Mele term as $\frac{\hbar}{m_{0}}\langle s|p\rangle$.
The next-nearest-neighbor $V_{ss\sigma}^{(2)}=0$ and $V_{sp\sigma}^{(2)}=0$\cite{Grosso G},
for the specific parameters, see the Refs.\cite{Min H,Liu C C,Guzmán-Verri G G,Grosso G} and the references therein.
We can also know that the $H_{\sigma}\neq 0$ even in the Dirac-point unlike the $H_{\pi}$,
and don't relay on the effective mass of charge carriers but the rest mass $m_{0}$.
By defining $v_{F}{\rm sin}\theta=\frac{\sqrt{3}}{2}k_{x}a=\frac{\sqrt{3}}{2}k_{y}a$,
the above frame are also practicable for the graphene and polyaeetylene (when the dimerization energy gap and the $\pi$-band gap satisfys
$\Delta_{{\rm dimer}}=-\Delta_{\pi}$(see Ref.\cite{Su W P})).

In the basis of the perturbative ${\bf k}\cdot {\bf p}$ theory which is widely used for the semiconductor system,
with the twofold degenerate dispersion in the $\Gamma-$point which is comtributed by the $p$ orbits,
and with the bare wave function\cite{Liu C X}
\begin{equation} 
\begin{aligned}
H=\frac{\hbar}{m_{0}}{\bf k}\cdot{\bf p}=\frac{\hbar}{m_{0}}{\bf k}\cdot\langle p_{+}|-i\hbar \partial_{{\bf r}}|p'_{-}\rangle,
\end{aligned}
\end{equation}
with the center momentum formed by two electron states $p_{+}$ and $p_{-}$ with distinct angular momentums, and suffer a perturbation ${\bf k}$.
In the case of TRI, the momentum operator has ${\bf p}={\bf p}^{*}$.
For two sublattices in a unit cell, the momentum matrix element ${\bf p}_{ij}=\langle \psi_{A}({\bf k})|{\bf p}|\psi_{B}({\bf k})\rangle$ 
which is not zero since the inversion symmetry is broken,
and is related to the Wannier function as $\psi_{A}({\bf k})=\sum_{A}w({\bf r}-{\bf r}_{A})e^{i{\bf k}\cdot{\bf r}_{A}}$,
$\psi_{B}({\bf k})=\sum_{A}w({\bf r}-{\bf r}_{B})e^{i{\bf k}\cdot{\bf r}_{B}}$.

We represent the total Hamiltonian which under a perturbation (which origin from, i.e., a inhomogenerate electric field or electromagnrtic wave) as
\begin{equation} 
\begin{aligned}
H=H_{0}({\bf p})+\delta H(\partial_{{\bf r}})
\end{aligned}
\end{equation}
the perturbation tiled the spin order by a angle $\theta={\bf k}\cdot{\bf r}$ basis on a initial phase factor $\phi$ which defined above.
For the Zeeman field-induced perturbation, we can perfrom the the canonical transformation to the total Hamiltonian as
\begin{equation} 
\begin{aligned}
H\rightarrow e^{H_{M}}He^{-H_{M}},\\
H_{d}=\begin{pmatrix}[1.5] 0&-M_{z}\\
M_{z}^{\dag}&0\\
\end{pmatrix},
\end{aligned}
\end{equation}
and the SOC term $N_{2\times 6}$ has the below relation with the Zeeman effect\cite{Yao Y}
\begin{equation} 
\begin{aligned}
N_{2\times 6}=M_{z}H_{\sigma}-H_{\pi}M_{z},
\end{aligned}
\end{equation}
Then the above matrix element $T$ under the perturbation-induced rotation is 
$T({\bf k},\partial_{{\bf r}})=\sum_{{\bf k}}\mathcal{R}_{z}^{\dag}T({\bf k})\mathcal{R}_{z}e^{i{\bf k}\cdot {\bf r}}$ with the rotation arounds the
$z$-axis as\cite{Farrell A,Yao Y}
\begin{equation} 
\begin{aligned}
R_{z}&=e^{i\theta}=\begin{pmatrix}[1.5]
{\rm cos}\phi&-{\rm sin}\phi&0\\
{\rm sin}\phi&{\rm cos}\phi&0\\
0&0&1
\end{pmatrix},
&\theta^{{\rm odd}}=\begin{pmatrix}[1.5]
0&i&0\\
-i&0&0\\
0&0&0
\end{pmatrix},
&\theta^{{\rm even}}=\begin{pmatrix}[1.5]
1&0&0\\
0&1&0\\
0&0&0
\end{pmatrix},\\
e^{i{\bf k}\cdot {\bf r}}{R}_{z}=&\begin{pmatrix}[1.5]
-{\rm cos}\phi\ {\rm sin}({\bf k}\cdot{\bf r})&-{\rm sin}\phi&{\rm cos}\phi\ {\rm cos}({\bf k}\cdot{\bf r})\\
-{\rm sin}\phi\ {\rm sin}({\bf k}\cdot{\bf r})&{\rm cos}\phi&{\rm sin}\phi\ {\rm cos}({\bf k}\cdot{\bf r})\\
-{\rm cos}({\bf k}\cdot{\bf r})&0&-{\rm sin}({\bf k}\cdot{\bf r})
\end{pmatrix}.\\
\end{aligned}
\end{equation}

\section{Appendix C: Vanishing of the sign problem in the tight-binding Hubbard model}
To prevent the sign problem in the QMC simulation and keeps the fermion determinant to be positive defined,
Ref.\cite{Wu C3} put forward a class of models which the four-fermion interaction (four independent fermion bilinears) 
term can be decomposed with the TRI and particle-hole symmetry
through the Hubbard-Stratonovich (HS) decoupling with the SU(2) symmetry and replaced the normal local interaction Hamiltonian term $H_{{\rm loc}}=U(n_{\uparrow}-1/2)(n_{\downarrow}-1/2)$
into the anisotropic interaction term $U(n_{1}-n_{2}+n_{3}-n_{4})$ with the redefined four spin components $1,2,3,4$ (four-flavor fermions)
obtained by rotating the spin direction around the $x$ and $y$ axis with a angle $\pi$.
Although this breaks the spin rotation symmetry (and thus breaks the spin dihedral symmetry of Haldane model,
the symmetry eigenvalue of the different spin states keeps.
The method in Ref.\cite{Wu C3} is by using a diagonal $n\times n$ matrix that has $\Theta H_{d}\Theta^{-1}=H_{d}$
which is the same as the relation happen in the Kramers degeneracy points as mentioned above but the $H_{d}$ don't have to be Hermitian,
e.g., for the time-evolution laser-disturbance energy non-conserved topological system which is non-linear and non-Hermitian\cite{Harari G},
and has\cite{Wu C3}
\begin{equation}   
\begin{aligned}
H_{d}|\psi\rangle=&\Theta H_{d}\Theta^{-1}|\psi\rangle=\lambda|\psi\rangle,\\
H_{d}|\psi\Theta\rangle=&\Theta H_{d}\Theta^{-1}\Theta|\psi\rangle=\lambda^{*}\Theta|\psi\rangle,\\
{\rm det}H_{d}=&\prod|\lambda|^{2}\ge 0,
\end{aligned}
\end{equation}
where the eigenvalue $\lambda$ is complex.


\end{large}
\renewcommand\refname{References}


\clearpage

\section{Tables}

Table 1:Three kinds of distorted structure of silicene which are transferred into the insulating phase with unambiguous band gaps
compare to the pristine one.
The corresponding bond angles between e-bond and f-bond and the (average) bulkling distances $\overline{\Delta}$ are also shown.
The corresponding schematic of this displacing (distorting) was shown in the Fig.4(e).
\begin{table}[!hbp]
\centering
\begin{tabular}{cccccccccc}
\hline
\hline  
structure                   &  a (\AA)   &b (\AA)&c (\AA)&d (\AA)&e (\AA)&f (\AA)&$\overline{\Delta}$ (\AA) &Bond angle&Band gap (eV)\\
\hline    
pristine &2.280 &2.280 &2.280 &2.280 &2.280  &2.280 &0.51&115.96$^\text{o}$& $5.45\times 10^{-7}$\\
distorted 1    &  2.245    &2.245&2.263&2.263&2.246&2.247&0.47 &116.821$^\text{o}$&1.609\\
distorted 2    &  2.245    &2.245&2.263&2.263&2.245&2.246 &0.45&117.119$^\text{o}$&1.610\\
distorted 3    &  2.245    &2.245&2.264&2.263&2.245&2.245 &0.42&117.247$^\text{o}$&1.612\\
\hline
\hline     
\end{tabular}
\end{table}
\clearpage
\section{Figures}
Fig.1
\begin{figure}[!ht]
   \centering
   \begin{center}
     \includegraphics*[width=0.5\linewidth]{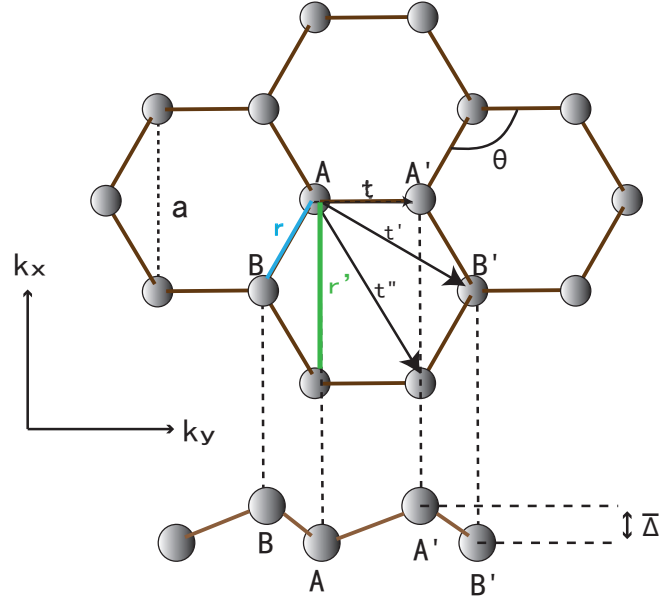}
\caption{(Color online) Top view and side view of the silicene.
with four sites (sublattices) $A,B,A',B'$ in unit cell. 
 The bond-angle $\theta$ and the buckling distance $\Delta$ were marked.
The three dashed lines with $t,t',t''$ denotes the
nearest-, second nearest-, and third nearest-neighbor hopping, respectively.
The blue and green solid lines denotes the hopping in $r$ direction and $r'$ direction respectively,
where $r'$ contains the three hopping directions which goven by the phase $\phi$ and $r$ contains the three ones which not goven by the phase $\phi$.}
   \end{center}
\end{figure}

Fig.2
\begin{figure}[!ht]
\subfigure{
\begin{minipage}[t]{0.15\textwidth}
\includegraphics[width=1\linewidth]{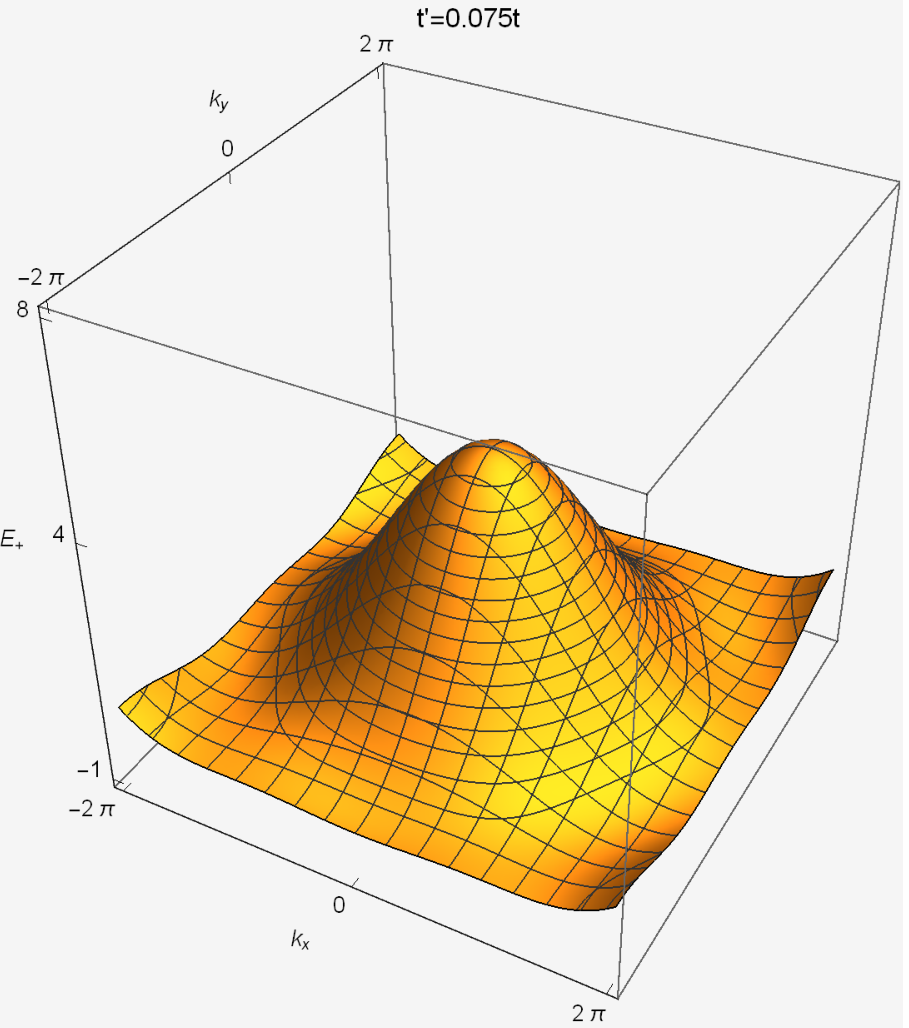}
\label{fig:side:a}
\end{minipage}
}
\subfigure{
\begin{minipage}[t]{0.15\textwidth}
\includegraphics[width=1\linewidth]{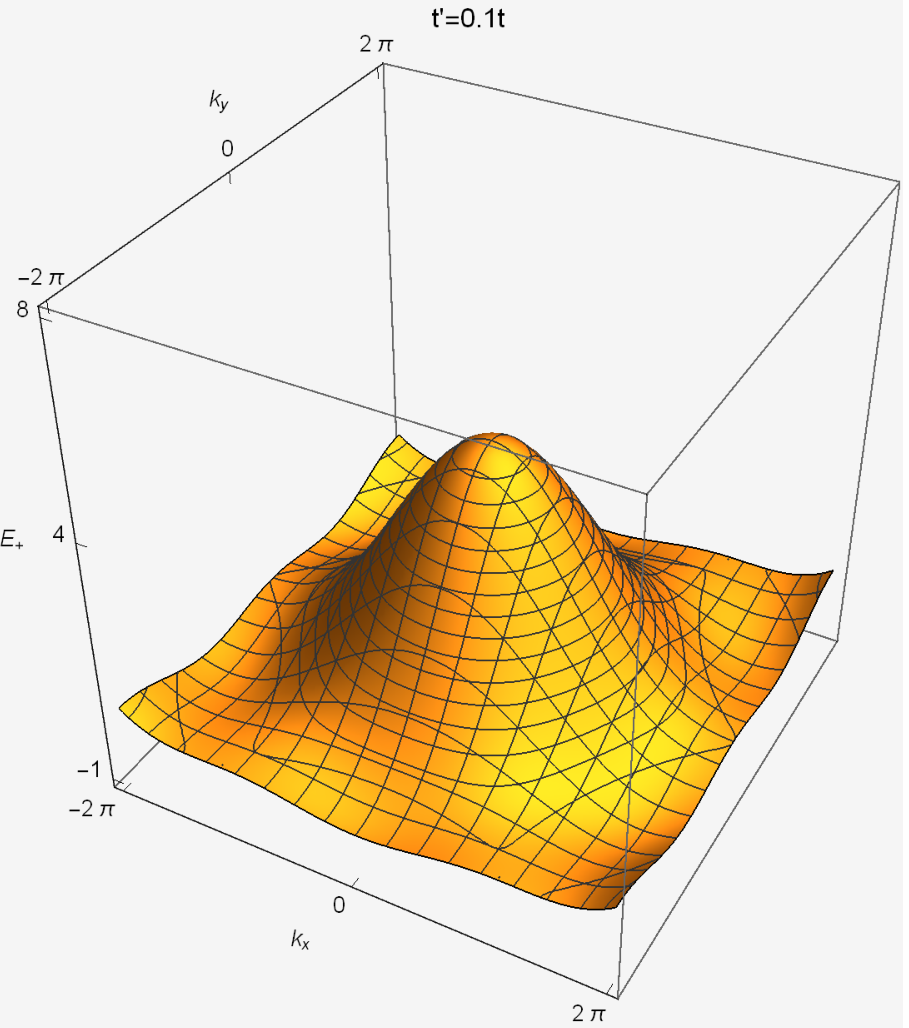}
\label{fig:side:b}
\end{minipage}
}
\subfigure{
\begin{minipage}[t]{0.15\textwidth}
\includegraphics[width=1\linewidth]{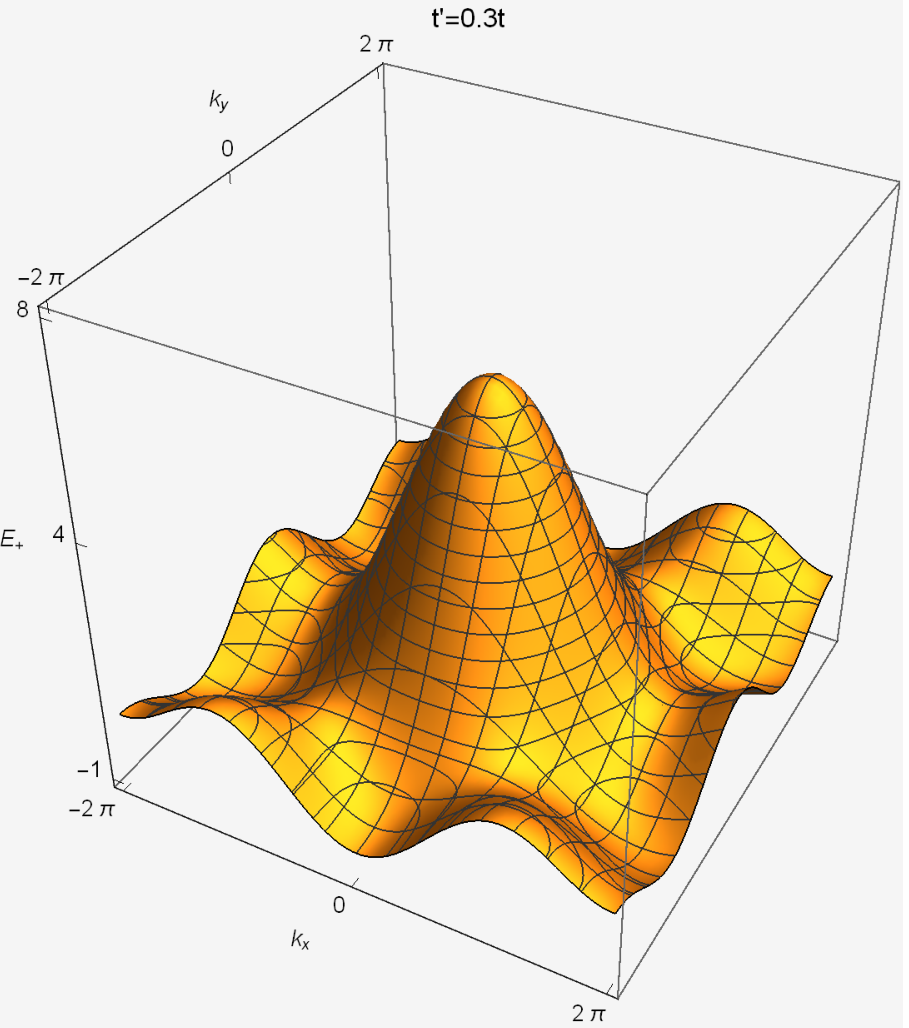}
\label{fig:side:a}
\end{minipage}
}
\subfigure{
\begin{minipage}[t]{0.15\textwidth}
\includegraphics[width=1\linewidth]{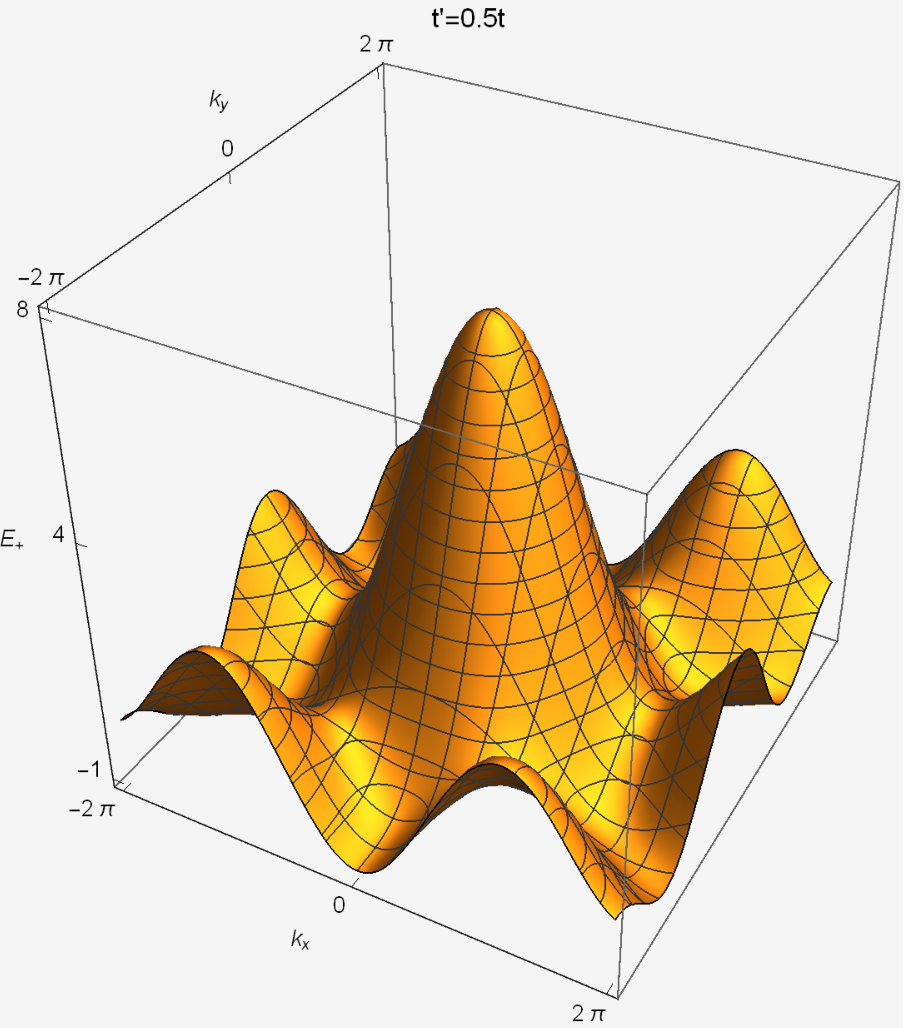}
\label{fig:side:b}
\end{minipage}
}\\
\subfigure{
\begin{minipage}[t]{0.15\textwidth}
\includegraphics[width=1\linewidth]{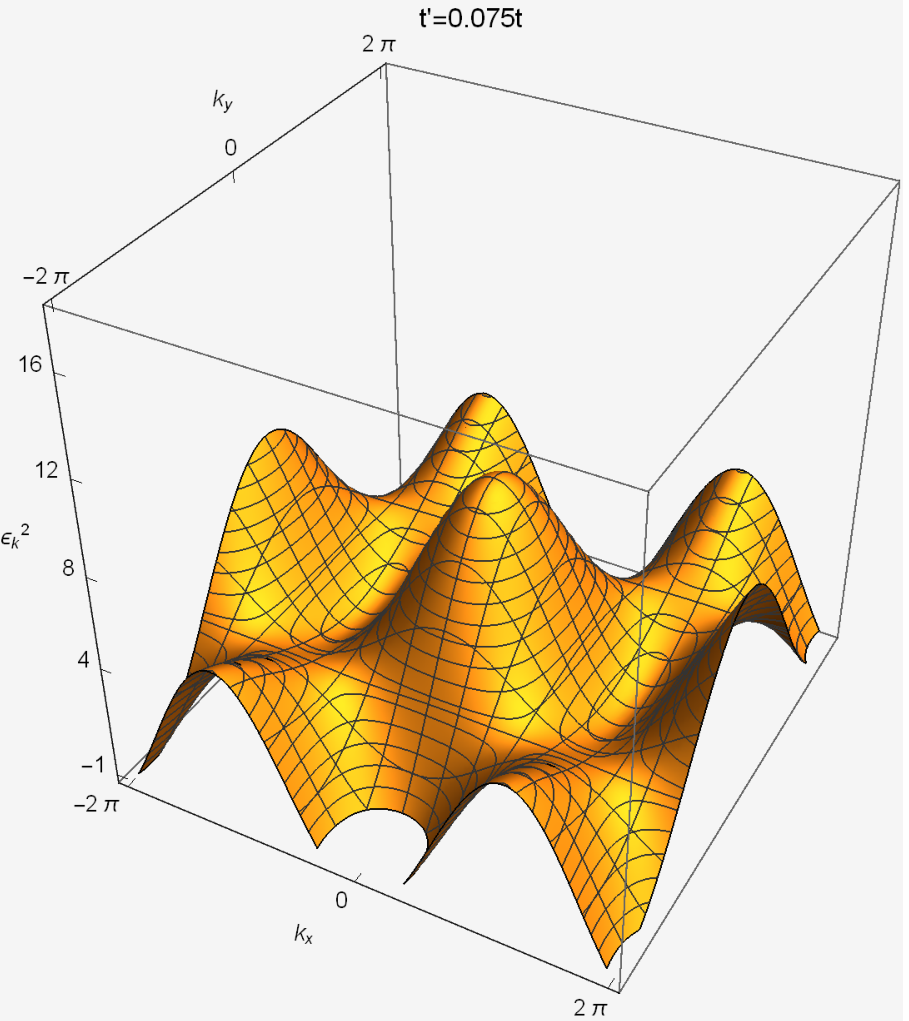}
\label{fig:side:a}
\end{minipage}
}
\subfigure{
\begin{minipage}[t]{0.15\textwidth}
\includegraphics[width=1\linewidth]{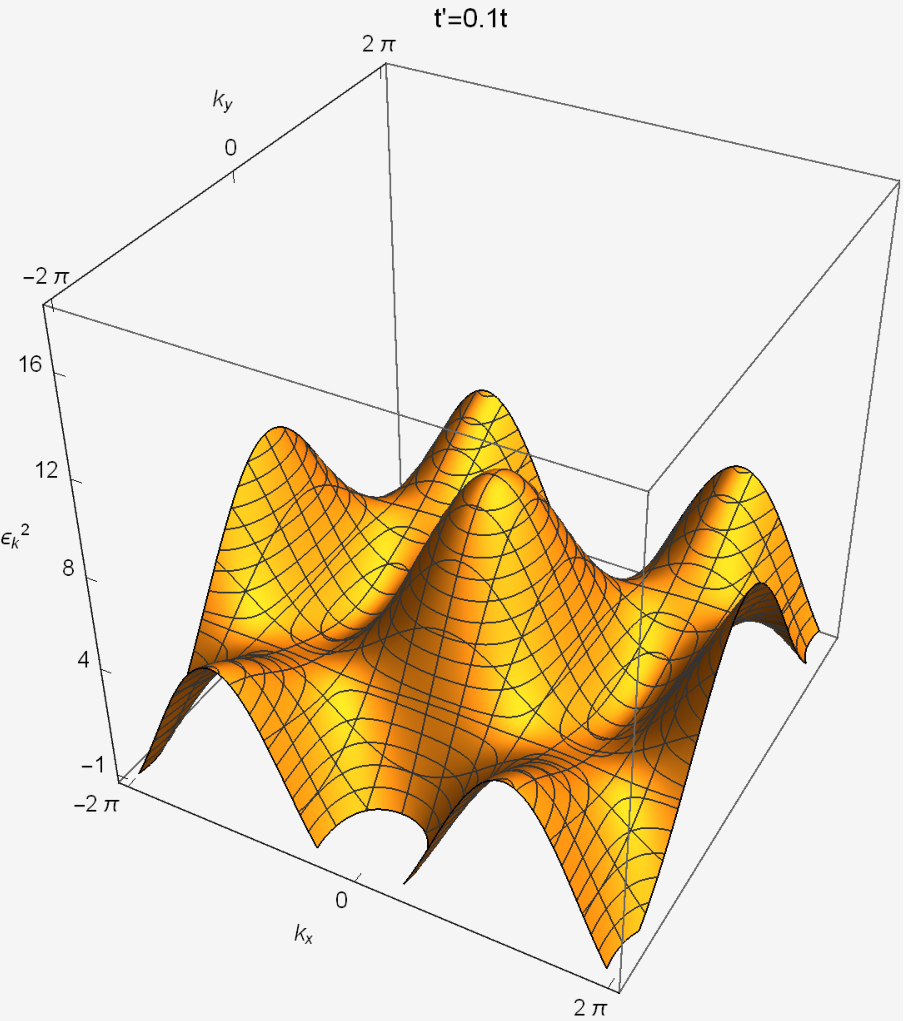}
\label{fig:side:b}
\end{minipage}
}
\subfigure{
\begin{minipage}[t]{0.15\textwidth}
\includegraphics[width=1\linewidth]{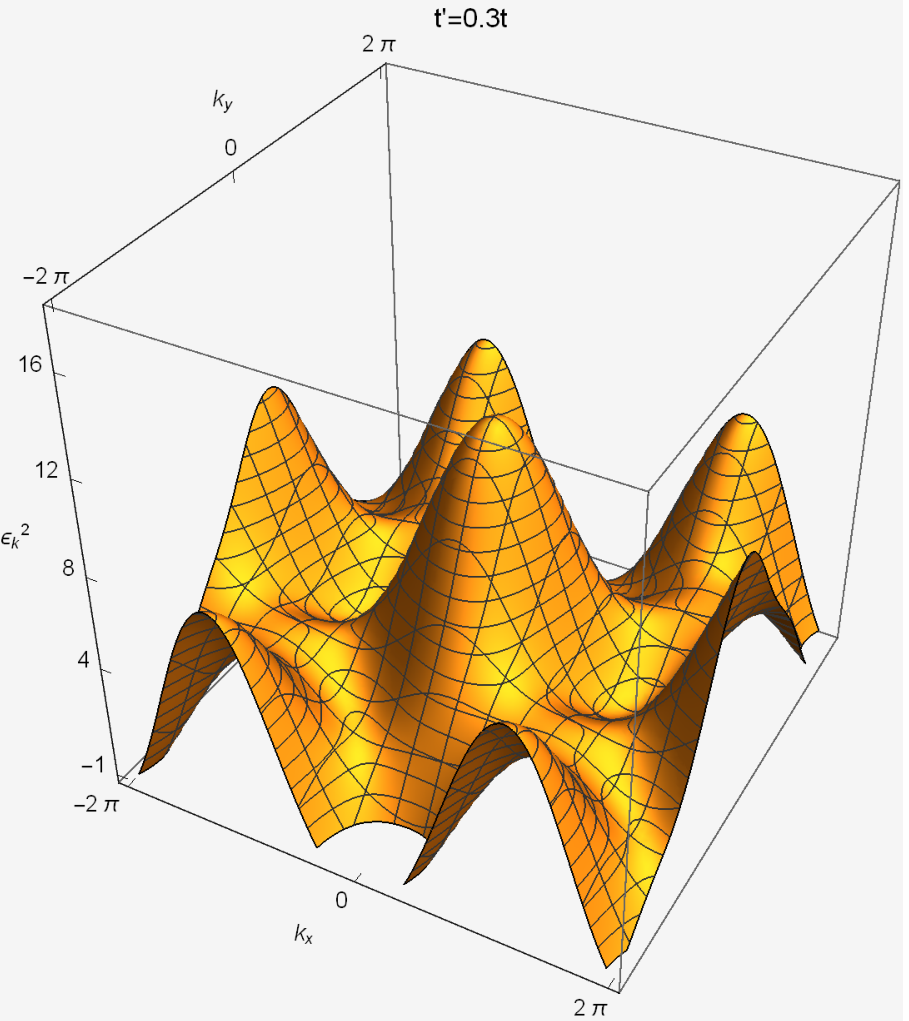}
\label{fig:side:a}
\end{minipage}
}
\subfigure{
\begin{minipage}[t]{0.15\textwidth}
\includegraphics[width=1\linewidth]{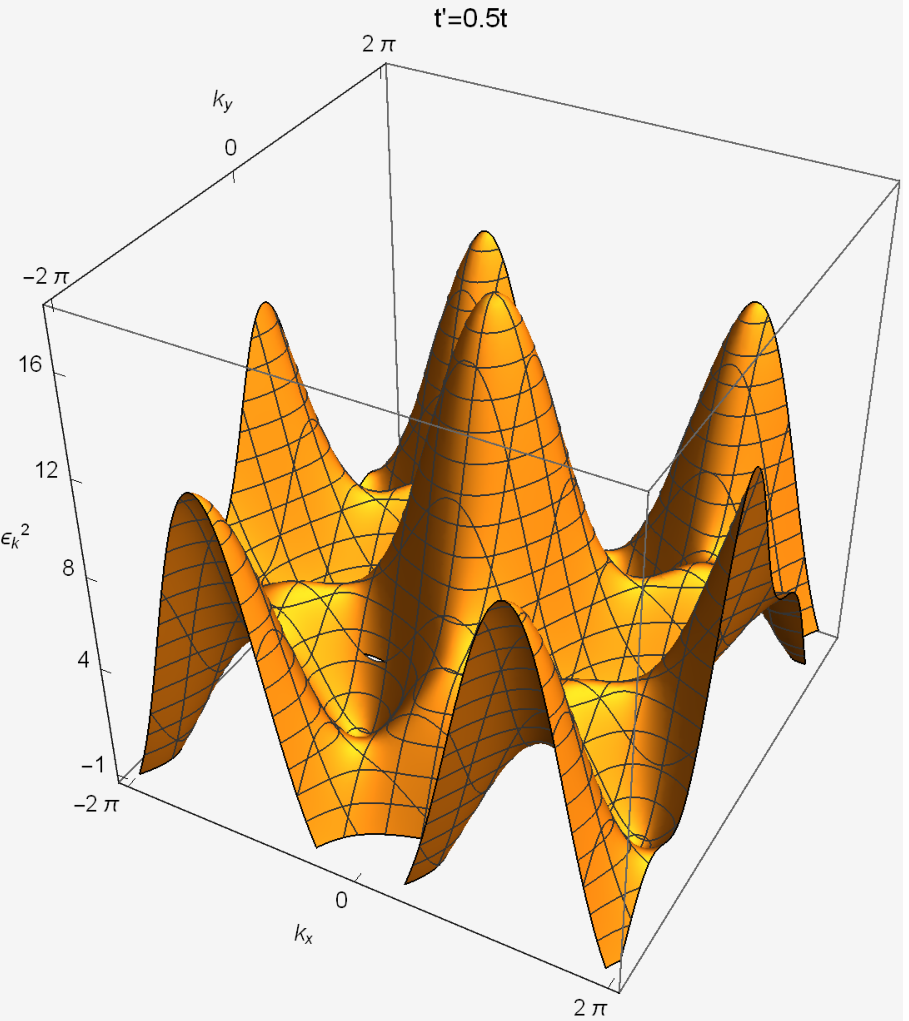}
\label{fig:side:b}
\end{minipage}
}
\caption{(Color online) 3D Schematic diagram of the band structures of monolayerd silicene in momentum space.
(upper panel) upper bands energy and (lower panel) energy in a single-particle picture with different $t'$.
The on-site energy is setted as 1 here.}
\end{figure}

\clearpage

\begin{figure}[!ht]
   \centering
   \begin{center}
     \includegraphics*[width=0.8\linewidth]{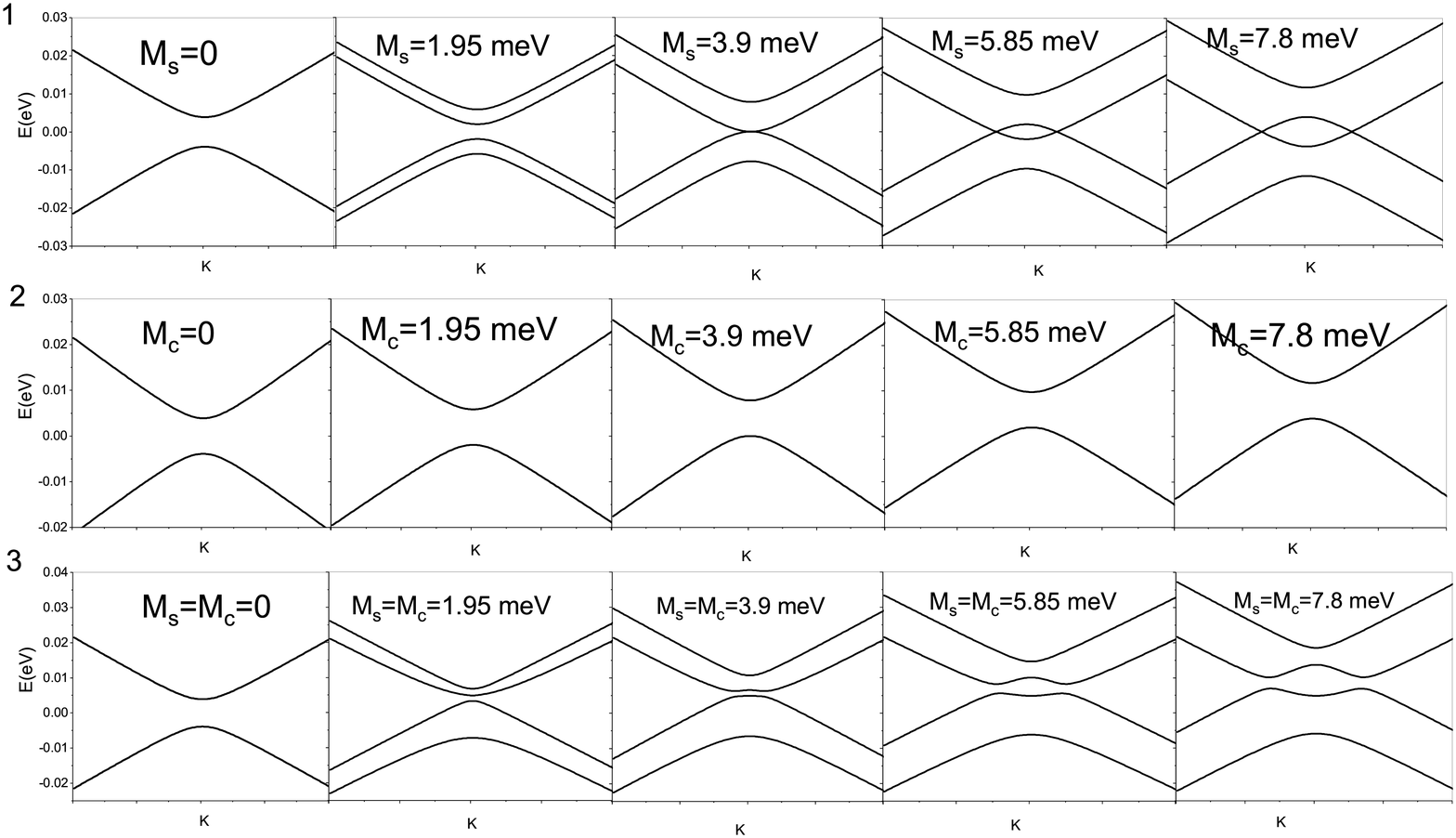}
\caption{Band gap evolution in valley K under the effect of $M_{s}$ (group 1), $M_{c}$ (group
2) and both of them (group 3) with different strength which are labeled in the figures.
Both the electric field and the on-site intraction U is setted as zero.}

   \end{center}
\end{figure}

\clearpage

Fig.4
\begin{figure}[!ht]
   \centering
   \begin{center}
     \includegraphics*[width=0.8\linewidth]{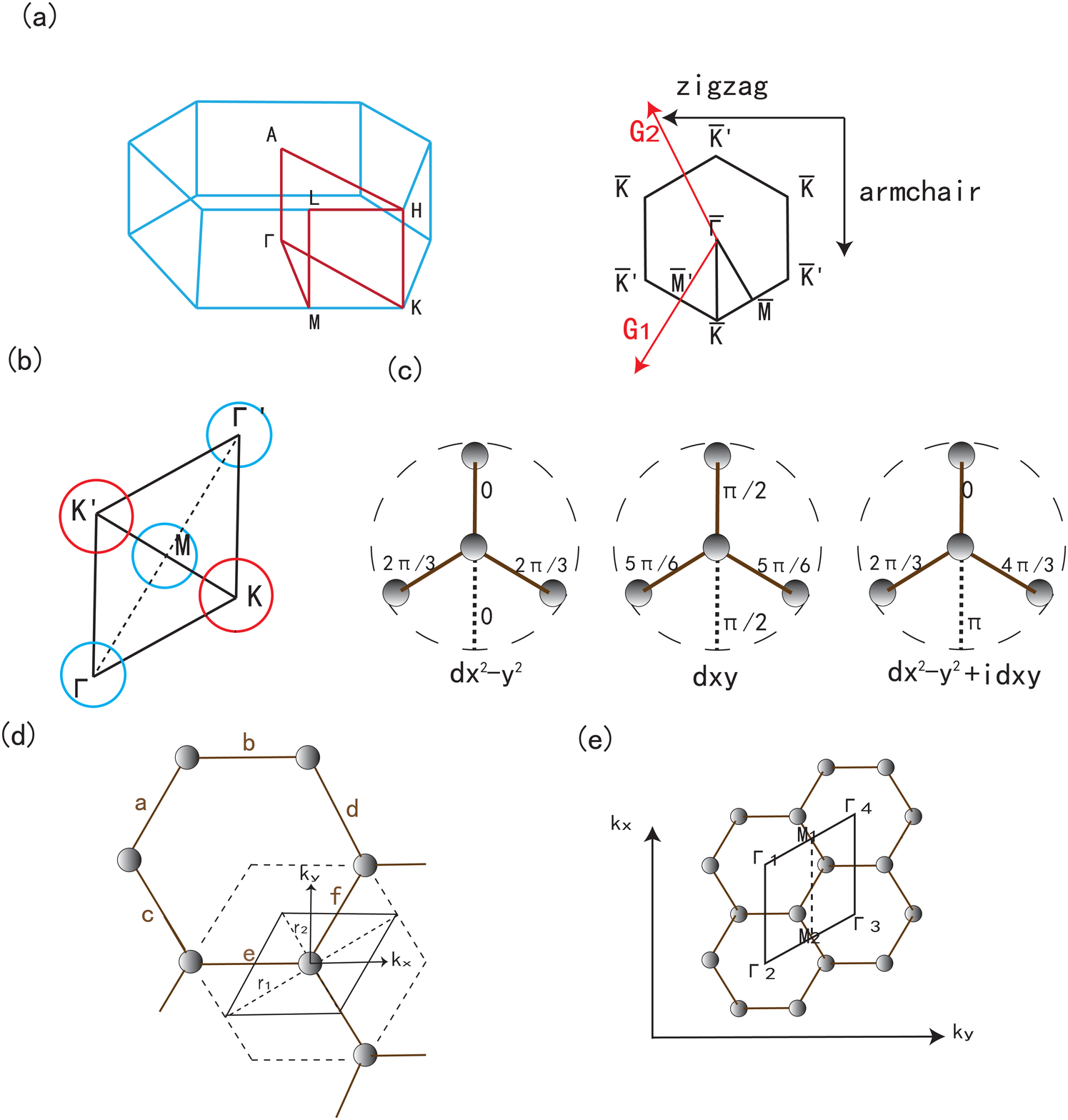}
\caption{(Color online)
(a)Oblique view and top view (2D BZ of the projected (111) surface) of the first Brillouin zone (the $k$-space) with the high symmetry points.
The Red vector in the right panel is the reciprocal lattice vector ${\bf G}_{1}=(\frac{-2\sqrt{3}\pi}{3a},-\frac{2\pi}{a}),
{\bf G}_{2}=(\frac{-2\sqrt{3}\pi}{3a},\frac{2\pi}{a})$.
(b)The correct zone with the electron pockets (red) and the hole pockets (blue) annular FS.
(c)Phase of the $d_{x^{2}-y^{2}},\ d_{xy},\ d_{x^{2}-y^{2}}+id_{xy}$ pairing symmetries (left to right) of silicene in real space.
(d)Schematic diagram of the displacement of the center Si atom. The labels corresponds the Table.A.
(e)The two TRI nodes is presented by the $M_{1}$ and $M_{2}$, which corresponds to the two TRI spin fluxs $\ell=0$ and $h/2e$
(i.e., the magnetic flux $\Phi_{0}$ when without the artific gauge field), respectively,
and the two TRI momentums. The directions of $s_{z}$-independent periodic boundary condition ($k_{y}$) and $s_{z}$-dependent open boundary condition ($k_{y}$)
are marked.}
   \end{center}
\end{figure}
\clearpage

Fig.5
\begin{figure}[!ht]
   \centering
   \begin{center}
     \includegraphics*[width=0.8\linewidth]{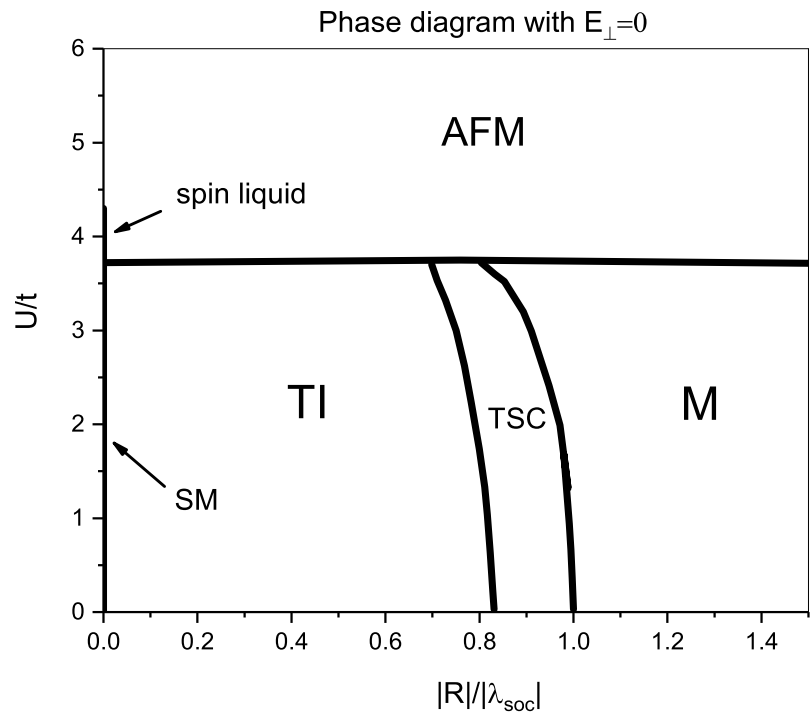}
\caption{Phase diagram of the KMH model without any external field.
The regions of the AFM Mott insulator, TI, metal (M), topological semiconductivity (TSC), semimetal (SM) and spin liquid are contained.}
   \end{center}
\end{figure}

\clearpage

\begin{figure}[!ht]
\subfigure{
\begin{minipage}[t]{0.5\textwidth}
\centering
\includegraphics[width=0.9\linewidth]{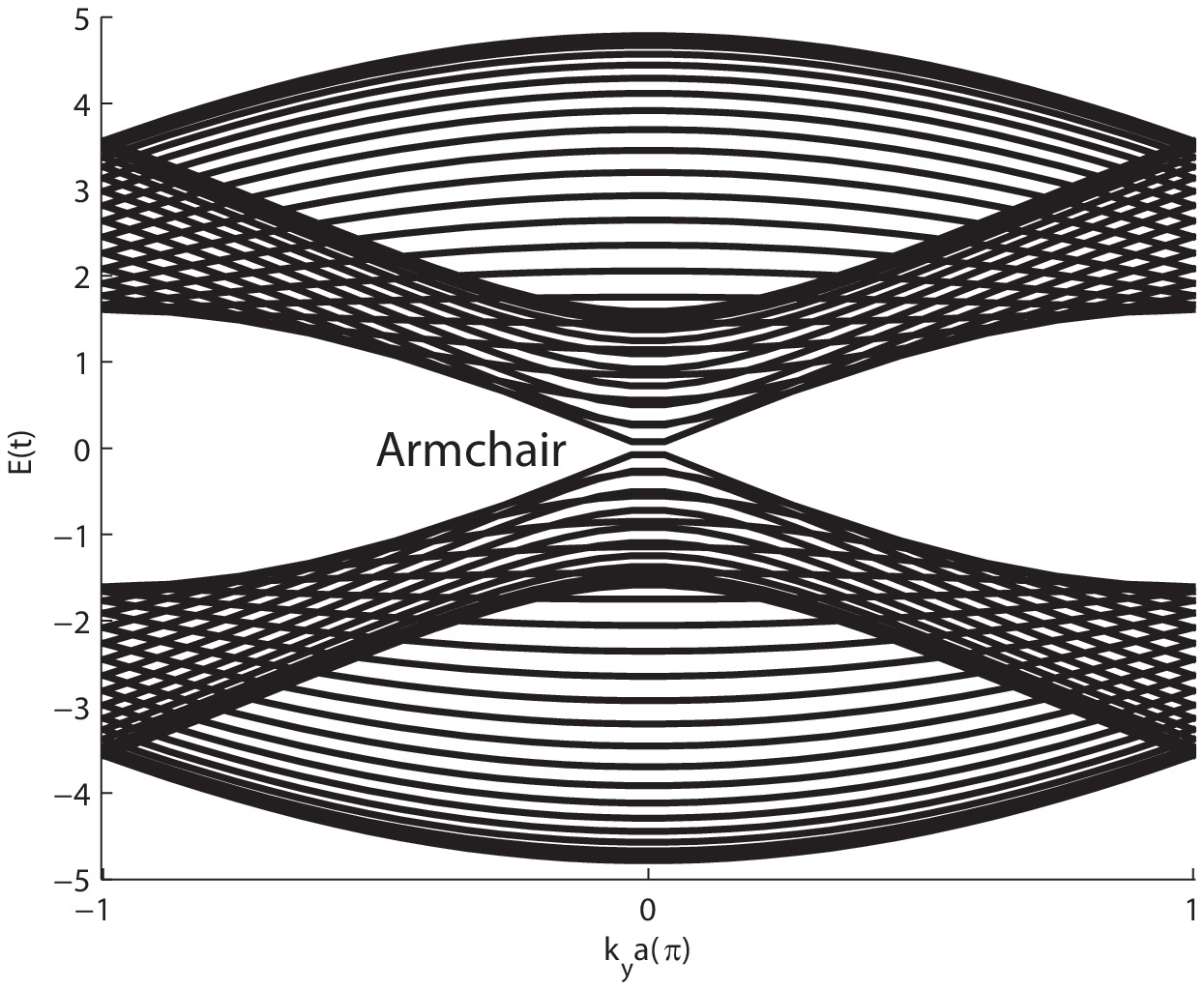}
\label{fig:side:a}
\end{minipage}
}
\subfigure{
\begin{minipage}[t]{0.5\textwidth}
\centering
\includegraphics[width=0.9\linewidth]{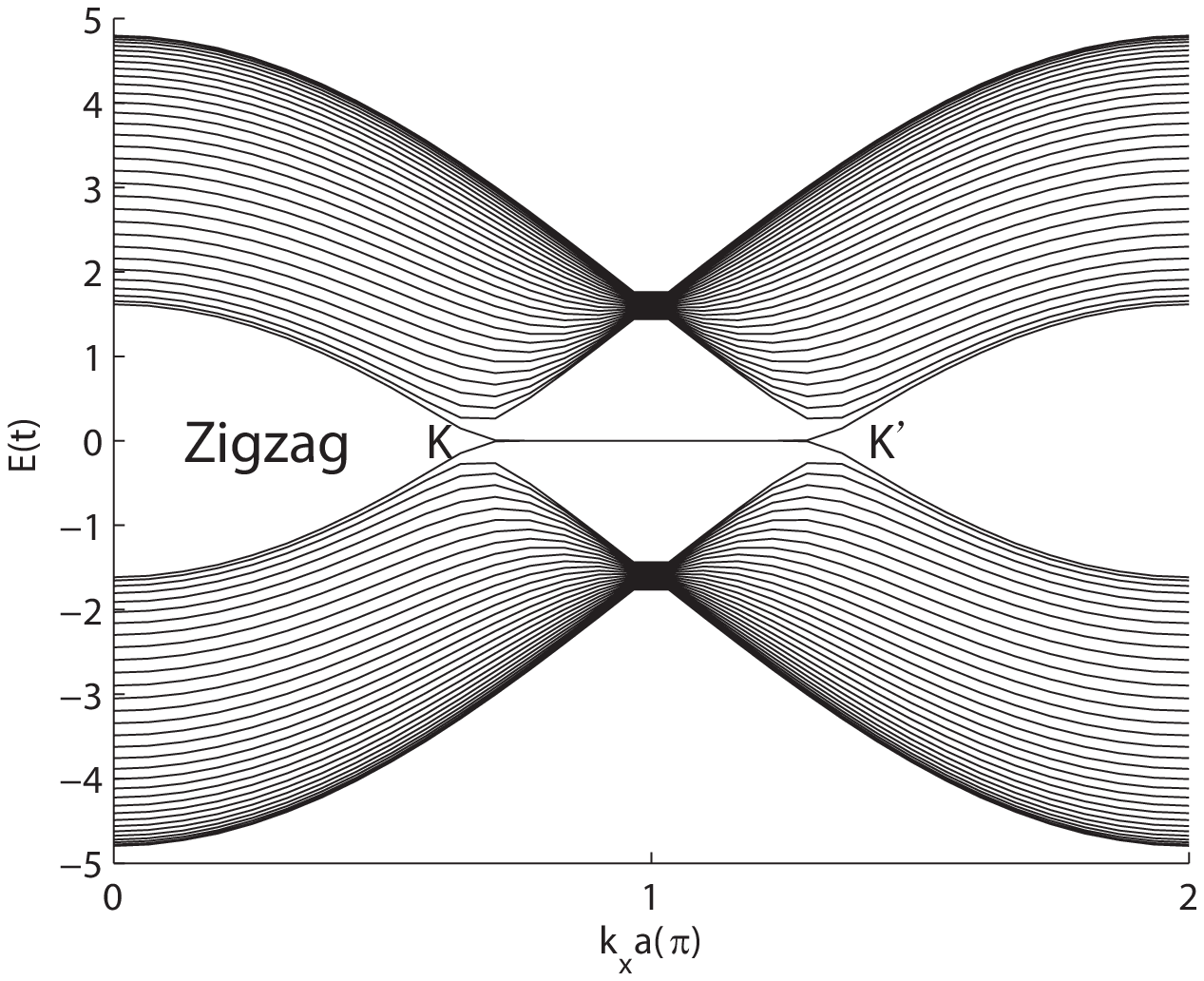}
\label{fig:side:b}
\end{minipage}
}\\
\subfigure{
\begin{minipage}[t]{0.5\textwidth}
\centering
\includegraphics[width=0.9\linewidth]{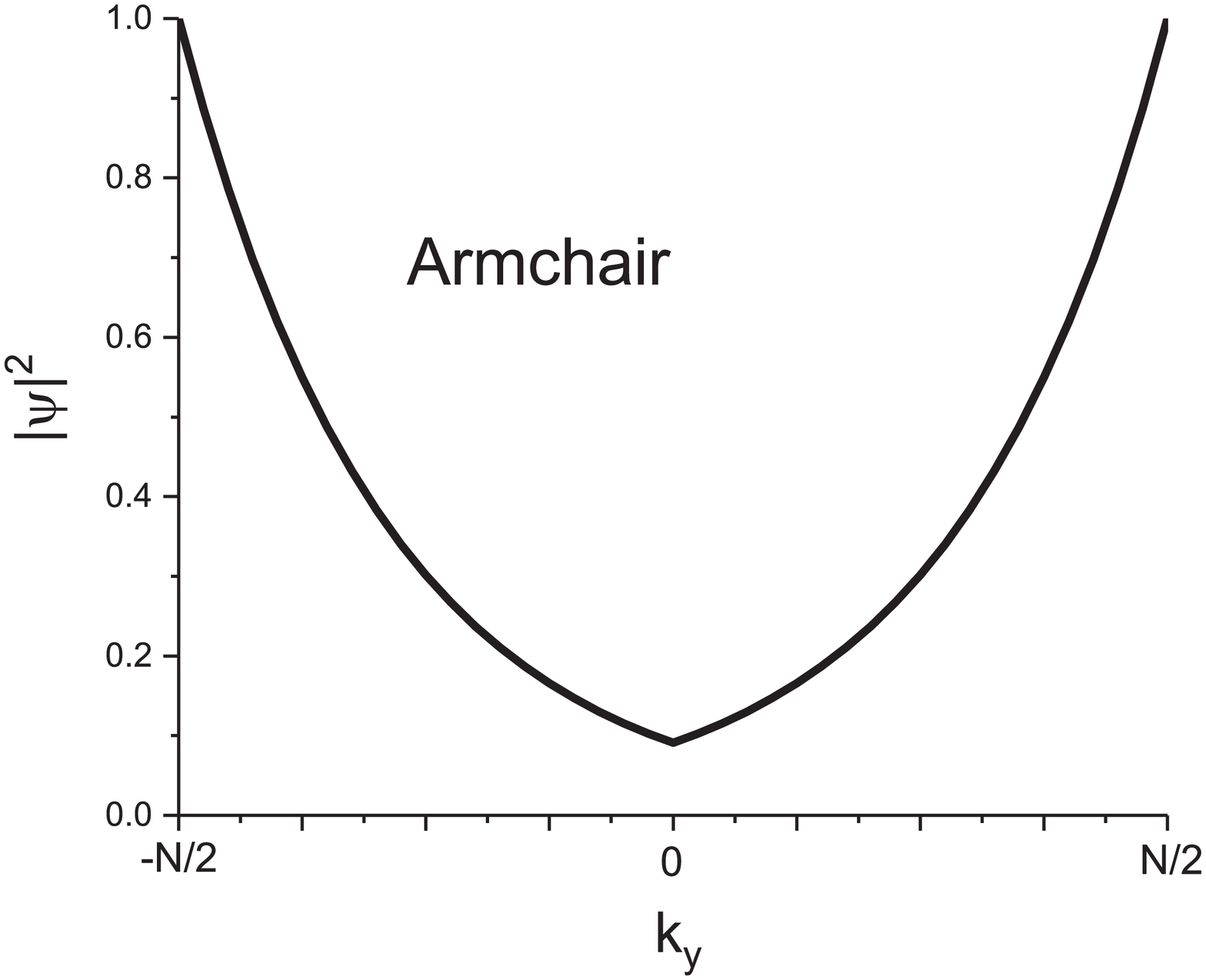}
\label{fig:side:b}
\end{minipage}
}
\caption{
Band structure of the (left) armchair silicene nanoribbon and
(middle) zigzag silicene nanoribbon in a strip geometry.
The energy is in unit of $t$.
The zigzag energy band is shown for the trivial phase with aero $\lambda_{SOC}$.
(right) The amplitude of wave function of the armchair silicene nanoribbon into the negative energy states.
The minimum period is setted as $N=40$ and set
$m_{D}=0.2t=0.32$ eV. The penetration length for the armchair nanoribbon is 
$l_{{\rm arm}}=3.272$ \AA\ and the and for the zigzag nanoribbon is $l_{{\rm zig}}=$0.656 \AA\ .}
\end{figure}

\clearpage
\begin{figure}[!ht]
   \centering
   \begin{center}
     \includegraphics*[width=0.8\linewidth]{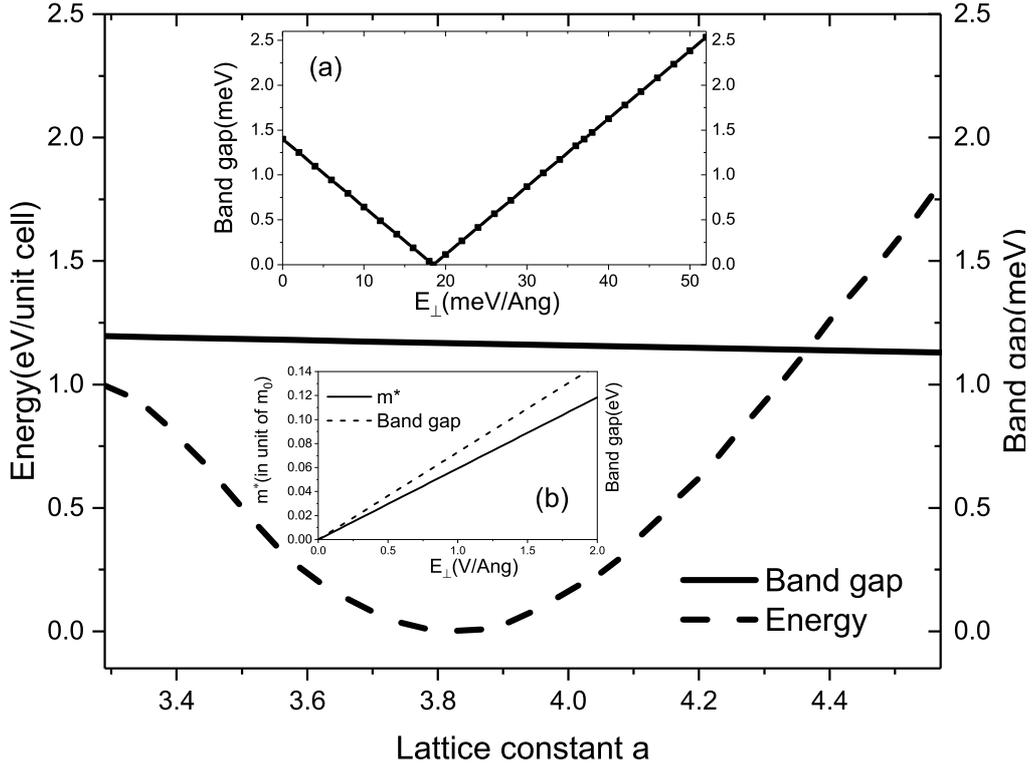}
\caption{The energy per unit cell (solid line) and the band gap (dashed line) as a function of the lattice constant $a$.
(upper inset (a)): The electric field-induced band gap as a function of the electric field $E_{\perp}$
(consider the effect of SOC).
The change of $\overline{\Delta}$ due to $E_{\perp}$ is ignored for it's very small ($<0.1$ \AA),
and hence the linear relation between the effective mass or band gap with the electric-field is obtained.
(lower inset (b)):
The effective mass $m^{*}$ and the electric field-induced band gap as a function of the electric field $E_{\perp}$
(don't consider the effect of SOC).}
   \end{center}
\end{figure}

\clearpage
\begin{figure}[!ht]
\subfigure{
\begin{minipage}[t]{0.6\textwidth}
\includegraphics[width=1\linewidth]{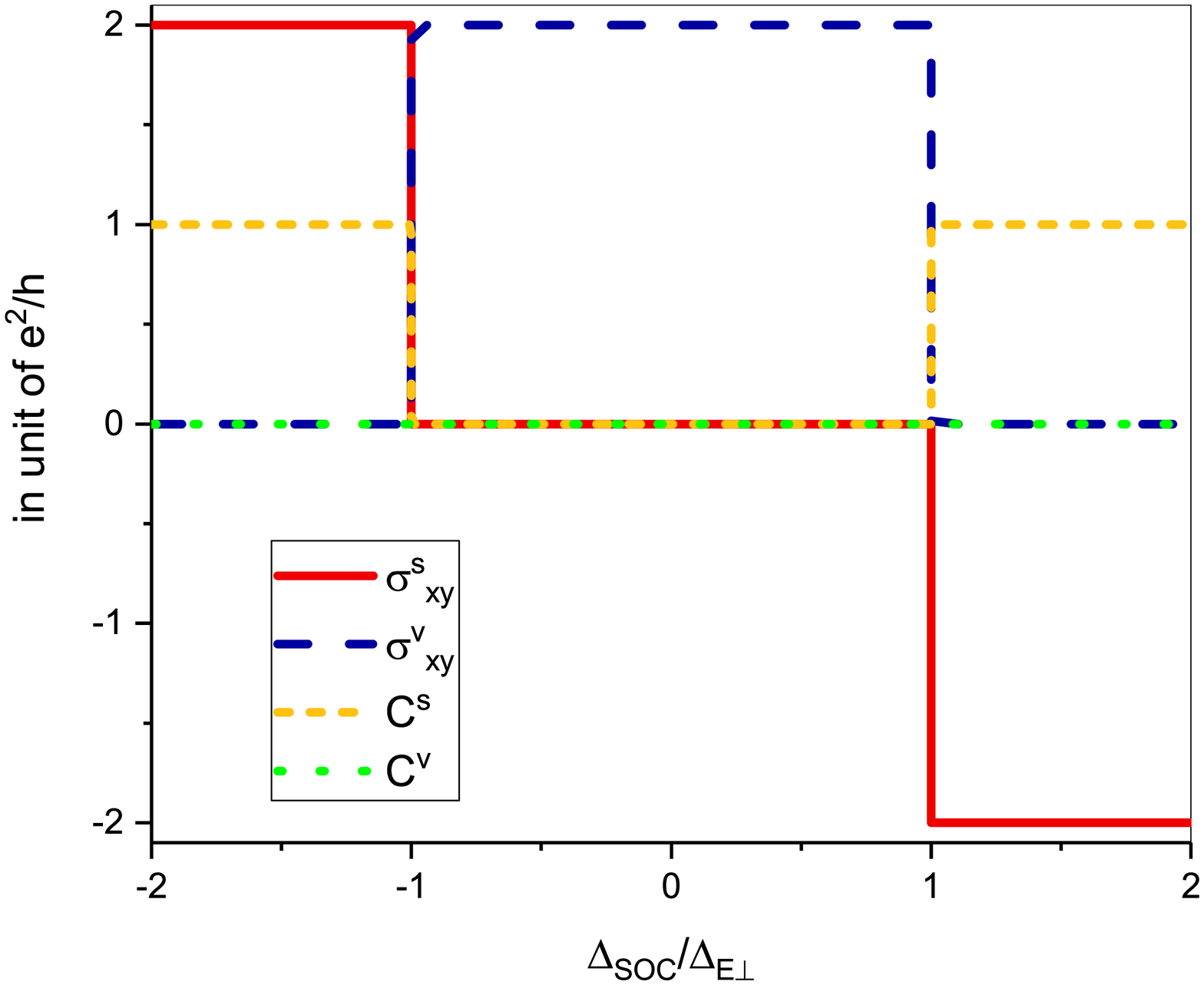}
\label{fig:side:a}
\end{minipage}
}\\
\subfigure{
\begin{minipage}[t]{0.2\textwidth}
\includegraphics[width=1\linewidth]{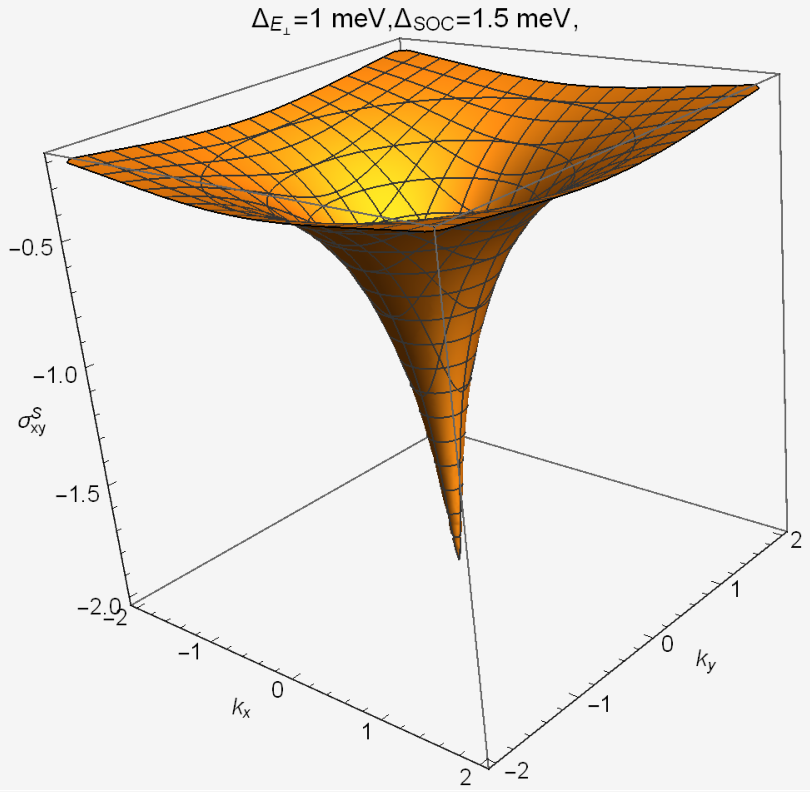}
\label{fig:side:a}
\end{minipage}
}
\subfigure{
\begin{minipage}[t]{0.2\textwidth}
\includegraphics[width=1\linewidth]{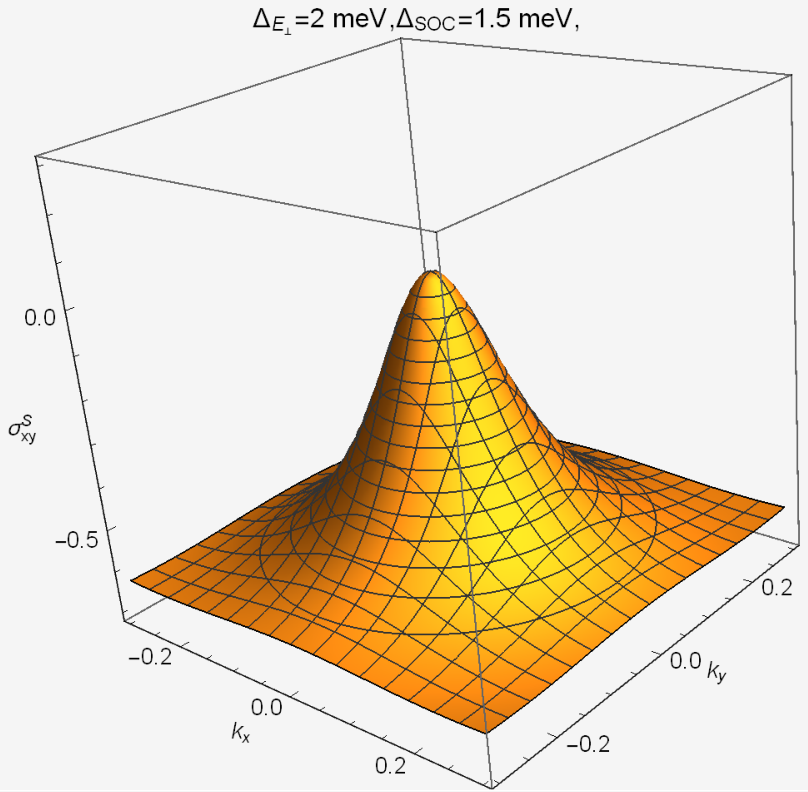}
\label{fig:side:a}
\end{minipage}
}
\subfigure{
\begin{minipage}[t]{0.2\textwidth}
\includegraphics[width=1\linewidth]{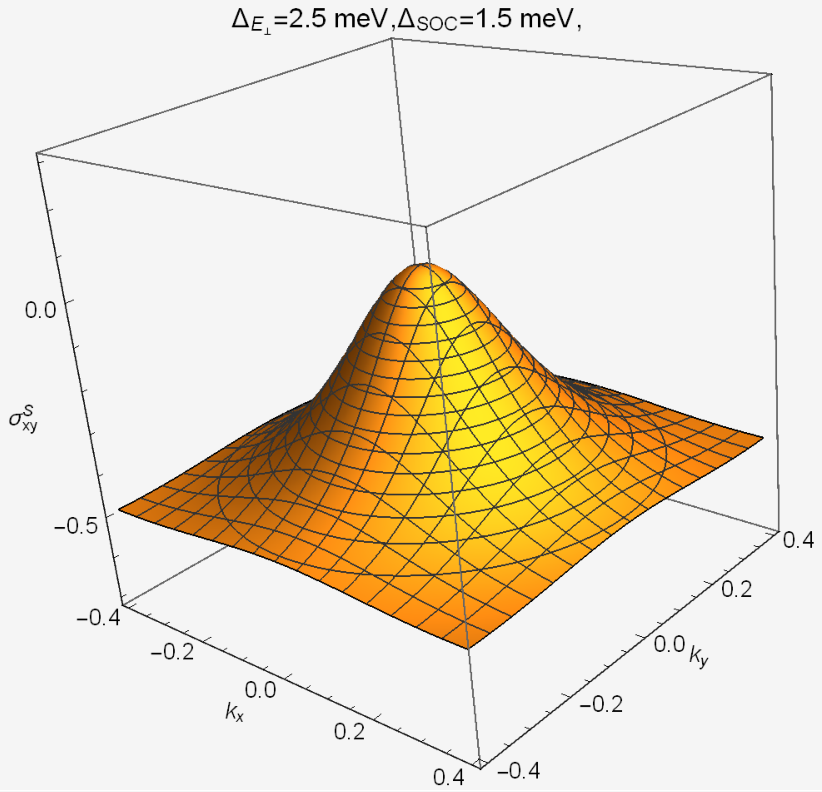}
\label{fig:side:b}
\end{minipage}
}\\
\subfigure{
\begin{minipage}[t]{0.2\textwidth}
\includegraphics[width=1\linewidth]{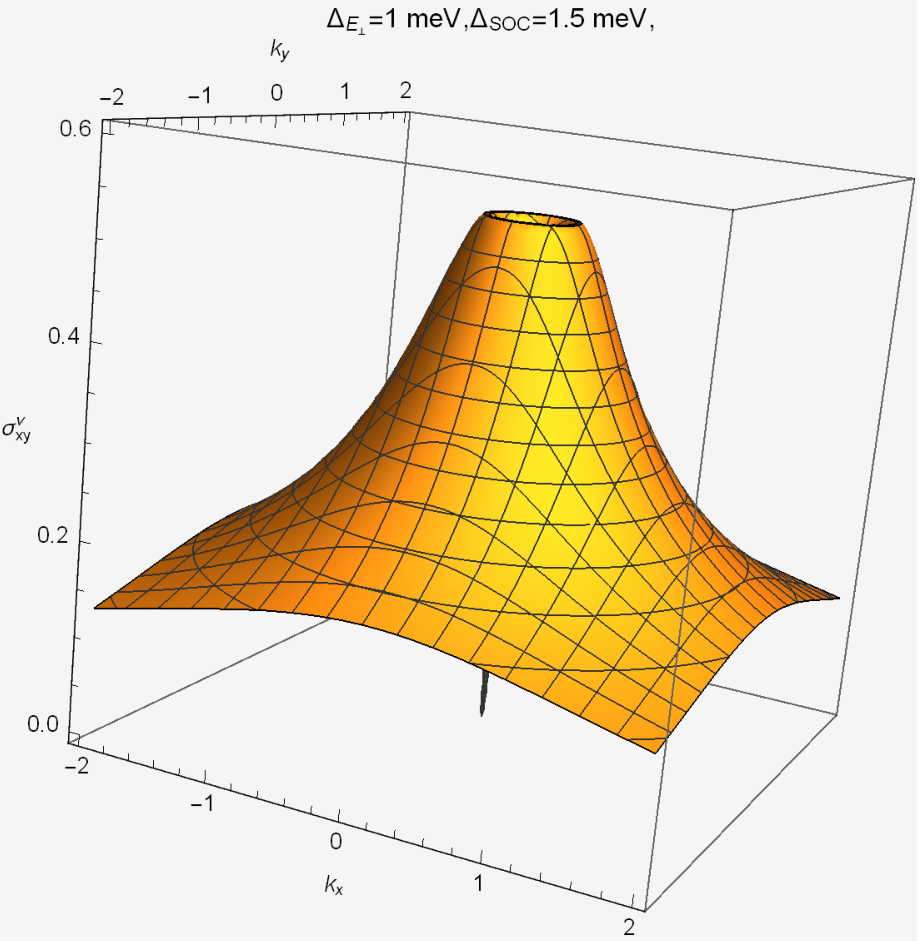}
\label{fig:side:a}
\end{minipage}
}
\subfigure{
\begin{minipage}[t]{0.2\textwidth}
\includegraphics[width=1\linewidth]{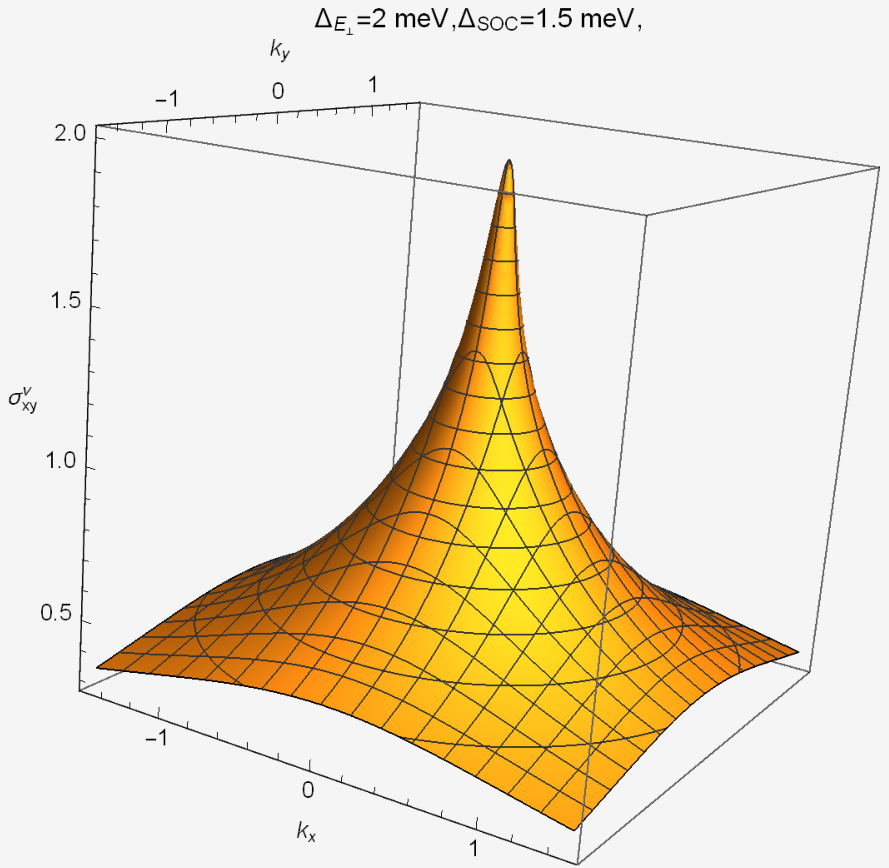}
\label{fig:side:a}
\end{minipage}
}
\subfigure{
\begin{minipage}[t]{0.2\textwidth}
\includegraphics[width=1\linewidth]{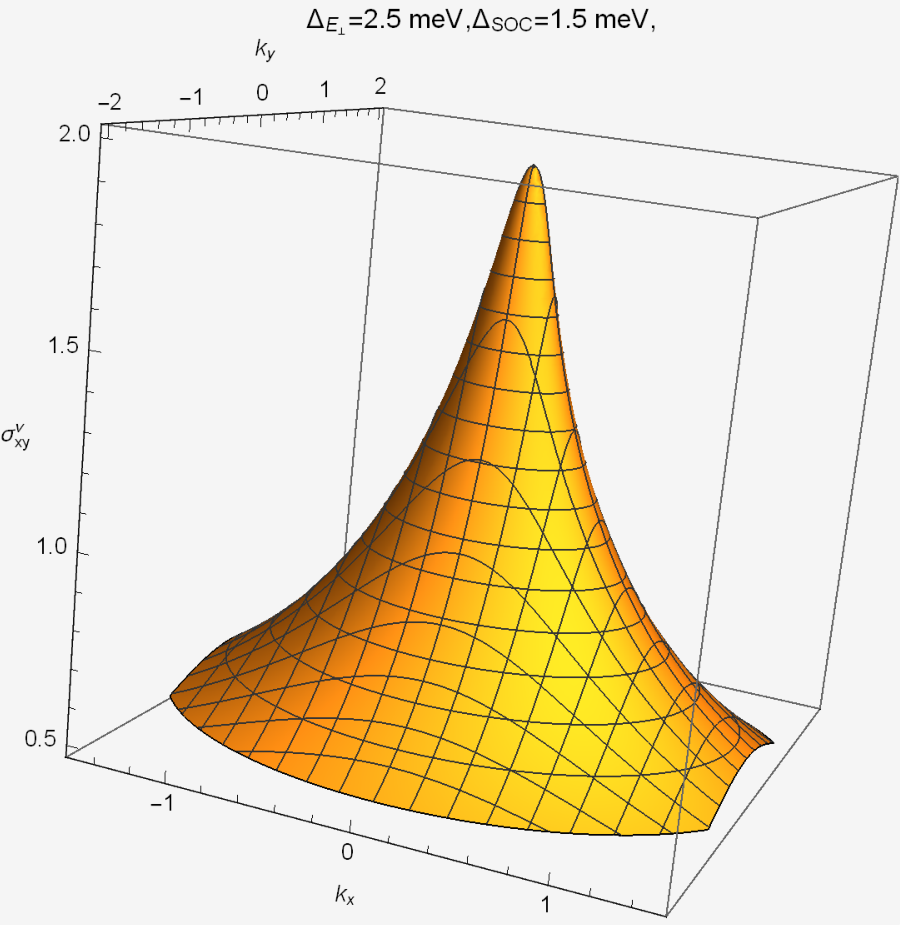}
\label{fig:side:b}
\end{minipage}
}
\caption{(Color online) Spin and valley Hall conductivity and the corresponding spin and valley Chern number $\mathcal{C}^{s}$ and $\mathcal{C}^{v}$ at zero temperature.
Top panel is for the case that the Fermi level in within the band gap.
The six subgraphs in the lower panel are the spin and valley Hall conductivity in units of $e^{2}/h$ with Fermi level within the conduction band
and for the case of $\Delta_{E_{\perp}}=1,2,2.5$ meV. The $\hbar v_{F}$ is setted as 5.34 eV\AA\ here.}
\end{figure}

\clearpage

\begin{figure}[!ht]
   \centering
   \begin{center}
     \includegraphics*[width=0.8\linewidth]{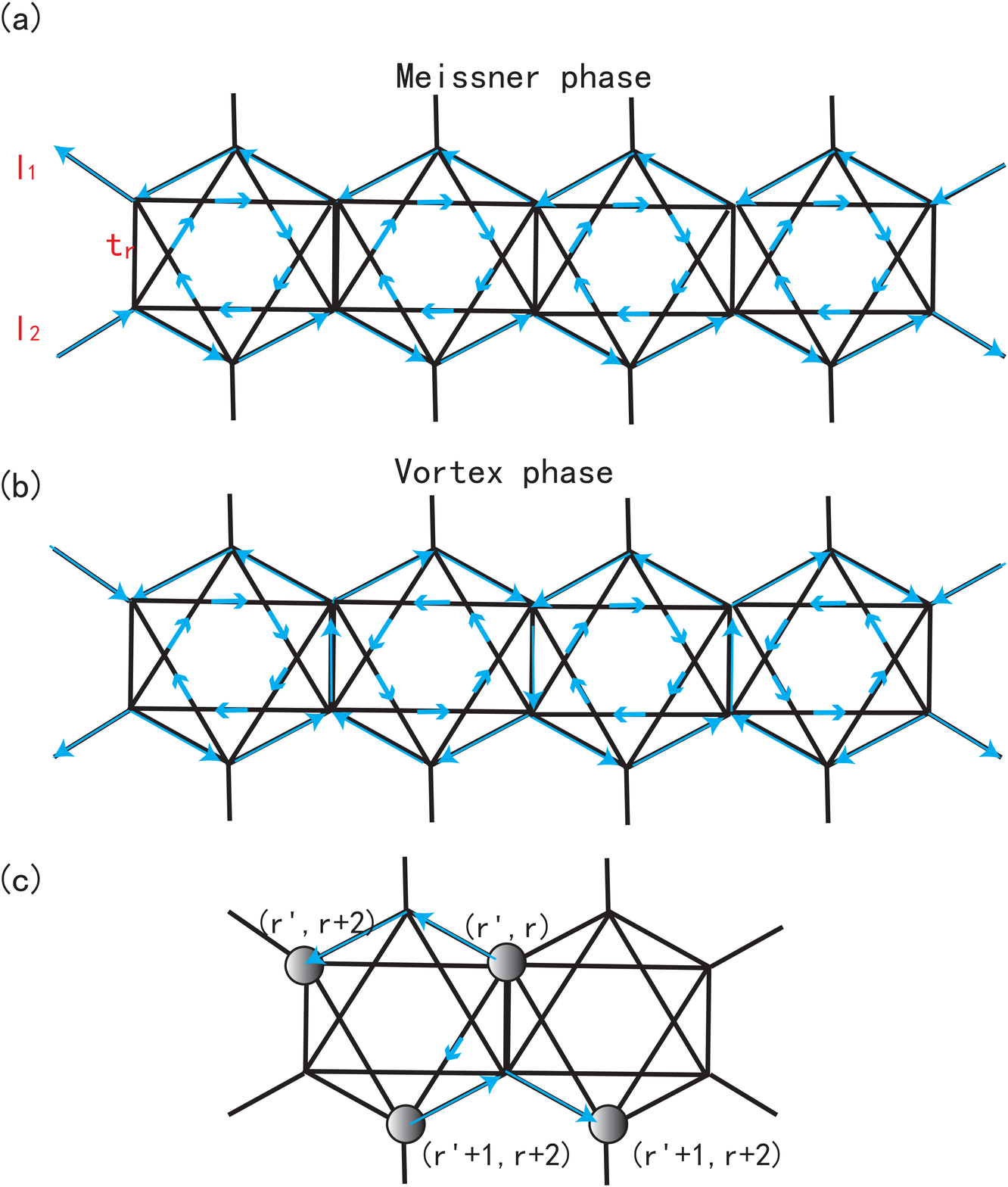}
\caption{(Color online)
Hopping currence in Meissner phase and vortex phase along the zigzag edges with open boundary condition under the artificial magnetic field.}
   \end{center}
\end{figure}

\clearpage

caption:(Color online)
Plots of the orbital magnetic moment (left-top panel) and the Berry curvature (right-top panel and bottom panels) for the conduction band in valley K'
(the another valley K which is antisymmetry with K is not shown).
The datas adoped are as following: the peak of $m(k)$ is center in the $k_{x}=2.092\ \pi/a$, $t=1.6$ eV, band gap $\Delta=0.01$ eV and $\Delta=0.2192$ eV
and $\Delta=-0.2192$ eV,
chemical potential $\mu=0.1$ eV. The unit of $\hbar=c=1$ are used.

\clearpage

\begin{figure}[!ht]
   \centering
   \begin{center}
     \includegraphics*[width=0.8\linewidth]{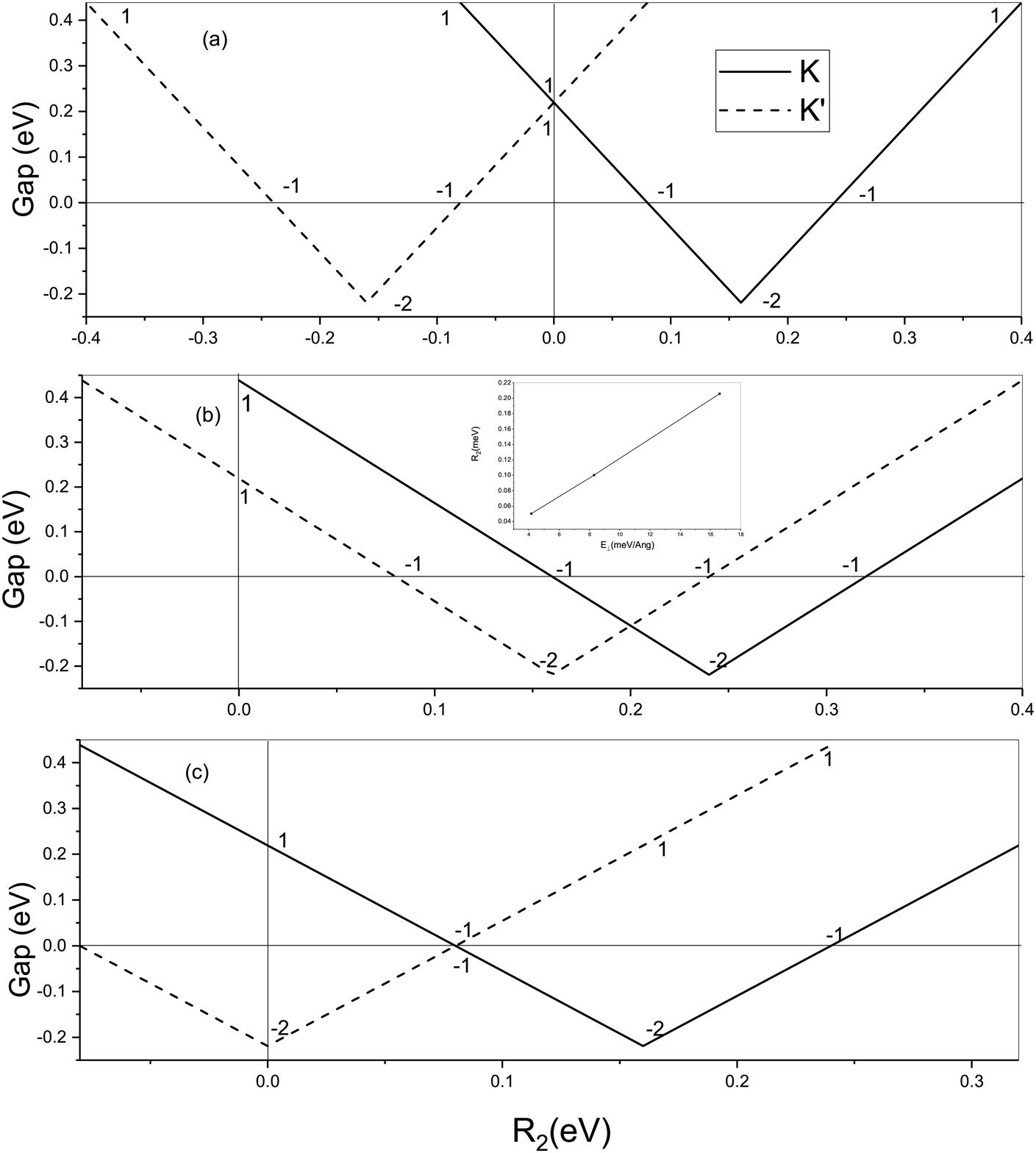}
\caption{The band gap as function of the electric-field-induced Rashba-coupling $R_{2}(E_{\perp})$ in several configurations.
Here the $R$ is setted as 0.7 meV.
The Chern-number of each valley $\mathcal{C}_{K}$ and $\mathcal{C}_{K'}$ are labeled in the plots.
The inset in the middle panel is mearsured $R_{2}(E_{\perp})$ as a function of the electric field and it's clearly to see the linear-dependence
with the slope as 0.012.}
   \end{center}
\end{figure}

\clearpage

\begin{figure}[!ht]
   \centering
   \begin{center}
     \includegraphics*[width=0.8\linewidth]{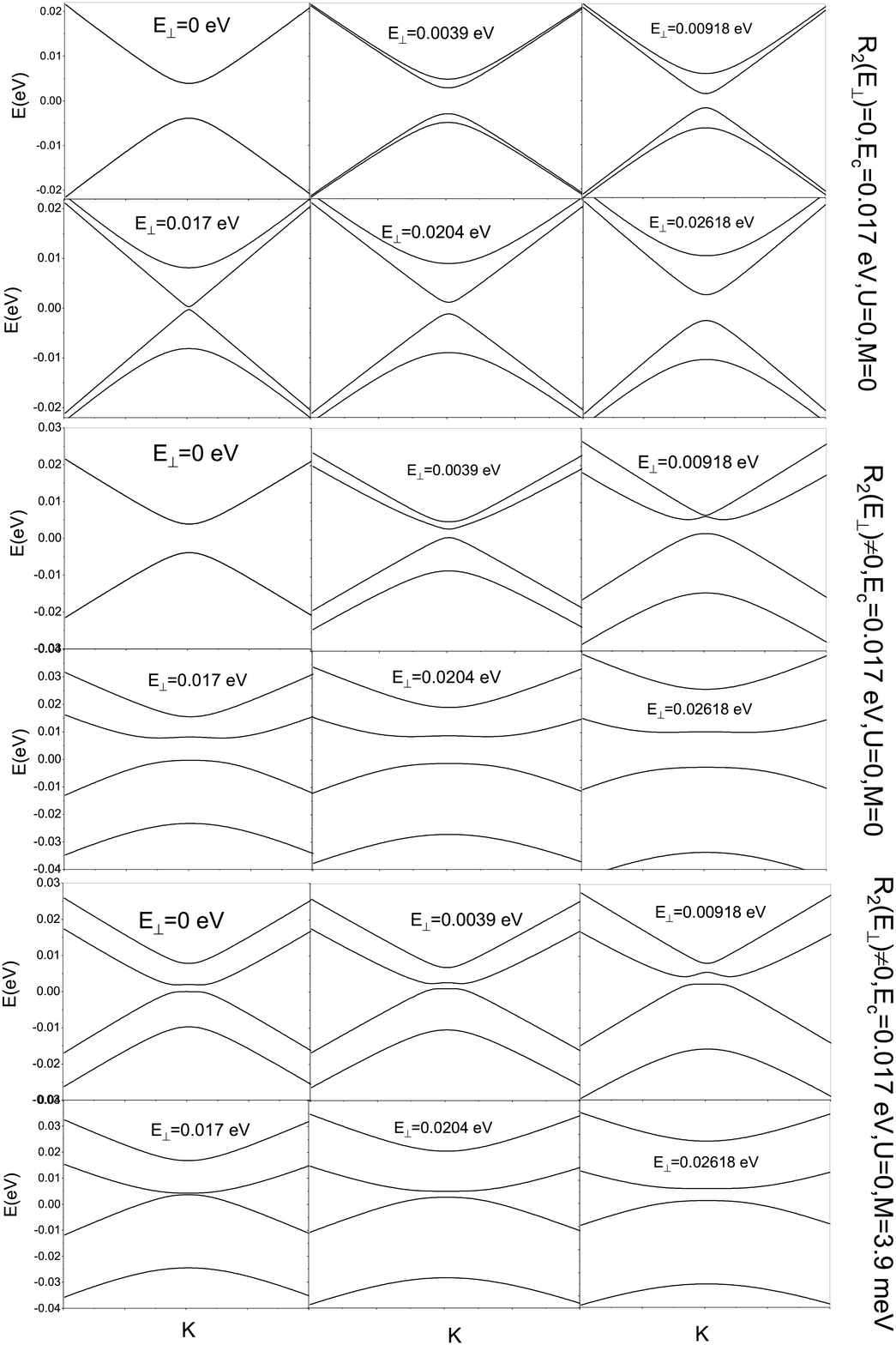}
\caption{Band gap evolution in valley K for the silicene nanoribbon under the effect of electric field and effective SOC.
The electric field-induced Rashba coupling is consider in the lower panel,
we can see obvious differece from the upper one.
The on-site interaction $U$ is setted as zero for simplicity here,
and the critical electric field is arounds 17 mev/\AA\ .
There are three groups from top to the bottom and with different conditions labeled in the right side.
For the first and second groups,
The divided branches of energy bands which close to the $E=0$ level is corresponds to the spin-up component, while the
other one corresponds to the spin-down one.}
   \end{center}
\end{figure}

\clearpage

\begin{figure}[!ht]
   \centering
   \begin{center}
     \includegraphics*[width=0.8\linewidth]{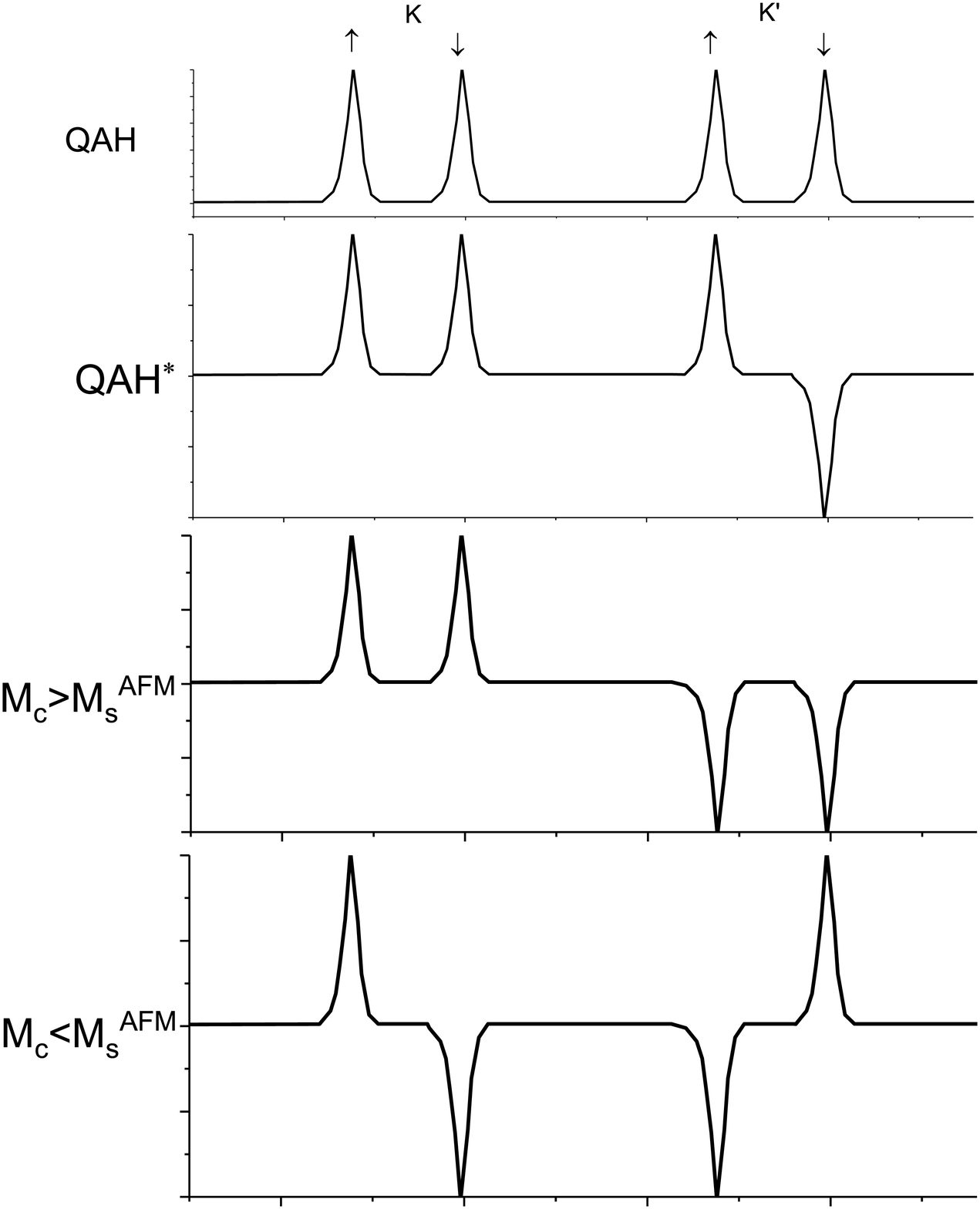}
  \caption{Berry curvature at two valleys for the QAH phase, QAH$^{*}$ phase, and the trivial band insulator dominated by the $M_{c}$ and $M^{*}_{s}$, respectively,
from top to bottom.}	
   \end{center}
\end{figure}

\clearpage
\begin{figure}[!ht]
   \centering
   \begin{center}
     \includegraphics*[width=0.8\linewidth]{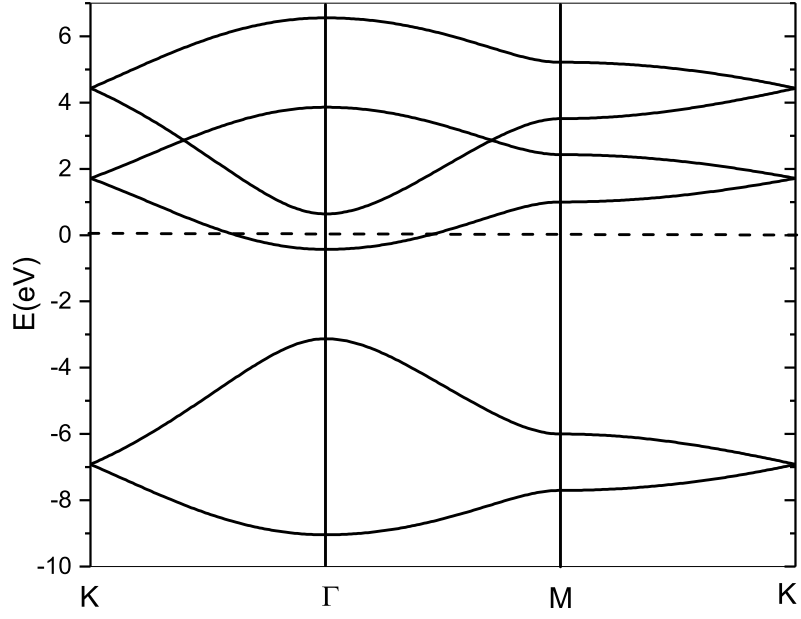}
\caption{Band structure of the silicene for the projected surface BZ (the irreducible one).
The contributions from both the $\pi$-band and $\sigma$-band are considered.
The follwing data are used:
$V^{(1)}_{ss\sigma}=-2.075$, $V^{(1)}_{sp\sigma}=2.48$, $V^{(1)}_{pp\sigma}=2.7163$, $V^{(1)}_{pp\pi}=-0.715$, 
$\varepsilon_{s}=-4.2$, $\varepsilon_{p}=1.715$.}
   \end{center}
\end{figure}

\clearpage

caption:(Color online)
(a)The band structure of low-buckled silicene using the ultrasoft pseudopotentials with a band gap of 0.000545 meV, and
(b) is the PDOS corresponds to the band structure of (a).
The orbitals characters together with the valence $\pi$ and conductivity $\pi^{*}$ bands are marked in (a).
(c)The band structure of low-buckled silicene with LDA+SOC (with a band gap as 1.5 meV) using the norm-conserving pseudopotentials.
(d)The band structure of metallic-FM-state low-buckled silicene with LDA+U (U=1.48 eV and $\mu=1.7\mu_{B}$) using the ultrasoft pseudopotentials. 
(e)The same as (a) but for the planar silicene and with a band gap of 0.000272 meV.
(f)The same as (c) but for the planar silicene and with band gap as 0.142 meV.
(g)The same as (d) but for the planar silicene.
(h)-(j)The same as (e)-(g) but for the high-buckled silicene.
The magnetic moment used here is $1.7\mu_{B}$, but
we comment that the experiment reported in Ref.\cite{Zheng F} reveals that for 
the half-hydrogenated silicene and half-brominated silicene with a lower magnetic moment as $1\mu_{B}$
per unit cell.
%

\clearpage

caption(Color online)
Band structure and PDOS of low-buckled silicene with $50\%$ H-doping (a) and $25\%$ H-doping (b) and
$50\%$ S-doping (c).
The schematic diagram of these three doping-lattice are shown in (d), with
the gray atom stands SI, blue atom stands H, and orange atom stands S.

\clearpage

caption:Phonon spectrum (dispersion) obtained by the method of the energetics of finite displacements along the mode
eigenvectors for the undoped silicene in zero external field for the buckled silicene (left) and the planar silicene (right).

\clearpage
\begin{figure}[!ht]
   \centering
   \begin{center}
     \includegraphics*[width=0.8\linewidth]{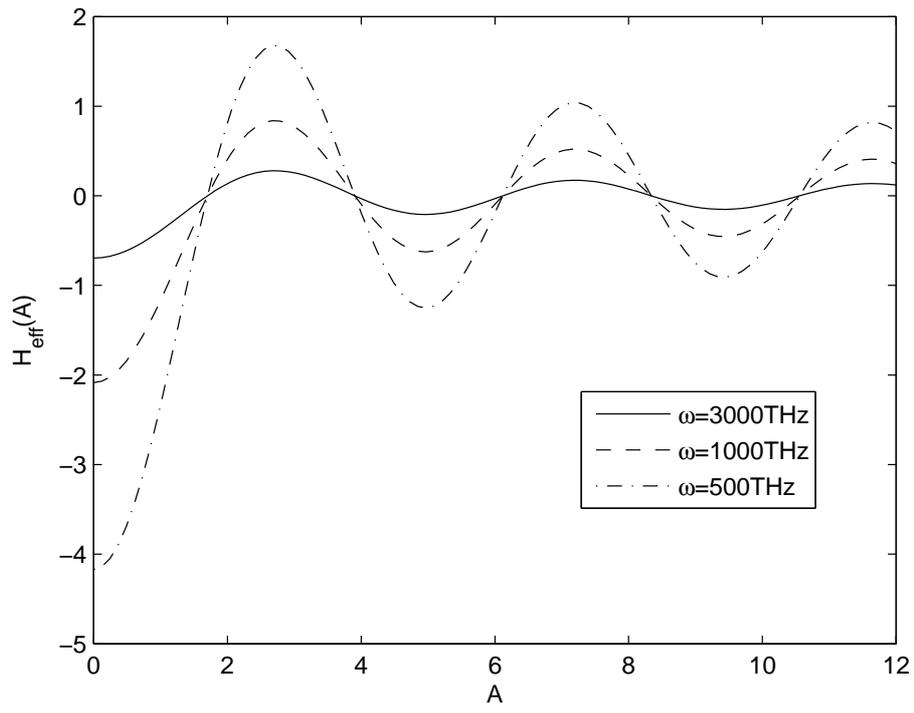}
\caption{The effective Hamiltonian given by the time-dependent periodic vector potential with the different light frequency.}
   \end{center}
\end{figure}

\clearpage

caption(Color online)Optical propertices of monolayer buckled silicene.
(a)optical absorbtion, (b)right-polarized optical conductivity (which consist of the longitudinal conductivity (diagonal) and the transverse one (non-diagonal) as 
$\sigma(\omega)=\sigma_{xx}(\omega)+i\sigma_{xy}(\omega)$), 
(c)energy loss function, (d)dielectric function, (e)Reflective index, Refractive index and 
Extinction coefficient. (f)The Lorentzian fit of the Raman spectrum of monolayer silicene.
Here we use the optical energy unit $\hbar\omega$ which is considered that much larger
than the thermodynamics one $k_{B}T$ here.

\clearpage
\begin{figure}[!ht]
   \centering
   \begin{center}
     \includegraphics*[width=0.8\linewidth]{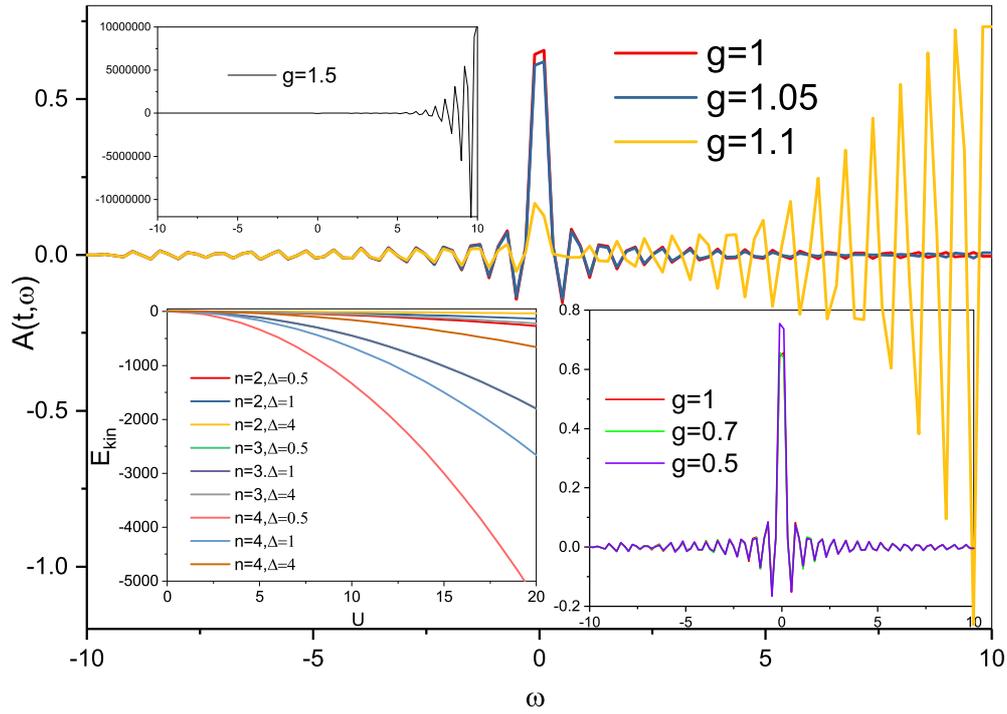}
\caption{(Color online)
The time-frequency-dependent spectral weigth function for different dimensionless reduced coulping $g=UN/t$\cite{Wu C H}.
The inital time is setted as $t_{i}=0$ and finial time setted as $t_{f}=10$.
The left-lower inset shows the kinetic energy as a function of the Hubbard U with different particle number and band gap.}
   \end{center}
\end{figure}

\clearpage
Fig.18
\begin{figure}[!ht]
\begin{minipage}[t]{0.5\textwidth}
\centering
\includegraphics[width=0.9\linewidth]{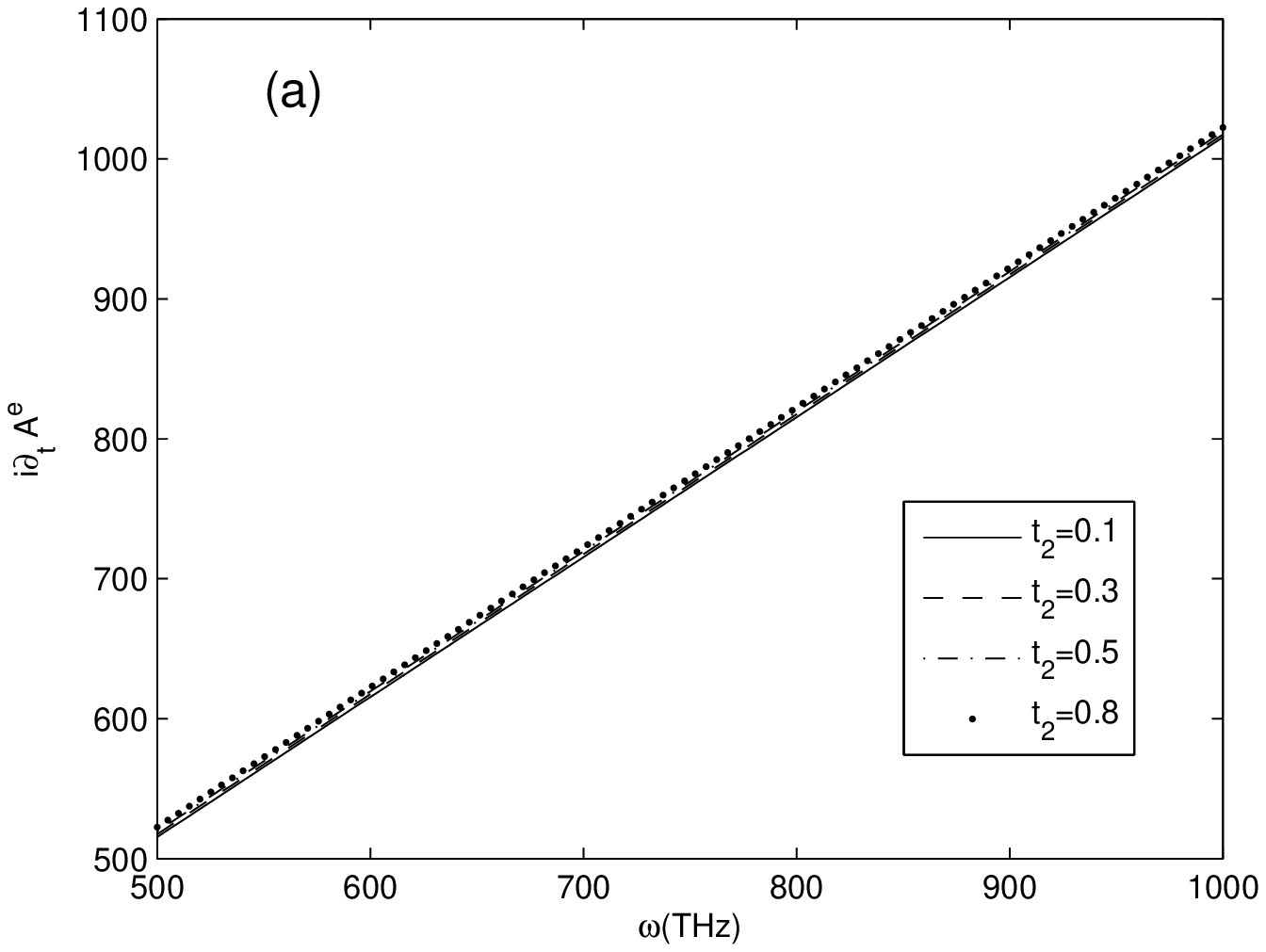}
\label{fig:side:a}
\end{minipage}
\begin{minipage}[t]{0.5\textwidth}
\centering
\includegraphics[width=0.9\linewidth]{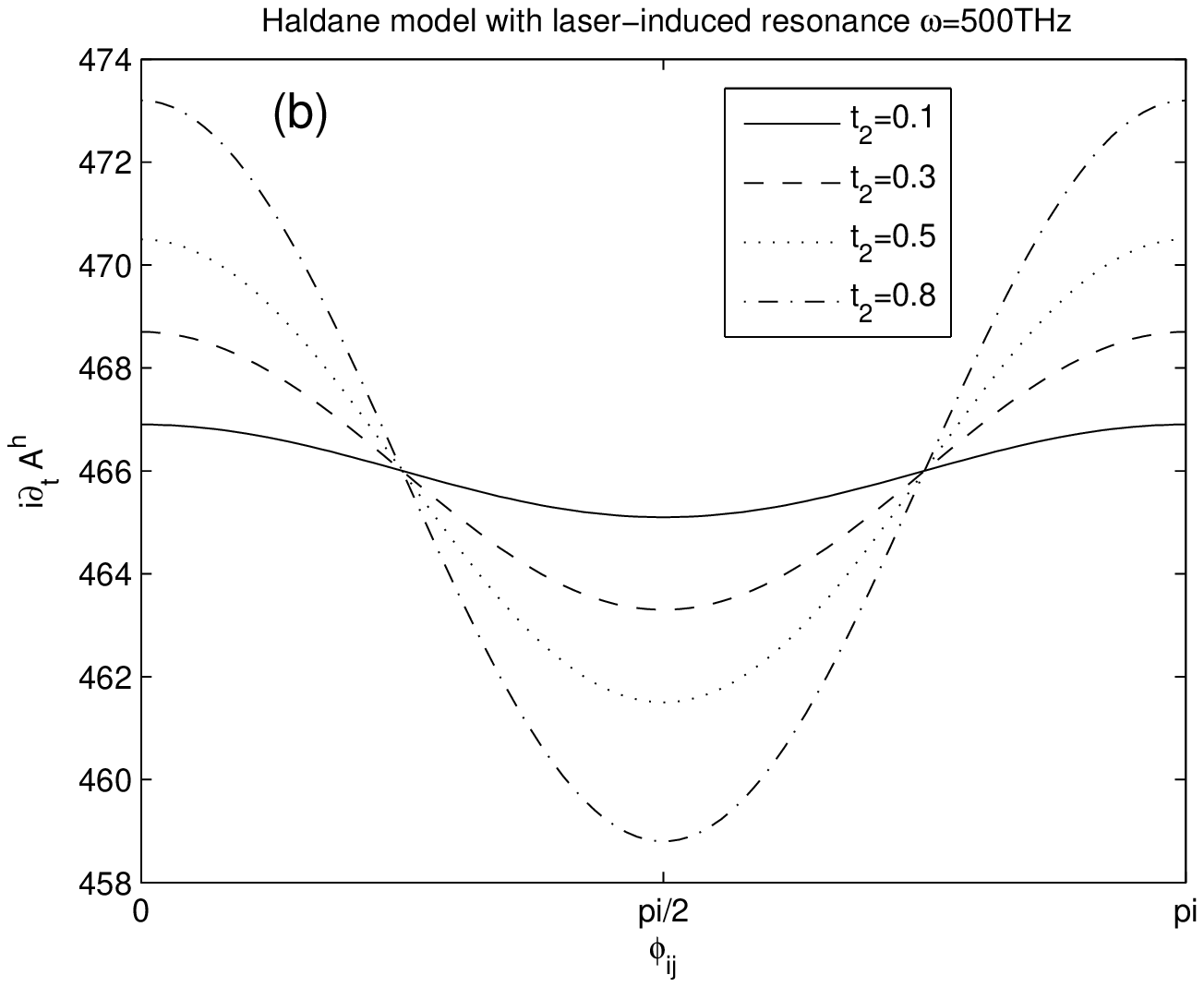}
\label{fig:side:b}
\end{minipage}
\caption{The equation of motion of amplitude for particle (a) and hole (b) 
in Haldane model in non-linear TI with laser-indeced on-site resonance frequency $\omega=500$ THz
with different $t_{2}$.
Here the next-nearest-neighbor hopping is complex as $t'=t_{2}e^{-i\phi_{ij}}$ and $t=1.6$ eV. 
In (b), a phase transition happen from the trivial band insulator (Haldane flux $\phi_{ij}=0,\pi$) to the TI (Haldane flux $0<\phi_{ij}<\pi$).}
\end{figure}

Fig.19
\begin{figure}[!ht]
\subfigure{
\begin{minipage}[t]{0.5\textwidth}
\centering
\includegraphics[width=0.9\linewidth]{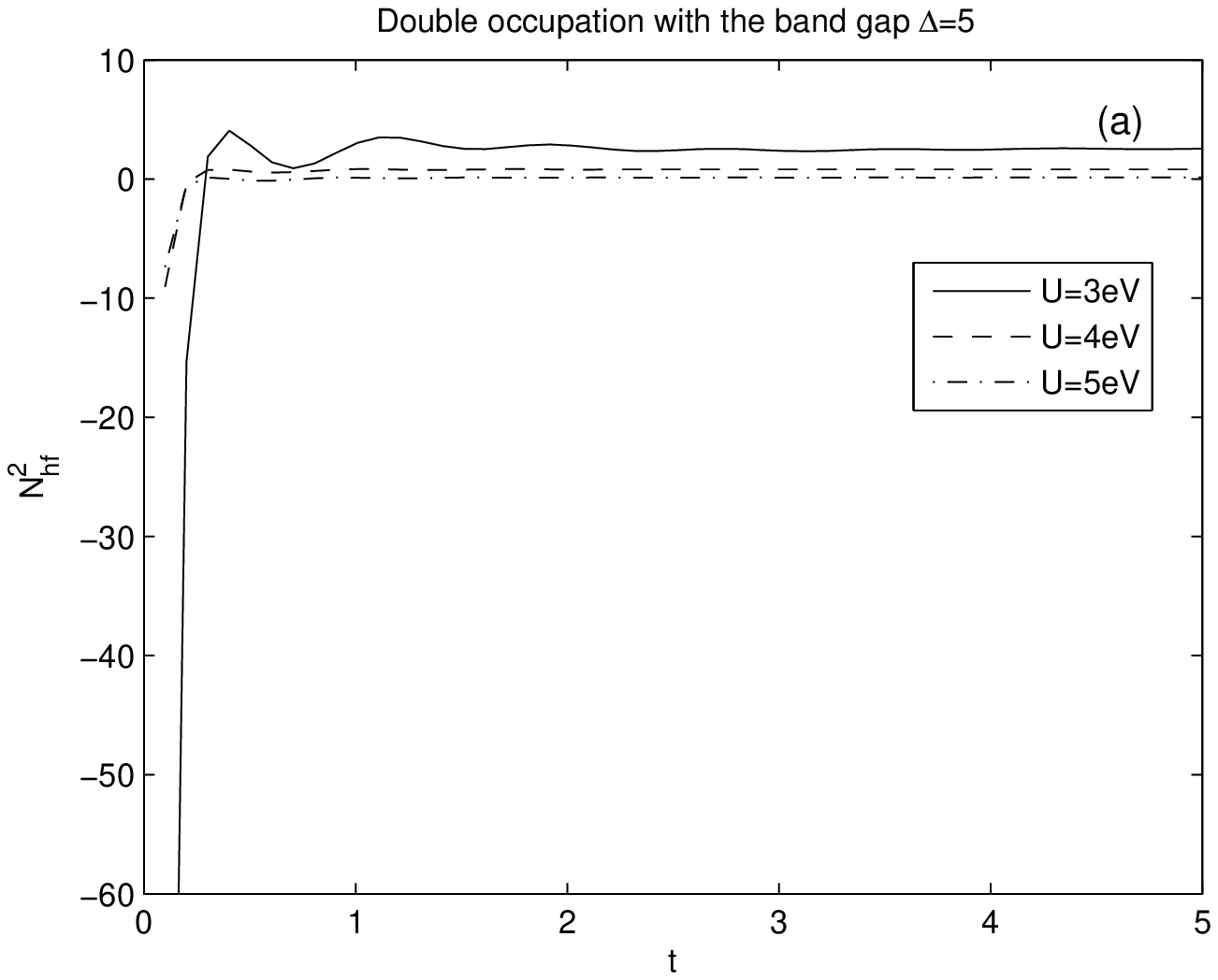}
\label{fig:side:a}
\end{minipage}
}
\subfigure{
\begin{minipage}[t]{0.5\textwidth}
\centering
\includegraphics[width=0.9\linewidth]{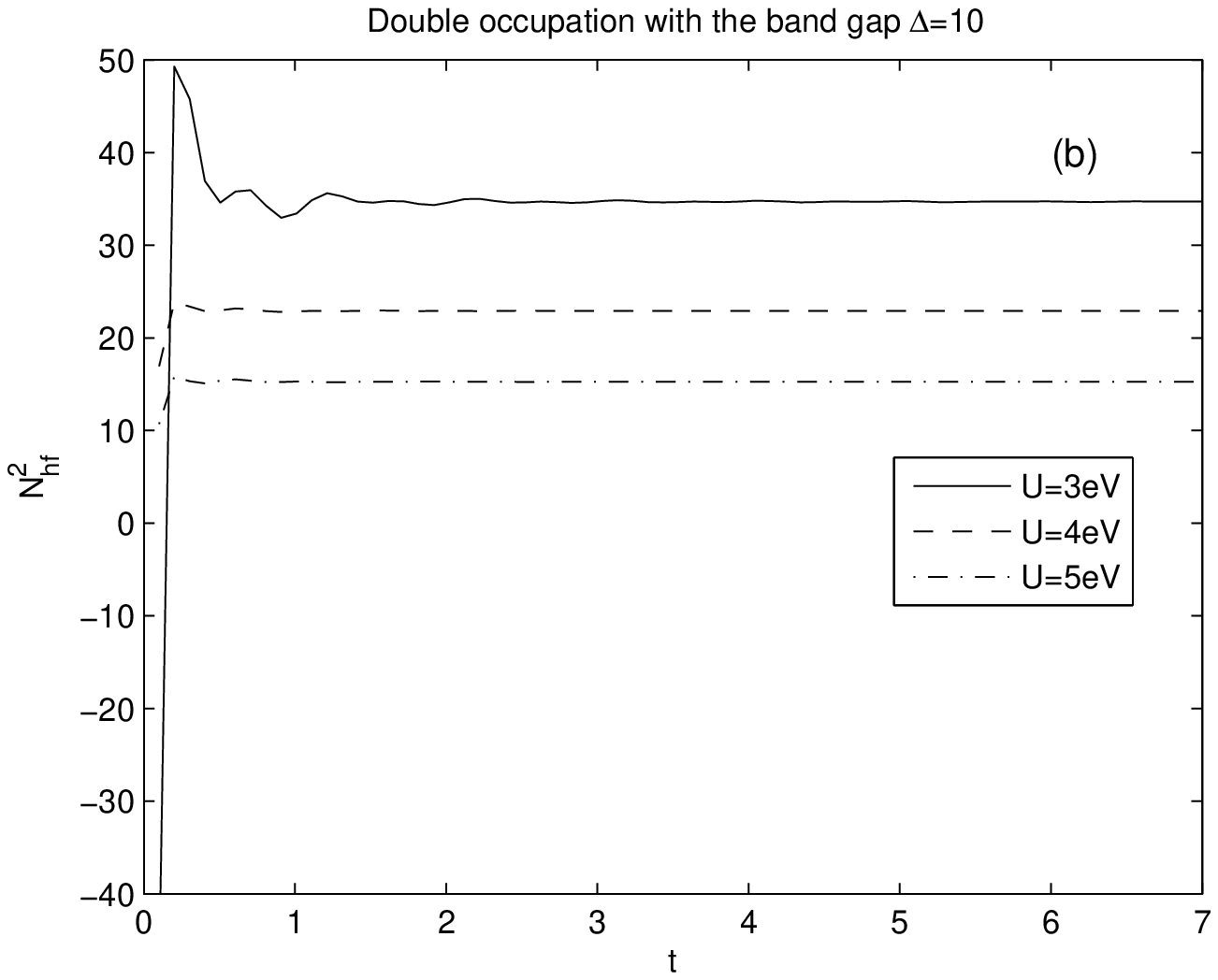}
\label{fig:side:b}
\end{minipage}
}
\subfigure{
\begin{minipage}[t]{0.5\textwidth}
\centering
\includegraphics[width=0.9\linewidth]{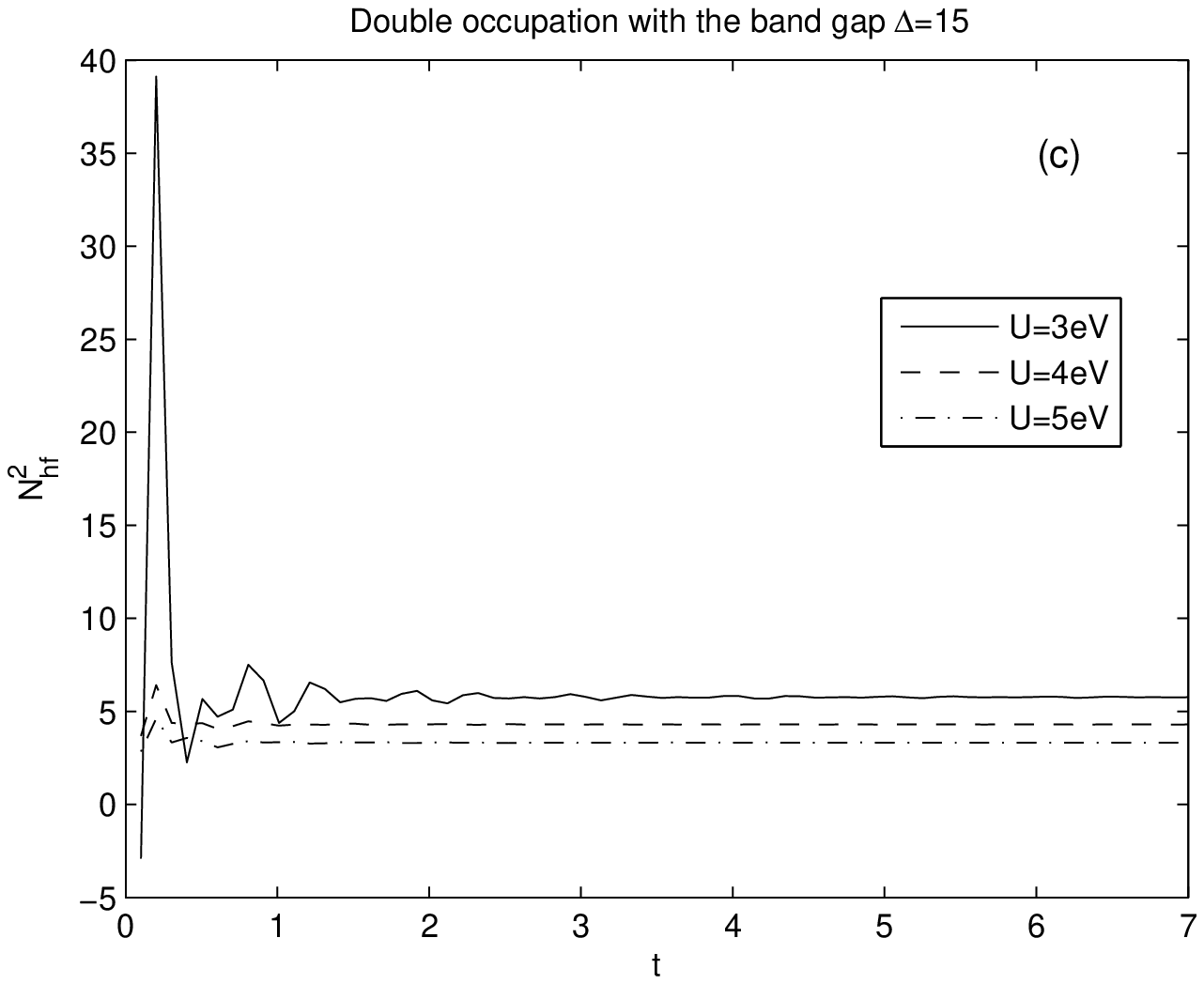}
\label{fig:side:b}
\end{minipage}
}
\subfigure{
\begin{minipage}[t]{0.5\textwidth}
\centering
\includegraphics[width=0.9\linewidth]{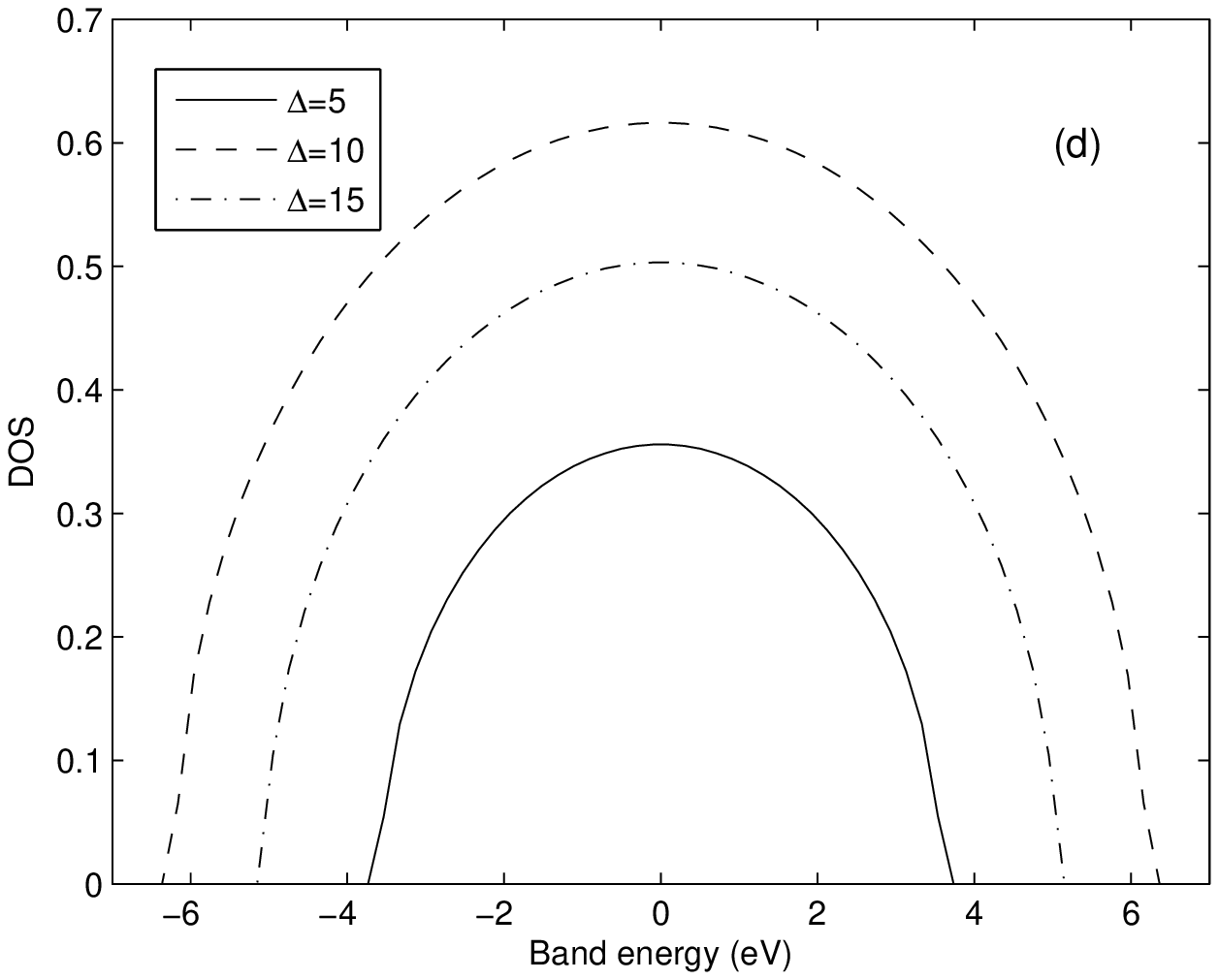}
\label{fig:side:b}
\end{minipage}
}
\caption{The double occupation in $\nu_{f}=1/2$ $(1/r)$-Hubbard model for the on-site interaction quench from 0 to $U$ 
with different bandgaps (bangwidths) $\Delta$ in DMFT (a-c) and the single-site DOS as a function of the time-dependent band energy(d).}
\end{figure}

\clearpage

caption:(Color online)
(left)DOS as a function of the energy where we set $|\lambda_{SOC}+M|=0.1\ {\rm eV},R=0,E_{\perp}=0$.
(right)
The negative orbital susceptibility as a function of the energy spectrum in unit of $m_{D}$
under the zero-temperature limit and a series of finite temperatures.

\end{document}